\documentclass[aip,graphicx]{revtex4-1}

\usepackage{graphicx}
\usepackage{float}
\usepackage{array}
\usepackage{verbatim}
\usepackage{amsmath} 
\usepackage{amssymb}  
\usepackage[utf8]{inputenc}
\usepackage{placeins}
\usepackage{times}
\usepackage{multirow}
\usepackage{amsmath}
\usepackage{graphicx}
\usepackage{lineno}
\usepackage[export]{adjustbox}
\usepackage{amsmath}
\usepackage{epstopdf}
\usepackage{xcolor}
\definecolor{amber}{rgb}{1.0, 0.75, 0.0}
\usepackage[utf8]{inputenc}
\usepackage{placeins}
\usepackage{booktabs,threeparttable}
\usepackage{lipsum}
\usepackage{times}
\usepackage{multirow}
\usepackage{graphicx}
\usepackage{lineno}
\usepackage{bm}
\usepackage[export]{adjustbox}
\usepackage{amsmath,amssymb}
\usepackage{epstopdf}
\usepackage{xcolor}
\definecolor{amber}{rgb}{1.0, 0.75, 0.0}
\definecolor{brickred}{rgb}{0.7960, 0.2550, 0.3290}
\definecolor{dartmouthgreen}{rgb}{0.05, 0.5, 0.06}

\newcommand{\subfigimgtwo}[3][,]{%
  \setbox1=\hbox{\includegraphics[#1]{#3}}
  \leavevmode\rlap{\usebox1}
  \rlap{\hspace*{-5pt}\raisebox{\dimexpr\ht1-1\baselineskip}{#2}}
  \phantom{\usebox1}
}

\usepackage{makecell}
\usepackage{tikz}

\DeclareMathOperator*{\argmax}{\arg\!\max}
\DeclareMathOperator{\EX}{\mathbb{E}}

\begin{document}


\title{Wind farm yaw control set-point optimization under model parameter uncertainty} 



\author{Michael F. Howland}
\email[]{mhowland@mit.edu}
\affiliation{Civil and Environmental Engineering, Massachusetts Institute of Technology, Cambridge, MA 02139}
\affiliation{Graduate Aerospace Laboratories (GALCIT), California Institute of Technology, Pasadena, CA 91125, U.S.A.}


\date{\today}

\begin{abstract}
Wake steering, the intentional yaw misalignment of certain turbines in an array, has demonstrated potential as a wind farm control approach to increase collective power.
Existing algorithms optimize the yaw misalignment angle set-points using steady-state wake models and either deterministic frameworks, or optimizers which account for wind direction and yaw misalignment variability and uncertainty.
Wake models rely on parameterizations of physical phenomena in the mean flow field, such as the wake spreading rate.
The wake model parameters are uncertain and vary in time at a wind farm depending on the atmospheric conditions, including turbulence intensity, stability, shear, veer, and other atmospheric features.
In this study, we develop a yaw set-point optimization approach which includes model parameter uncertainty, in addition to wind condition variability and uncertainty.
The optimization is tested in open-loop control numerical experiments using utility-scale wind farm operational data for which the set-point optimization framework with parameter uncertainty has a statistically significant impact on the wind farm power production for certain wind turbine layouts at low turbulence intensity, but the results are not significant for all layouts considered nor at higher turbulence intensity.
The set-point optimizer is also tested for closed-loop wake steering control of a model wind farm in large eddy simulations of a convective atmospheric boundary layer.
The yaw set-point optimization with model parameter uncertainty improved the robustness of the closed-loop wake steering control to increases in the yaw controller update frequency.
Increases in wind farm power production were not statistically significant due to the high ambient power variability in the turbulent, convective ABL.
\end{abstract}


\maketitle 

\section{Introduction}
\label{sec:intro}

The intentional yaw misalignment of leading wind turbines to deflect the energy deficit wake region away from downwind generators \cite{howland2016wake}, termed wake steering, has emerged as a promising collective control strategy to increase wind farm power production \cite{gebraad2016wind}.
The initial approach to wake steering in field experimental studies leverages open-loop control, where optimal yaw-misalignment set-points are computed offline for each turbine as a function of the low-pass filtered wind conditions and are provided to the wind turbines in a discrete lookup table format \cite{fleming2017field}.
While open-loop wake steering control has demonstrated potential in large eddy simulations (LES) \cite{gebraad2016wind} and field experiments \cite{fleming2017field, fleming2019initial, howland2019wind, doekemeijer2021field} to increase wind farm power production, control methodologies which are designed for variations in the atmospheric boundary layer (ABL) wind conditions require further development to reliably increase annual energy production (AEP)\cite{van2020expert}.

Wind conditions, including wind speed, wind direction, turbulence intensity, and static stability, change as a function of time in the ABL.
During the diurnal evolution of the ABL, modifications to the surface heat flux by solar heating alter the dominant flow mechanisms \cite{wyngaard2010turbulence}.
Further, with fixed boundary conditions, the turbulent nature of the ABL results in chaotic flows which depend on the initial conditions, requiring ensemble averages to converge statistical quantities based on the instantaneous states, including turbine power production.
As a result, even within narrow wind condition bins in an open-loop lookup table, a variety of wind farm power outcomes will occur.
Beyond variability within a wind condition bin, high frequency variations in the wind speed and direction occur above the low-pass filter cut-off frequency and optimal yaw misalignment set-point calculations should consider these variability contributions rather than utilizing deterministic mean wind conditions \cite{quick2017optimization, campagnolo2020wind}.
Aside from natural condition variability, wind turbine sensors are inherently noisy and imperfect\cite{johnson2005controls, fleming2014field}, introducing further wind condition uncertainty in active wake control.

Yaw misalignment set-point optimization is typically performed with steady-state wake models which represent time averaged flow behavior \cite{gebraad2016wind}.
The steady-state wake models estimate $P_\infty$, the infinite time average of the power production of the wind farm with fixed wind conditions, including wind speed and wind direction, and fixed turbine control decisions, including yaw misalignment. 
This modeling approach assumes an inherent scale separation between turbine induced flow adjustment and atmospheric condition changes.
Initial wake steering experiments used steady-state wake models to optimize the yaw set-points with deterministic, fixed wind conditions \cite{gebraad2016wind}.
In a wake steering application, the wind conditions may have high-frequency, turbulent variations and low-frequency atmospheric condition variations.
Such variations, are not directly modeled using a steady-state wake model with deterministic wind conditions, which may introduce model bias when applying a steady-state wake model to low-pass filtered (e.g. 10 minute averaged) wind conditions.

Recent studies have extended model-based yaw set-point optimization to maximize the expected value of the power production given wind condition or turbine control system variations and uncertainty.
Quick {\it et al.} (2017) \cite{quick2017optimization} used a steady-state wake model and formulated the wake steering yaw set-point calculation as an optimization under uncertainty with yaw deviations from the set-point value.
Rott {\it et al.} (2018) \cite{rott2018robust} formulated the set-point calculation as an optimization under wind direction uncertainty (termed robust optimization), rather than with fixed, deterministic incident winds. 
The high-frequency wind direction variations above the yaw controller low-pass filter cut-off (5 minutes in Rott {\it et al.} (2018) \cite{rott2018robust}) were modeled by a Gaussian probability density function based on field measurements.
Simley {\it et al.} (2020) \cite{simley2020design} extended the robust set-point optimization to include natural yaw misalignment variability due to slowly evolving yaw control systems.
Finally, Quick {\it et al.} (2020) \cite{quick2020wake} used polynomial chaos expansion to solve the yaw set-point optimization under uncertainty problem with the addition of stochastic turbulence intensity and shear and found that the uncertain wind direction had the largest impact on the set-point optimization results.

Steady-state wake models introduce a number of assumptions and parameterizations in order to analytically represent time averaged wake behavior.
The power production prediction from the resulting wake models rely on the empirical calibration of the parameters which represent key features of the flow, including the wake spreading rate \cite{niayifar2016analytical}.
However, empirical calibrations using idealized LES or wind tunnel experiments introduce error and uncertainty in field deployments, where the flow physics exhibits different forcings, such as Coriolis forces, stratification, and terrain complexity.
Here, we introduce a difference between wind conditions, such as wind direction and speed, and wake model parameters, such as the wake spreading rate, which parameterize physical phenomena.
The wake model parameters depend on the inflow conditions, but the exact functional dependence is not known.
Wake model parameters have been tuned using LiDAR field data \cite{zhan2020optimal} and neutral ABL LES flow fields\cite{doekemeijer2019tutorial}.
Recent work has optimized the wake model parameters using only wind farm power data and analytic gradients \cite{howland2019wind}, a novel calibration procedure\cite{teng2020calibration}, genetic algorithms \cite{howland2019wake}, and Kalman filtering \cite{howland2020optimal} and demonstrated that assimilating operational wind farm data into the wake model improves its predictive capability.
Zhang \& Zhao (2020)\cite{zhang2020quantification} used sampling to approximate the Bayesian posterior distributions of wake model parameters given LES data as the ground truth.
Using the wake model parameter posteriors, a stochastic wake model based on uncertain model parameters was proposed which improved predictions compared to wake modeling with deterministic model parameters.
Wake model error correction terms have also been proposed \cite{schreiber2020improving} and learned using operational data, which improved wake model predictions.
Yaw set-point optimization which is robust to model parameter uncertainty becomes more critical when applying closed-loop control due to limited statistical averaging and additional wind condition uncertainty \cite{howland2020optimal}.

In this paper, we extend yaw misalignment set-point optimization methods to include model parameter uncertainty, where the wake model is optimized based on a probability distribution of wake model parameters rather than deterministic values.
We first develop a yaw set-point optimization based on stochastic programming\cite{birge2011introduction} including wake model parameter and wind condition uncertainty.
Then we develop a simple methodology to estimate the probability distribution of the uncertain wake model parameters without requiring computationally expensive Bayesian posterior sampling methods such as Markov chain Monte Carlo (MCMC) \cite{brooks2011handbook}, although the yaw set-point optimization can be used with arbitrary parameter distribution estimate methods.
The estimated model parameter distributions are used in tandem with the wind condition distributions in the stochastic programming approach.

To test the performance of the stochastic approach, two distinct numerical experiments are performed.
Using Supervisory Control and Data Acquisition (SCADA) data from a utility-scale wind farm, numerical wake model experiments are performed to represent open-loop wake steering.
The wind farm operational data is used offline to construct yaw-set point lookup tables as a function of the wind conditions for deterministic and optimization under uncertainty.
The approaches are tested using the wake model as a surrogate wind farm, where the model is locally fit to time series SCADA data, to test open-loop control performance.
This experiment can be viewed as an idealized, perfect model setting, where the same wake model is used for yaw set-point lookup table construction and for simulations.
Therefore, there is no inherent model error, only variations in wind conditions and model parameters in the time series.
Additionally, the set-point optimization method is tested using closed-loop wake steering control in LES of an unstable, convective ABL.
The LES has inherent wind condition variations due to the turbulent, convective ABL.
These experiments include potential model discrepancy when applying the steady-state wake model to predict LES power production, compared to the perfect model setting open-loop experiments.

The contributions of this paper are:
\begin{enumerate}
    \item The extension of robust wake steering yaw misalignment set-point optimization\cite{quick2017optimization,rott2018robust, simley2020design} to include wake model parameter uncertainty
    \item The development of a simple method to approximate the wake model parameter probability distributions without requiring computationally expensive posterior sampling
    \item Testing of the set-point optimization under model parameter uncertainty in a utility-scale wind farm open-loop numerical experiment and in closed-loop control of a wind farm in LES of a convective ABL
\end{enumerate}
The yaw misalignment set-point optimization formulation under wind condition variability and model parameter uncertainty is given in \S \ref{sec:contol}.
The wake model is discussed in \S \ref{sec:model} and the simple parameter probability distribution estimation methodology is introduced in \S \ref{sec:enkf}.
In \S \ref{sec:farm}, wake steering numerical case studies of a utility-scale wind farm using operational data are performed.
LES simulations of closed-loop control of a model wind farm in convective ABL conditions are performed in \S \ref{sec:les}.
Conclusions are given in \S \ref{sec:conclusions}.

\section{Yaw set-point optimization under model parameter uncertainty}
\label{sec:contol}

The goal of wake steering is to maximize wind farm power production through the use of intentional yaw misalignment.
For open-loop control, the goal is to calculate the optimal yaw misalignment set-points for each wind condition bin, specified by wind speed, turbulence intensity, and wind direction\cite{fleming2019initial}.
For wake model-based closed-loop control (e.g. the method proposed by Howland {\it et al.} (2020) \cite{howland2020optimal}), the goal is to calculate the optimal yaw set-points for the wind farm over the finite control update time horizon.
In both approaches, the set-point optimization can be considered over wind condition probabilities.
The wind condition probabilities are pre-tabulated in the open-loop setting.
In closed-loop control, the wind condition probabilities are collected online, as wind condition and power measurements over the previous finite time control update horizon.

Given variability and uncertainty in the wind conditions $\bm{c}$ and wake model parameters $\psi$, this optimization goal is to select the yaw misalignment set-points $\gamma_s$ which maximize the expected value of the power production
\begin{equation}
\gamma_s^*(\bm{c}, \psi) = 
\argmax_{\gamma_s} \EX \left[\mathcal{G}(\bm{c}, \psi, \gamma_s) \right],
\label{eq:opti_param}
\end{equation}
where $\mathcal{G}(\bm{c}, \psi, \gamma_s)$ is the wind farm power production.
The optimal yaw misalignment set-points over the wind turbines in the farm is $\gamma_s^*$.
The expected value of the power production is given by an integration over the wind condition and wake model parameter spaces
\begin{equation}
\EX\left[\mathcal{G}(\bm{c}, \psi, \gamma_s) \right] = 
\idotsint
f(\bm{c}) f(\psi) \mathcal{G}(\bm{c}, \psi, \gamma_s) d\bm{c}\,d\psi,
\label{eq:expected_value_param_uncertainty}
\end{equation}
where $f(\psi)$ is the probability density function over the wake model parameter space $\psi$ and $f(\bm{c})$ is the probability density function over the wind condition space $\bm{c}$.
The computationally efficient analytic gradient-based optimizer developed by Howland {\it et al.} (2019) \cite{howland2019wind} is extended to the stochastic programming problem in Eq. \ref{eq:opti_param} and is used in this study.
For open-loop control, Eq. \ref{eq:expected_value_param_uncertainty} will be solved for each wind condition bin, while for closed-loop control, Eq. \ref{eq:expected_value_param_uncertainty} will be solved at each control update step in an online fashion\cite{howland2020optimal}.

The probability distributions of the wake model parameters can be estimated using approximate Bayesian inference methods, such as Markov chain Monte Carlo \cite{brooks2011handbook, zhang2020quantification}.
However, sampling based methods are computationally intensive, requiring $\mathcal{O}(10^5)$ forward model evaluations to perform the Bayesian inverse problem analysis\cite{brooks2011handbook}.
This requirement becomes challenging for on-the-fly estimation of wake model probability densities in closed-loop control.
Instead, in this study (\S \ref{sec:enkf}), we will develop a simple method to estimate an approximate wake model parameter probability distribution to be used in Eq. \ref{eq:expected_value_param_uncertainty} based on power production data by leveraging parameter estimation techniques.

Previous approaches estimated the wake model parameters based on an average of the power within a wind condition bin \cite{howland2019wind}.
The SCADA operational data also has underlying uncertainty and variability, for example from physical effects such as dynamic wake meandering and statistical effects from finite time averaging, which corresponds to uncertainty and variability in the wake model parameters represented by the probability function $f(\psi)$.
The probability density functions will be defined specifically for each wind condition bin based on the utility-scale wind farm data in \S \ref{sec:condition_pdfs} and for LES data in \S \ref{sec:les}.

\begin{figure}
    \centering
    \includegraphics[width=\linewidth]{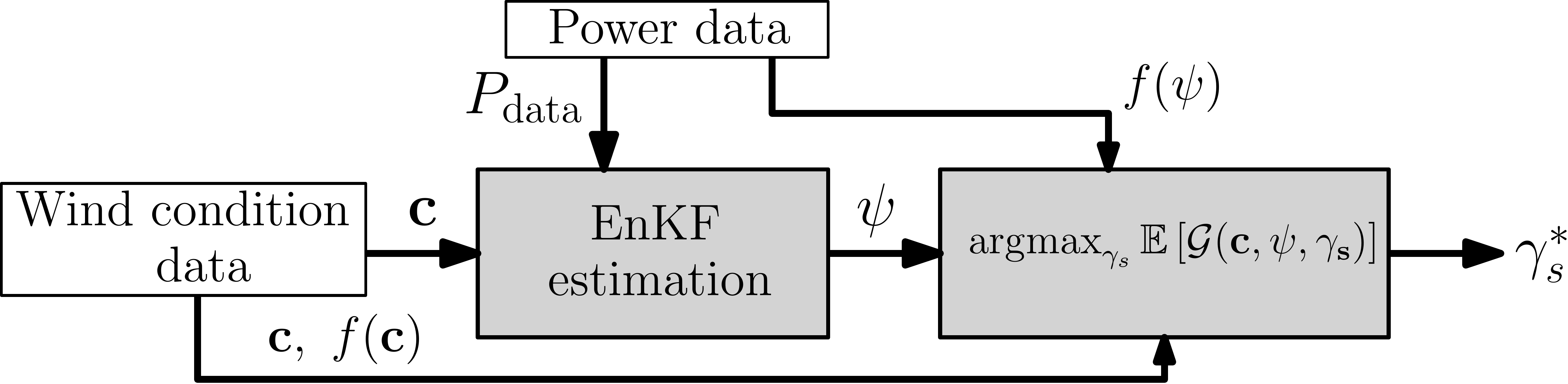}
    \caption{Set-point optimization under model parameter uncertainty.
    The wake model parameters $\psi$ are estimated by the ensemble Kalman filter (EnKF) for a set of wind conditions $\bm{c}$ and averaged power data $P_{\mathrm{data}}$.
    The probability distribution of the parameters $f(\psi)$ is derived from the power data and used in tandem with the condition probability distributions $f(\bm{c})$ to compute $\gamma_s^*$ with set-point optimization (Eq. \ref{eq:opti_param}).}
    \label{fig:robust_opti}
\end{figure}

The optimization framework is shown in Figure \ref{fig:robust_opti}.
In practice, Eq. \ref{eq:expected_value_param_uncertainty} is discretized to solve using the analytic wake model \cite{quick2017optimization, rott2018robust}.
While other expectation calculation methods could be used, such as polynomial chaos expansion \cite{quick2020wake}, direct quadrature is used in this study for simplicity and due to the computational efficiency of the gradient-based set-point optimization \cite{howland2019wind}. 
The discretization, along with the prescribed wind condition uncertainty \cite{rott2018robust}, become additional hyperparameters.
In \S \ref{sec:cases}, we will analyze the sensitivity of wake steering to these hyperparameters.

\subsection{Steady-state wake model}
\label{sec:model}

The nonlinear wake model used in the present study, denoted by $\mathcal{G}$, is the lifting line model \cite{shapiro2018modelling}, however, the methods described below can be used for arbitrary steady-state wake models.
The wake model represents the time averaged wake region trailing a yaw misaligned or yaw aligned turbine and was validated against experimental data \cite{shapiro2018modelling}. 
The model has been used in subsequent wake steering power optimization studies in LES \cite{howland2020optimal} and field experiments \cite{howland2019wind}.
The wake model predicts wind farm power for arbitrary inflow wind conditions and wind turbine layout given two physics-based parameters, $k_w$ and $\sigma_0$, which denote the wake spreading rate and the proportionality constant for the presumed Gaussian wake profile.
The area averaged velocity deficit at a downwind turbine $j$ as a result of the wake of an upwind turbine $i$ is
\begin{equation}
\Delta u_{i,j}(x) = \frac{\sqrt{2 \pi} \delta u_i(x) d_{w,i}(x) D}{16 \sigma_{0,i}} \left [ \mathrm{erf} \left(\frac{y_T+D/2 - y_{c,i}(x)}{\sqrt{2} \sigma_{0,i} d_{w,i}(x)}\right) - \mathrm{erf} \left(\frac{y_T-D/2 - y_{c,i}(x)}{\sqrt{2} \sigma_{0,i} d_{w,i}(x)}\right) \right],
\label{eq:delta_u_ij}
\end{equation}
where $D$ is the turbine diameter, $x$ is the streamwise direction, $d_w$ is the normalized wake diameter, $y_c$ is the lateral centroid of the wake, $y_T$ is the lateral turbine centroid of turbine $j$, and $\delta u$ is the streamwise velocity deficit \cite{shapiro2018modelling, howland2021influence}.
The wake diameter is $d_{w}(x) = 1 + k_{w} \log \left(1+\exp[2(x/D - 1)]\right)$.
The wake centroid $y_{c,i}$ is a function of the yaw misalignment of the upwind turbine $i$.
Details on the wake deflection model are given in Shapiro {\it et al.} (2018)\cite{shapiro2018modelling}.
Modified linear wake superposition\cite{niayifar2016analytical} is used since its performance is similar to the more computationally expensive momentum conserving superposition \cite{howland2021influence, zong2020momentum}.
The rotor area averaged velocity at turbine $j$ is
\begin{equation}
u_{e,j} = u_\infty - \sum_i^{N_f} \Delta u_{i,j},
\label{eq:linear_ue}
\end{equation}
where $N_f$ is the number of upwind turbines and the power production is $\hat{P}_j=\frac{1}{2} \rho A C_P u_{e,j}^3$.
In this study, $C_P = C_P(\gamma=0) \cdot \cos^{P_p}(\gamma)$ is used although the recently developed blade-element model which accounts for wind velocity profiles could be used in future studies with rotational turbine models \cite{howland2020yaw}.
For the utility-scale wind farm numerical experiments, $P_p=2$, which represents a reasonable first order approximation of $C_p(\gamma)$ as shown in a yaw misalignment field experiment at the same wind farm\cite{howland2020yaw}.
For the LES experiments, $P_p$ was set to $2.5$ based on empirical tuning.
The secondary steering model proposed by Howland \& Dabiri (2021)\cite{howland2021influence} is used.
Aside from the wake model parameters, the predicted power production of a wind farm depends on the turbine layout\cite{bossuyt2017measurement}, the incident wind speed $u_\infty$, the wind direction, and the turbine yaw misalignment set-points $\gamma_s$.
Additionally, there may be deviation from the set-point due to yaw error\cite{quick2017optimization} which is denoted as $\gamma$. 
The wind and turbine control conditions are collected into the vector $\bm{c} = [u_\infty, \alpha, \gamma]$, with wind direction given by $\alpha$.
The effects of turbulence intensity (TI) are included implicitly in the wake model parameters $\psi$ (see \S \ref{sec:enkf}).
Therefore, TI is not included explicitly in $\bm{c}$. 

The wake model power predictions are collected into a vector and are denoted compactly as $\mathcal{P}(\bm{c}, \psi, \gamma_s) = [\hat{P}_1,...,\hat{P}_{N_t}] \in {\rm I\!R}^{N_t}$, where $N_t$ is the number of turbines in the farm, $\hat{P}_i$ is the wake model power estimate for turbine $i$, and
\begin{equation}
\mathcal{G}(\bm{c}, \psi, \gamma_s) = \sum_{i=1}^{N_t}\mathcal{P}_i(\bm{c}, \psi, \gamma_{s}).
\end{equation}
The wake model predictions depend on two types of inputs: the wind conditions $\bm{c}$ and the wake model parameters $\psi$,
\begin{equation}
\psi = [k_{w,1},...,k_{w,N_t-1},\sigma_{0,1},...,\sigma_{0,N_t-1}].
\label{eq:psi}
\end{equation}

\subsection{Wake model parameter probability distribution estimation}
\label{sec:enkf}

In this study, the wake model parameters are estimated using the ensemble Kalman filter (EnKF) \cite{evensen2003ensemble} and SCADA power production data.
Previous studies have used the EnKF to estimate the mean wake model parameters\cite{shapiro2019wake, howland2020optimal} and the wake model state \cite{doekemeijer2017ensemble}.
For parameter estimation, the EnKF can be viewed as an approximation of gradient-based optimization without requiring direct gradient calculations \cite{schillings2017analysis}.
The wind farm power production SCADA data is measured in finite-time averages $\tilde{P}\in {\rm I\!R}^{N_t}$.
Steady-state wake models predict $P_\infty$, the infinite time average of the power. 
However, $P_\infty$ is not available in measurements and normally distributed measurement noise in the form of $\varepsilon=N(0,\Sigma_\varepsilon)$ appears in finite-time averages of power due to the central limit theorem, where $\Sigma_\varepsilon$ is the power data covariance matrix, giving
\begin{equation}
\tilde{P} = P_\infty + N(0,\Sigma_P),
\end{equation}
where $N$ denotes a normal distribution with zero mean and covariance $\Sigma_P$.
In practice, the infinite time average of the power production $P_\infty$ has a functional dependence on the incident wind conditions.
In field data, this effect is approximated using conditional averaging techniques\cite{howland2019wind}, where the data are grouped by wind condition bins and then the power data samples are averaged in each bin.
Since the steady-state wake model predicts $P_\infty$, we will use the underlying variability in finite time average SCADA power data samples $\tilde{P}$ to estimate the probability distributions of the wake model parameters.

In order to approximate the wake model probability distribution $f(\psi)$, the wake model parameters are estimated using the EnKF for various values of turbine power $\tilde{P} / \tilde{P}_1$ within the empirically measured SCADA power probability distribution, where $\tilde{P}_1$ is the finite time average of power for a freestream turbine at the farm.
The probability distribution of $\tilde{P} / \tilde{P}_1$ is assumed to be Gaussian with an empirically measured mean and standard deviation.
The Gaussian assumption is a reasonable approximation through the central limit theorem, although it will not be exact due to the influence of the varying state of the atmospheric conditions such as thermal stability or heterogeneous flow field effects, which are not accounted for in the standard conditional averaging techniques \cite{fleming2019initial, howland2019wind}.
The Gaussian approximation will tested using field data in \S \ref{sec:condition_pdfs}.
The distribution is discretized between $\bar{P}\pm \sigma_p$, where $\bar{P}=\left<\tilde{P}\right>$ is the mean of instances within the wind condition bin and $\sigma_p$ is the corresponding standard deviation.
The wake model parameters $k_w$ and $\sigma_0$ are estimated using the EnKF for each of the discrete values of power and the probability mass for each point is computed by integrating the Gaussian distribution with the empirically measured standard deviation in the power data.

Each turbine in the farm, except for the furthest downwind, has distinct values for the model parameters, since the values depend on the turbine inflow conditions and turbine layout \cite{stevens2015coupled}.
For $N_t$ wind turbines, the model parameters are $\psi = [k_{w,1},...,k_{w,N_t-1},\sigma_{0,1},...,\sigma_{0,N_t-1}]$ (Eq. \ref{eq:psi}).
The wake model parameters for the last turbine downwind do not affect the power prediction and are therefore neglected.
The errors associated with the modeling and measurement are given by $\chi=[\chi_{k_w}^T, \chi_{\sigma_0}^T]^T \in {\rm I\!R}^{2(N_t-1)}$ and $\varepsilon \in {\rm I\!R}^{N_t}$, respectively.
The errors have prescribed variances and zero mean.
The modeling errors have variances of $\sigma_{k_w}^2 = 0.0009$ and $\sigma_{\sigma_{0}}^2=0.0009$.
The parameter variances for the Gaussian prior distribution are hyperparameters and were selected based on tuning experiments \cite{howland2020optimal}.
Perturbations are added to the SCADA power production data for each ensemble $\xi^{(i)}  = \tilde{P}_s + \varepsilon^{(i)},$
where $^{(i)}$ denotes ensemble $i$ and $\tilde{P}_s$ is a sample from the empirical distribution of power data $\tilde{P}$.
The perturbed power production ensemble and noise matrix are given by
\begin{align}
\Xi &= [\xi^{(1)},...,\xi^{(N_e)}], 
&
\Sigma &= [\varepsilon^{(1)},...,\varepsilon^{(N_e)}], 
\end{align}
where $N_e$ is the number of ensembles.

The ensemble of wake model parameters is
\begin{equation}
\Psi = [\psi^{(1)},...,\psi^{(N_e)}] \in {\rm I\!R}^{2 (N_t-1) \times N_e}.
\end{equation}
The wake model ensemble power predictions are collected in the matrix $\hat{\Pi} \in {\rm I\!R}^{N_t \times N_e}$.
The wake model parameter and power production ensemble means are
$\overline{\Psi} = \Psi \mathbf{1}_{N_e},$ and  
$\overline{\hat{\Pi}} = \hat{\Pi} \mathbf{1}_{N_e},$ 
with $\mathbf{1}_{N_e}\in {\rm I\!R}^{N_e \times N_e}$ is a matrix with all entries as $1/N_e$.
Perturbation matrices are given by
$\Psi^\prime = \Psi - \overline{\Psi}$ and $\hat{\Pi}^\prime = \hat{\Pi} - \overline{\hat{\Pi}}.$

The intermediate forecast step is
\begin{align}
\Psi_{+} &= [\psi^{(1)}+B\chi^{(1)},...,\psi^{(N_e)}+B\chi^{(N_e)}] \\
\hat{\Pi}_{+} &= [\mathcal{P}(\psi^{(1)}+B\chi^{(1)}),...,\mathcal{P}(\psi^{(N_e)}+B\chi^{(N_e)})].
\end{align}
with $B\in {\rm I\!R}^{2(N_t-1)\times2(N_t-1)}$ prescribed as the identity matrix.
The measurement analysis step is
\begin{equation}
\Psi_{p} = \Psi_{+} +
\Psi_{+}^\prime \hat{\Pi}_{+}^{\prime T} ( \hat{\Pi}_{+}^{\prime} \hat{\Pi}_{+}^{\prime T} + \Sigma \Sigma^{ T} )^{-1} \cdot (\Xi - \hat{\Pi}_{+} ).
\label{eq:analysis}
\end{equation}
The EnKF estimated values of $\sigma_0$ and $k_w$, or $\psi$, are the columns of $\overline{\Psi}_{p}$.

In summary, the EnKF is used to estimate the wake model parameters in the vector $\psi$ for a given realization from the probability distribution of finite time averaged SCADA data $\tilde{P}$.
The EnKF estimation is used to map the probability distribution of power production samples $f(\tilde{P})$ to the distribution of wake model parameters $f(\psi)$, which is required for the expected value calculation in Eq. \ref{eq:expected_value_param_uncertainty}.
The wake model parameter distribution method proposed here neglects structural model form bias and uncertainty, which should be considered in future work.

\section{Utility-scale wind farm case study}
\label{sec:farm}

\begin{figure}
  \centering
  \begin{tabular}{@{}p{0.45\linewidth}@{\quad}p{0.45\linewidth}@{}}
    \subfigimgtwo[height=2.0in,valign=t]{(a)}{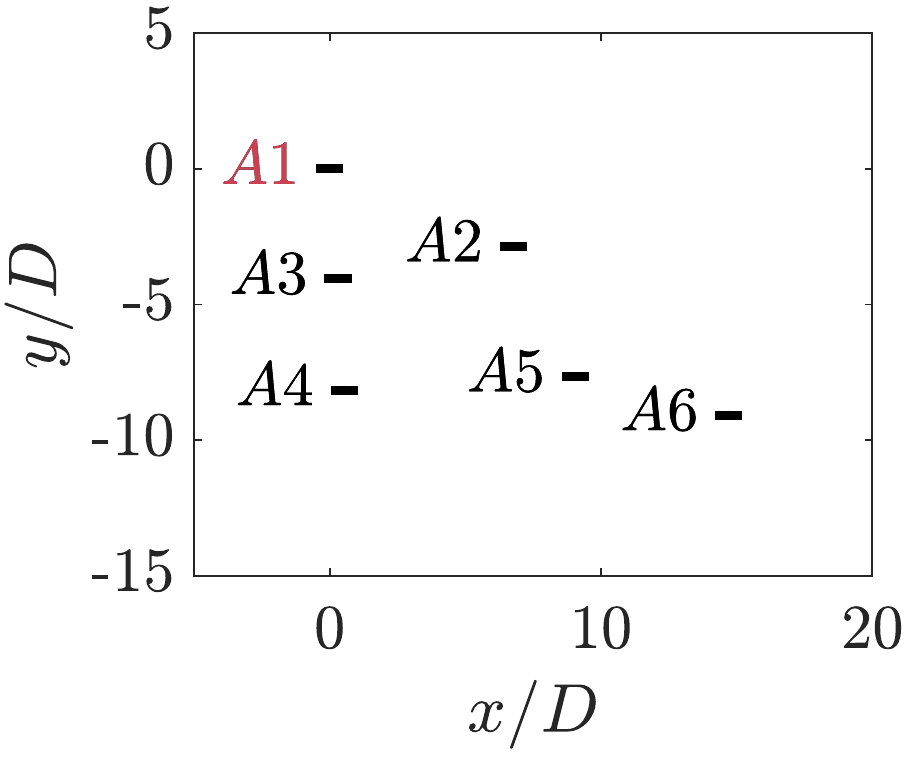} &
    \subfigimgtwo[height=2.0in,valign=t]{(b)}{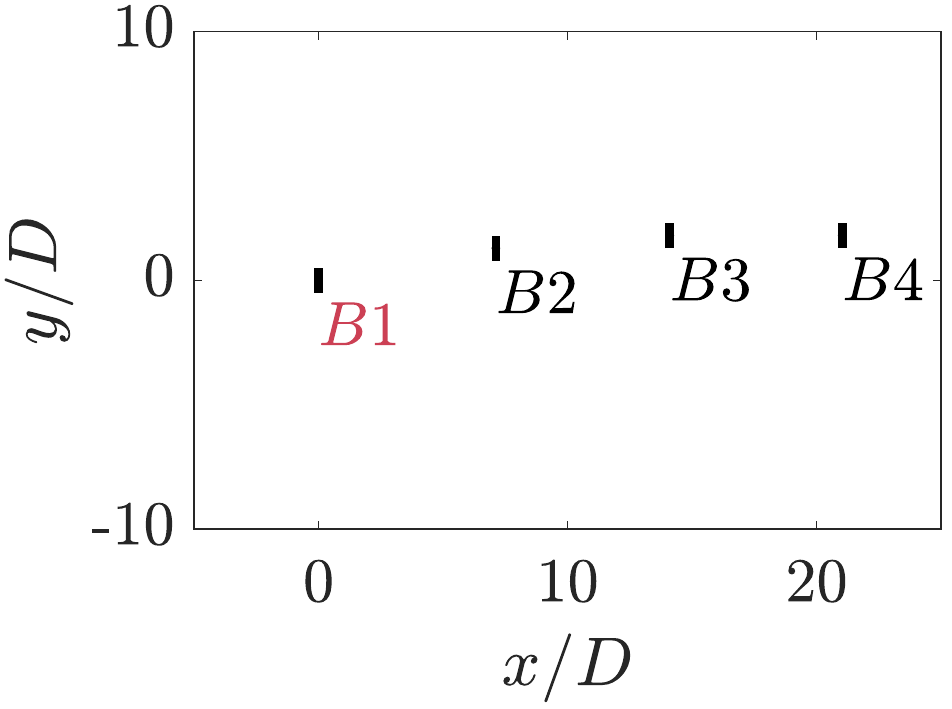}
  \end{tabular}
  \caption{(a) Utility-scale wind farm layout for Cluster A. The turbines are oriented for flow from the north ($\alpha=0^\circ$) and the coordinates are normalized by the wind turbine diameter $D$.
  (b) Cluster B layout oriented for flow from the west ($\alpha=270^\circ$).
  The wind turbine which is used as a wind condition reference is shown in red.
  The $x$ and $y$ axes correspond to easting and northing directions.
        }
    \label{fig:layouts}
\end{figure}

The optimization under model parameter uncertainty will be analyzed for open-loop control using a utility-scale wind farm case study.
In \S \ref{sec:condition_pdfs}, field measurements from the utility-scale wind farm are used to inform the wind condition probability distributions in Eq. \ref{eq:expected_value_param_uncertainty}.
The yaw misalignment set-point lookup table is synthesized in \S \ref{sec:param_pdf}.
Wake model-based numerical experiments are performed in \S \ref{sec:cases}.
Since the same wake model is being used to construct and test the lookup table in these experiments, this can be viewed as an idealized, perfect model setting to test the stochastic programming performance with no structural wake modeling error.
In \S \ref{sec:les}, the optimization framework is tested in closed-loop control in LES.

The wind farm is located in northwest India.
The wind turbines have approximately 100 meter diameters and hub heights and the terrain is flat with no significant complexity.
The rated power for the turbines is approximately 2 MW.
Two clusters of turbines are considered as shown in Figure \ref{fig:layouts}.
For Cluster A, the flow of interest is centered at $\alpha \approx 0^\circ$, where $0^\circ$ corresponds to north and proceeds clockwise, such that turbines $A3$ and $A4$ experience wake losses from turbine $A1$.
For Cluster B, the focus for wake steering is $\alpha \approx 270^\circ$, where turbines $B2$, $B3$, and $B4$ experience wake losses.
One-minute averaged SCADA data is recorded from the wind farm.

In open-loop control, there is a set of hyperparameters associated with the wind condition bin widths for the yaw set-point lookup table.
Contemporary wind turbine control systems are computational operation count and memory limited, which restricts the quantity of wind condition bins in the lookup table.
In the present study, we will use $u\pm 1$ m/s, $\alpha \pm 2.5^\circ$, and $TI \pm 2.5\%$, which is based on the memory limitations of a utility-scale wind turbine control system and similar to previous studies \cite{fleming2019initial}.
The methods developed in this study apply to arbitrary wind condition lookup table selections, although the specific probability distributions for each wind condition may be changed with different bin width selections.

\subsection{Probability distributions for wind conditions}
\label{sec:condition_pdfs}

\begin{figure}
  \centering
  \begin{tabular}{@{}p{0.45\linewidth}@{\quad}p{0.45\linewidth}@{}}
    \subfigimgtwo[width=\linewidth,valign=t]{(a)}{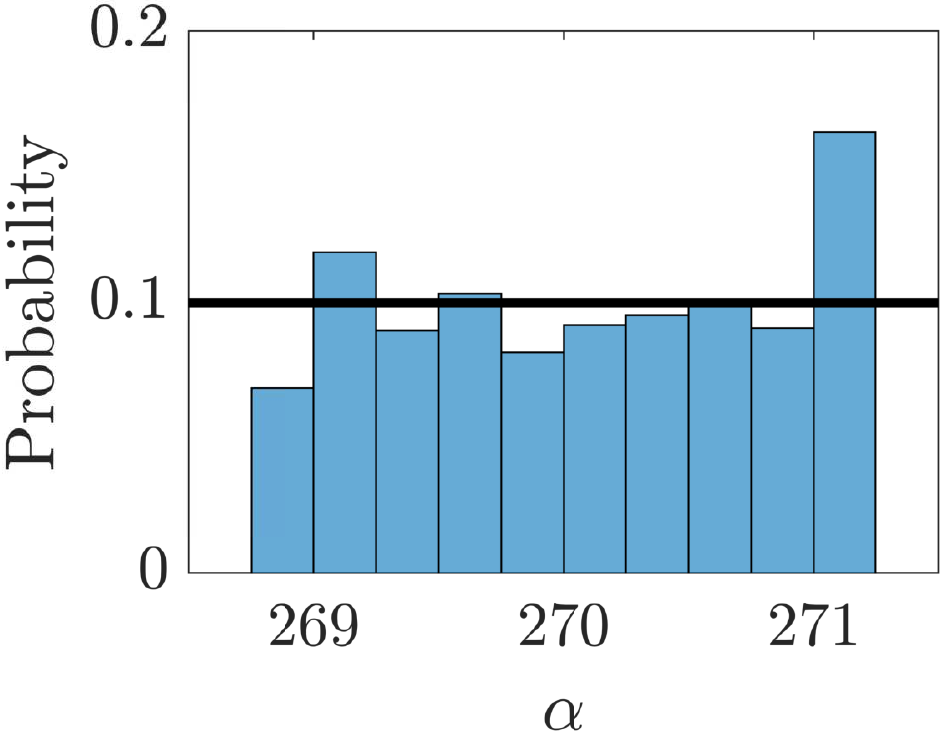} &
    \subfigimgtwo[width=\linewidth,valign=t]{(b)}{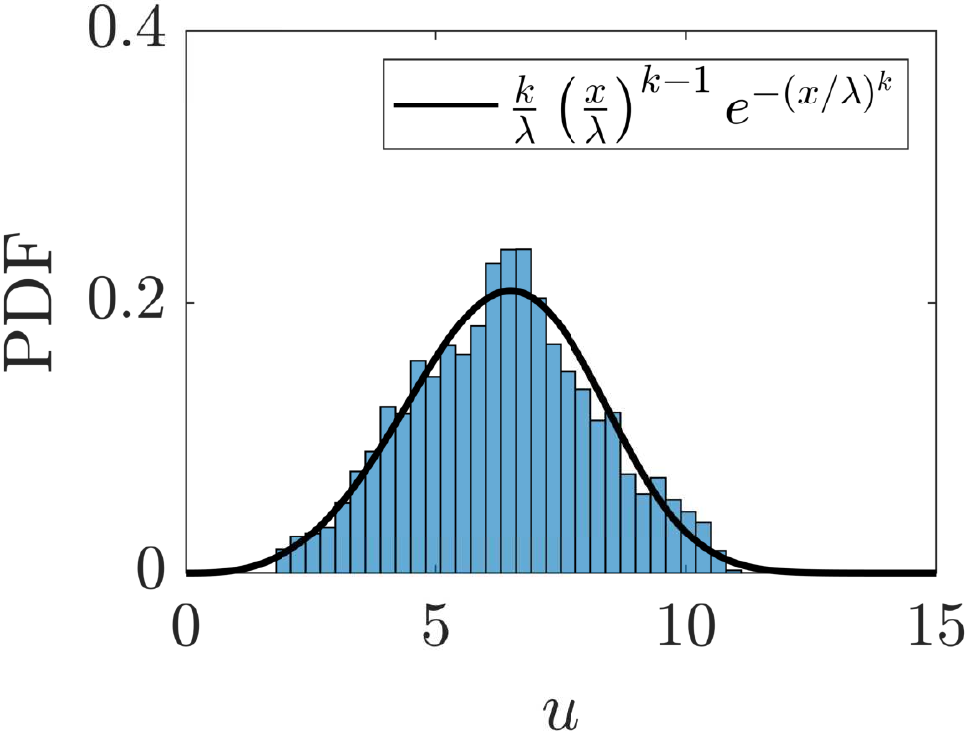}
  \end{tabular}
  \caption{(a) Wind direction probability distribution within a $268.75^\circ<\alpha<271.25^\circ$ wind direction bin.
  (b) Wind speed empirical probability density function.
  A best-fit Weibull distribution with $k=3.8$ and $\lambda=7$ is shown with a solid black line.
        }
    \label{fig:condition_pdf}
\end{figure}

In open-loop lookup table synthesis, given the finite width of the wind condition bins which is not infinitesimally small, there is a probability distribution associated with each wind condition within each bin.
For example, for the freestream wind speed bin $5<u<7$ m/s, there will be a probability distribution associated with the wind speed values in this restricted range.
The historical SCADA measurements for wind direction, speed, and yaw misalignment will be used to define empirical probability distributions used in Eq. \ref{eq:expected_value_param_uncertainty}.
The SCADA data can be used both to establish a likelihood for each wind condition bin in the lookup table and to establish wind condition probability distributions within each lookup table bin.
The field data probability distributions for the wind direction and speed can be seen in Figures \ref{fig:condition_pdf}(a,b).
The wind direction within the wind condition bin can be approximated by a uniform distribution.
As discussed thoroughly by Rott {\it et al.} (2018) \cite{rott2018robust} and Simley {\it et al.} (2020) \cite{simley2020design}, there is uncertainty associated with the measurements of the wind direction as well as higher frequency dynamics which occur within a one-minute averaged sample that could cause larger values of $\alpha$ deviations.
In order to account for these dynamics, we will consider $\alpha$ uncertainty of $\alpha\pm2.5^\circ$, $\alpha\pm5^\circ$, and $\alpha\pm10^\circ$ in the case studies presented in \S \ref{sec:param_pdf}.
The case of $\alpha\pm2.5^\circ$ corresponds to wind direction variability from the bin width, whereas higher uncertainty values of $\pm5^\circ$ and $\pm10^\circ$ consider other variability and uncertainty factors.
Previous approaches considered higher frequency content and used Gaussian distributions to model directional variability\cite{rott2018robust}.
In this open-loop experiment, we will focus on low-pass filtered wind condition variations within a lookup table bin since SCADA data was not available at higher frequency than one-minute averages.
Higher frequency content will be considered in the LES case study in \S \ref{sec:les}.

\begin{figure}
  \centering
  \begin{tabular}{@{}p{0.45\linewidth}@{\quad}p{0.45\linewidth}@{}}
    \subfigimgtwo[width=\linewidth,valign=t]{(a)}{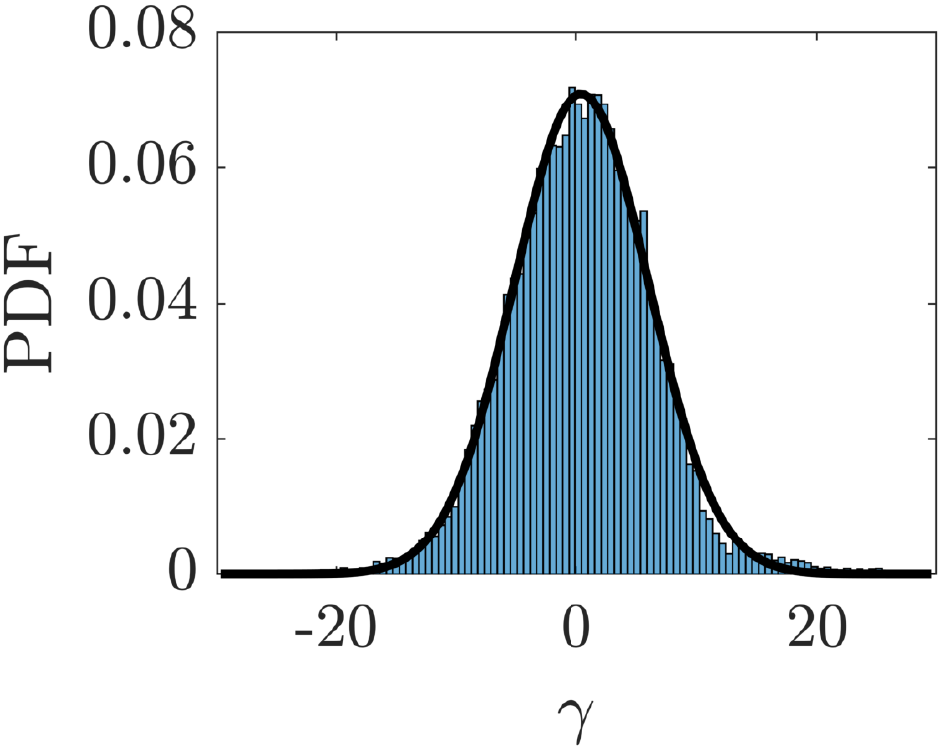} &
    \subfigimgtwo[width=\linewidth,valign=t]{(b)}{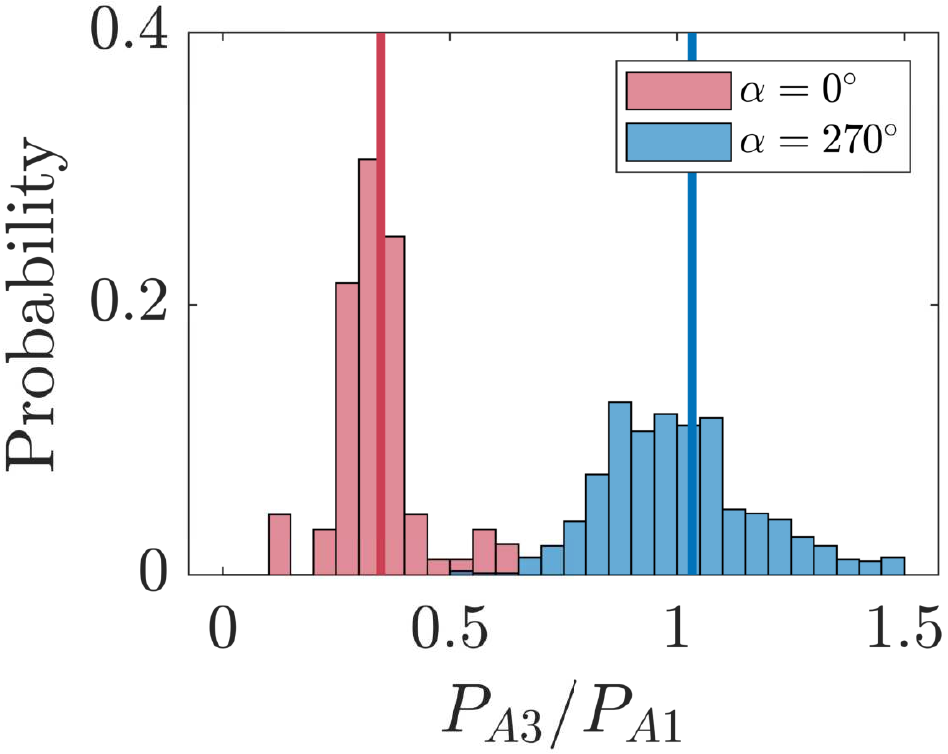}
  \end{tabular}
  \caption{(a) Yaw misalignment empirical probability density function with a best-fit Gaussian distribution shown with a solid black line.
    The mean and standard deviation of the Gaussian distribution are $\mu_\gamma=0.4^\circ$ and $\sigma_\gamma=5.6^\circ$.
    (b) Wind turbine power ratio probability distributions for wind direction bins centered at $\alpha=2.5^\circ \pm 2.5^\circ$ and $\alpha=270^\circ \pm 2.5^\circ$ for $u=7\pm 1$ m/s and $TI=5\%\pm2.5\%$.
    The vertical lines correspond the means of the respective power ratio distributions.
        }
    \label{fig:gamma_power_pdf}
\end{figure}

The wind speed in each wind condition bin approximately follows a Weibull distribution.
While narrowing the wind speed bins could result in an approximately uniform probability density function, the practical wind speed bins are $u\pm1$ m/s and the coefficient of power $C_p$ and coefficient of thrust $C_T$ for wind turbines depend strongly on $u$.
Therefore, the more representative Weibull fits will be used.
The yaw misalignment probability distribution for turbine $A1$ is shown in Figure \ref{fig:gamma_power_pdf}(a).
A Gaussian probability density function is used to approximate the distribution with $\sigma_\gamma=5.6^\circ$.

\subsection{Open-loop control yaw misalignment set-point lookup table synthesis}
\label{sec:param_pdf}

The wake model parameters are estimated using the normalized, one-minute averaged SCADA power production data, $\tilde{P}/\tilde{P}_1$, where $\tilde{P}_1$ is the power of the leading, freestream operating turbine in the array.
The wind turbine array power production data are sorted such that instances with wind conditions $u$, $\alpha$, and $TI$ are collated within a particular lookup table bin and matched with the corresponding one-minute averaged power productions. 
The power production empirical probability distributions for one-minute averaged $\tilde{P}_{A3}/\tilde{P}_{A1}$ for $\alpha=0^\circ \pm 2.5^\circ$ and $\alpha=270^\circ \pm 2.5^\circ$ are shown in Figure \ref{fig:gamma_power_pdf}(b).
The wind speed is filtered such that $u=7\pm 1$ m/s and $TI=5\%\pm2.5\%$.
For $\alpha=270^\circ \pm 2.5^\circ$, turbine $A3$ does not experience waked incident inflow, and therefore, $\tilde{P}_{A3}/\tilde{P}_{A1} \approx 1$ with variations due to spatiotemporal differences in the ABL turbulent inflow in one-minute averaged samples.
The resulting distribution for $\tilde{P}_{A3}/\tilde{P}_{A1}$ can be reasonably approximately by a Gaussian, centered around $\left<\tilde{P}_{A3}/\tilde{P}_{A1}\right> \approx 1$.
For $\alpha=0^\circ \pm 2.5^\circ$, turbine $A3$ experiences significant wake losses, with $\left<\tilde{P}_{A3}/\tilde{P}_{A1}\right> \approx 0.4$.
However, within the narrow wind direction and speed bins, $\tilde{P}_{A3}/\tilde{P}_{A1}$ is still reasonably Gaussian.

\begin{figure}
  \centering
  \begin{tabular}{@{}p{0.45\linewidth}@{\quad}p{0.45\linewidth}@{}}
    \subfigimgtwo[width=\linewidth,valign=t]{(a)}{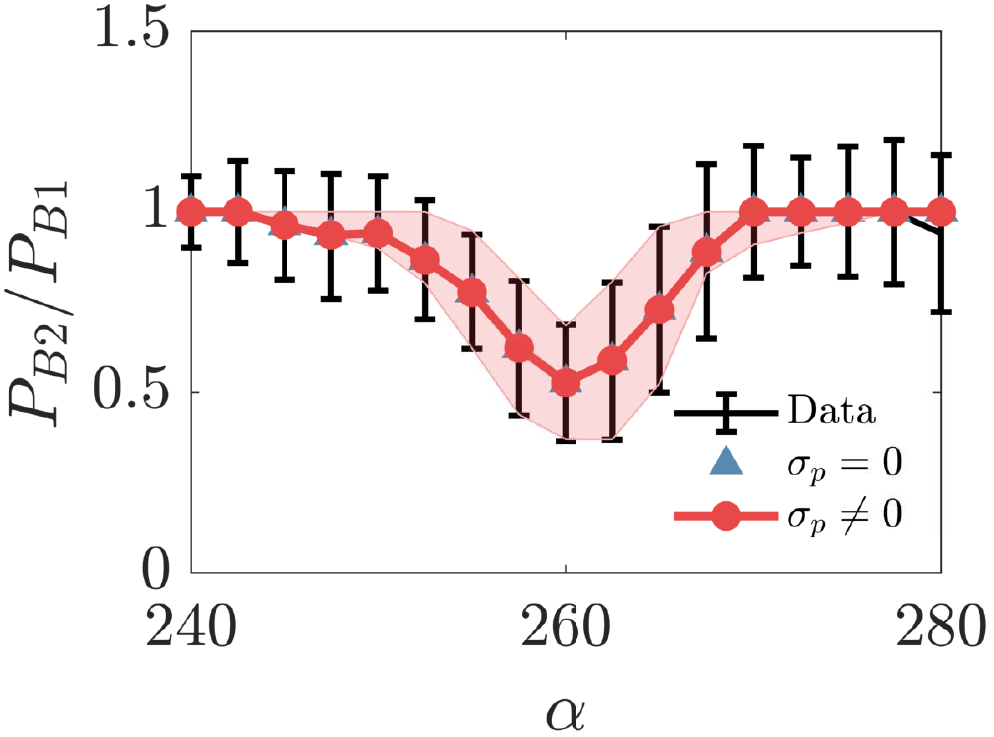} &
    \subfigimgtwo[width=\linewidth,valign=t]{(b)}{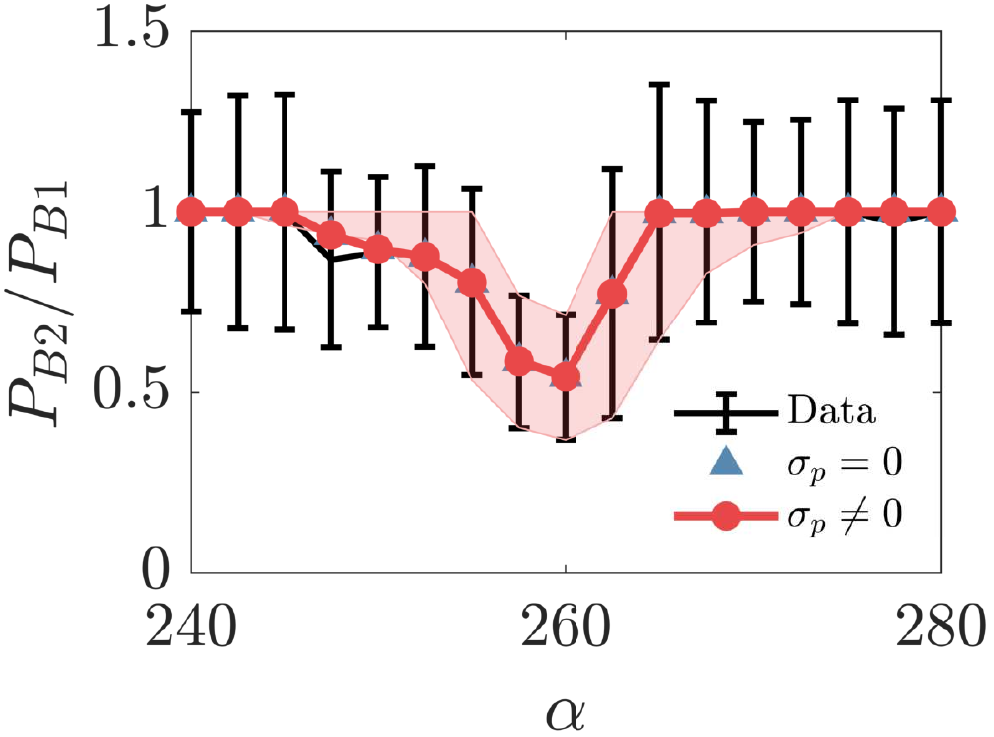}
  \end{tabular}
  \caption{Normalized power production for waked turbine $B2$ in Cluster B as a function of the incident wind direction for $u=7\pm1$ m/s and (a) $TI=5\pm2.5\%$ and (b) $TI=10\pm2.5\%$.
  Errorbars represent one standard deviation about the mean from the one-minute averaged SCADA data.
  The stochastic wake model predictions are represented by $\sigma_p \neq 0$.
  The shaded region corresponds to one standard deviation about the mean wake model power estimate.
        }
    \label{fig:pFit}
\end{figure}

The normalized power productions for turbine $B2$ are shown in Figure \ref{fig:pFit} for $TI=5\pm2.5\%$ and $TI=10\pm5\%$. 
The wind speed is $u=7\pm1$ m/s and various wind directions $\alpha$ are shown.
For the wind farm layouts shown in Figure \ref{fig:layouts}, the peak wake losses occur at direct wind turbine alignment, which is approximately $\alpha=260^\circ$ for turbine $B2$.
The mean SCADA power production data is shown with errorbars representing one standard deviation about the mean.
The wake model parameters are estimated using the SCADA data and the EnKF methodology.
The resulting wake model power production estimation is shown in Figure \ref{fig:pFit}, where the blue triangles estimate the wake model parameters only based on the mean $\overline{P} = \left<\tilde{P}\right>$.
In the red diamonds, the wake model estimates the model parameters for the distribution of $\tilde{P}$ for each wind condition bin.
The wake model estimates parameters $k_w$ and $\sigma_0$ to accurately capture the various power production targets, including $\overline{P}\pm\sigma_p$, except for wind directions where wake losses do not occur, such as $\alpha<250^\circ$ or $\alpha>270^\circ$ for turbine $B2$.
Wind directions such as $\alpha>270^\circ$ for turbine $B2$ will not be relevant to the optimization in Eq. \ref{eq:expected_value_param_uncertainty} since $\gamma^*_{s,B1}=0^\circ$ from geometry.
The model used in this study is only able to realize power production differences between the waked and freestream turbines which manifest from wake losses and cannot capture all variations which include effects from temporally dependent, stochastic, heterogeneous flow fields.
The model cannot estimate $P_{B2}>P_{B1}$ which occurs in freestream operation for $\overline{P}_{B2}+\sigma_p$ as a result of spatially and temporally varying ABL turbulence.
Future work can incorporate spatiotemporal inflow variations\cite{starke2020area} to more accurately model these effects.

\begin{figure}
  \centering
  \begin{tabular}{@{}p{0.4\linewidth}@{\quad}p{0.6\linewidth}@{}}
    \subfigimgtwo[height=2in,valign=t]{(a)}{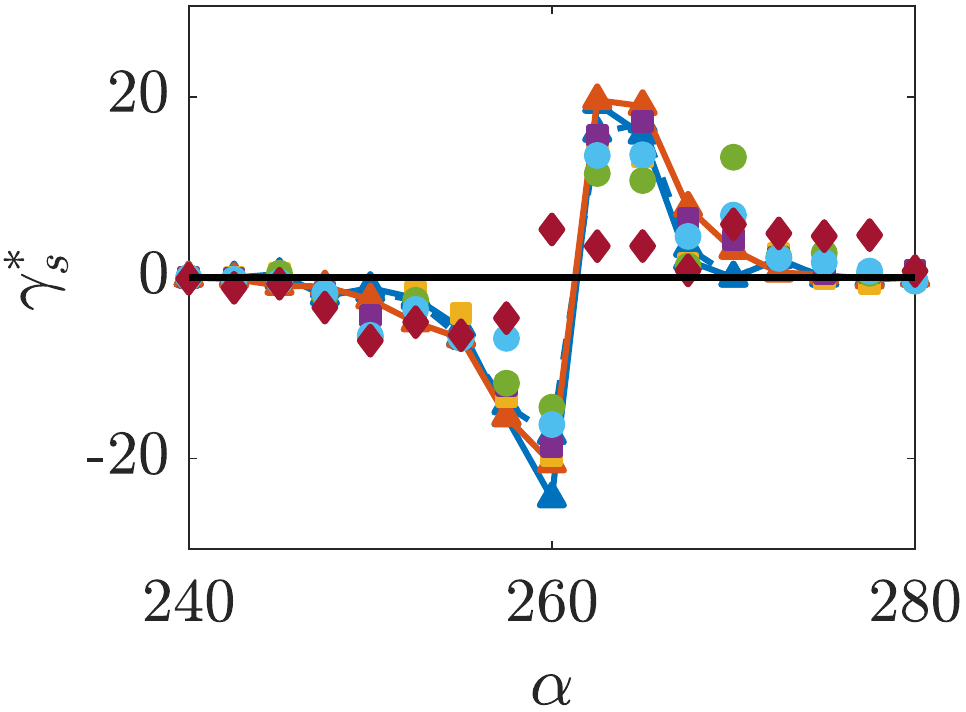} &
    \subfigimgtwo[height=2in,valign=t]{(b)}{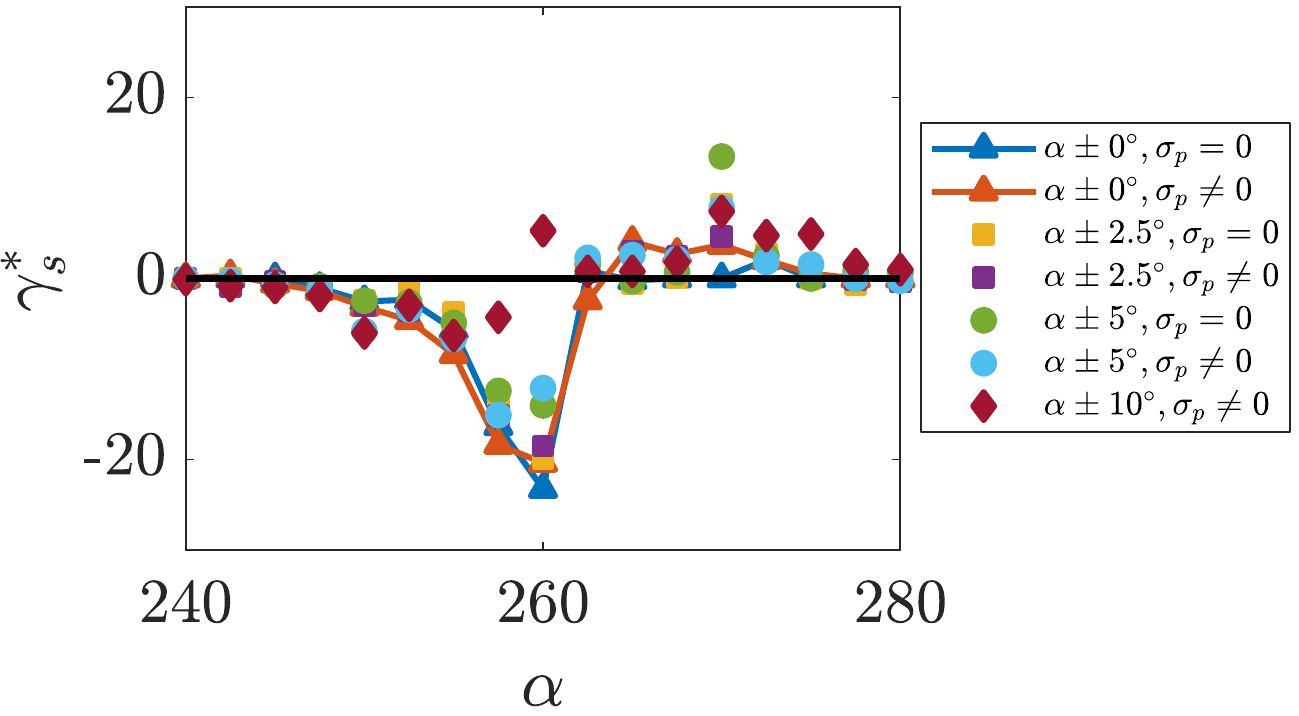}
  \end{tabular}
  \caption{Model-optimal yaw misalignment angles for turbine $B1$ as a function of the incident wind direction for $u=7\pm1$ m/s and (a) $TI=5\pm2.5\%$ and (b) $TI=10\pm2.5\%$.
  Optimal yaw set-points for various wind direction uncertainties are shown.
  Cases with $\alpha\pm0^\circ$ are deterministic set-point optimization with respect to the wind direction and $\sigma_p=0$ are deterministic with respect to the wake model parameters.
  The optimization with model parameter uncertainty are represented by $\sigma_p \neq 0$ cases.
        }
    \label{fig:yaw}
\end{figure}

The resulting optimal yaw misalignment angles $\gamma_s^*$ from various variability and uncertainty cases are shown for turbine $B1$ in Figure \ref{fig:yaw} for the two turbulence intensity cases.
Positive yaw misalignment corresponds to a counter-clockwise rotation viewed from above.
The optimization with model parameter uncertainty are represented by $\sigma_p \neq 0$ cases.
Cases with $\sigma_p\neq0$ optimize $\gamma_s$ with $9$ discrete sets of parameters $[k_w, \sigma_0]$ and the probability distribution is Gaussian, as discussed in \S \ref{sec:param_pdf}.
For the lower turbulence intensity inflow (Figure \ref{fig:yaw}(a)) and deterministic yaw optimization, the yaw set-points have a sharp inflection point around $\alpha\approx260^\circ$, where the sign of the optimal yaw switches.
The yaw misalignment values for turbine $B1$ are not symmetric about the inflection point since the wind turbine array is not directly aligned (see Figure \ref{fig:layouts}(b)).

The influence of wind direction uncertainty on optimal set-point values has been characterized in previous studies\cite{rott2018robust, quick2020wake}.
Given uncertainty or variability in the wind direction, small changes in $\alpha$ as a function of time would result in suboptimal $\gamma_s^*$ for instantaneous wind condition realizations.
With increasing $\alpha$ uncertainty, the $\gamma_s^*$ profile is generally smoothed around the inflection point.
Wind direction uncertainty of $\alpha\pm10^\circ$ results in dramatic changes compared to deterministic set-point optimization, where $\gamma_s^*(\alpha=260^\circ)$ becomes small and positive, instead of the deterministic negative value.
This model-optimal value is the result of a balance of weighting power production for $\pm10^\circ$ around the inflection point.

\begin{figure}
  \centering
  \begin{tabular}{@{}p{0.45\linewidth}@{\quad}p{0.45\linewidth}@{}}
    \subfigimgtwo[height=2in,valign=t]{(a)}{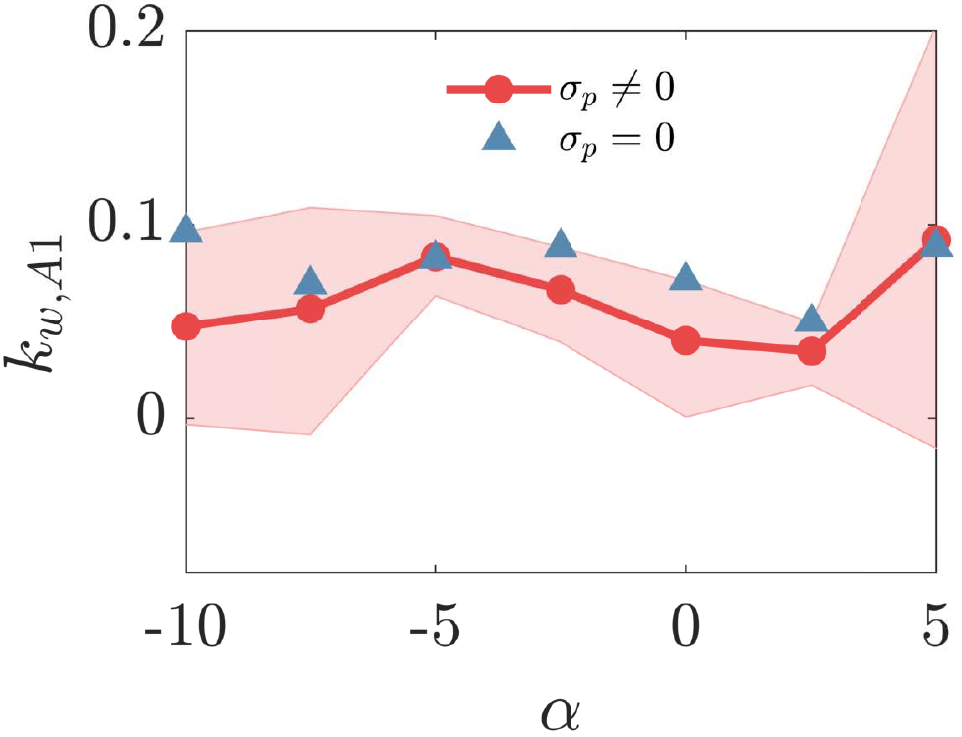} &
    \subfigimgtwo[height=2in,valign=t]{(b)}{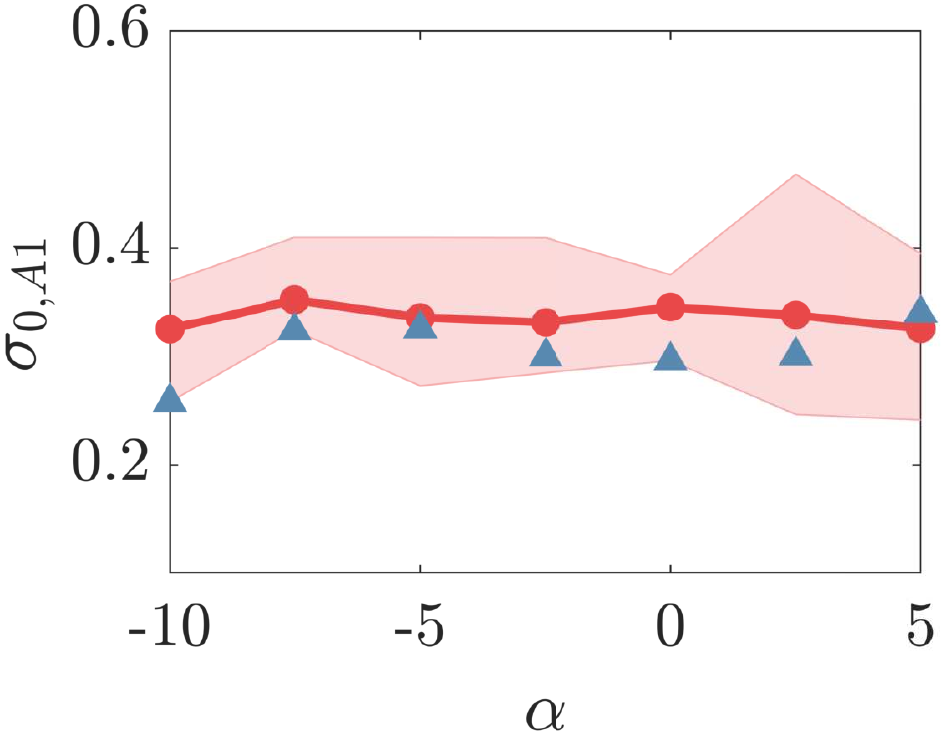}
  \end{tabular}
  \caption{Model parameter uncertainty for turbine $A1$ in Cluster A as a function of the incident wind direction for $u=7\pm1$ m/s and $TI=5\pm2.5\%$ for (a) $k_w$ and (b) $\sigma_0$.
  The shaded region corresponds to one standard deviation about the mean wake model parameter.
  Estimates of the mean wake model parameter based on the mean of the power data are given by $\sigma_p=0$, which corresponds to a wake model with deterministic wake model parameters.
        }
    \label{fig:cluster_A_params}
\end{figure}

The normalized SCADA and modeled power production for turbine $A3$, and associated optimal yaw angles for turbine $A1$, are shown in Figure \ref{fig:cluster_A}(a,b).
For turbine $A3$, the maximum wake loss inflection point occurs at approximately $\alpha=-2.5^\circ$.
For Cluster A, the incorporation of $\alpha$ or $\sigma_p$ uncertainty has a less pronounced impact on $\gamma_s^*$ due to the tight spacing in the streamwise direction ($4-5D$) and relatively small values of power production standard deviation within wind direction bins.
The uncertainties associated with the wake model parameters of turbine $A1$ are shown in Figure \ref{fig:cluster_A_params}, along with the deterministic model parameters resulting from the EnKF with $\left<\tilde{P}\right>$.
Some presence of non-physical wake model parameters exists in the distributions ($k_w<0$), which could be alleviated by tailoring the prior, using the more computationally expensive Bayesian sampling methods (e.g. MCMC) to approximately estimate parameter posteriors, or by increasing the time-averaging length of the SCADA data.

\subsection{Wake model-based open-loop wake steering case studies}
\label{sec:cases}

\begin{figure}
  \centering
  \begin{tabular}{@{}p{0.45\linewidth}@{\quad}p{0.45\linewidth}@{}}
    \subfigimgtwo[height=2in,valign=t]{(a)}{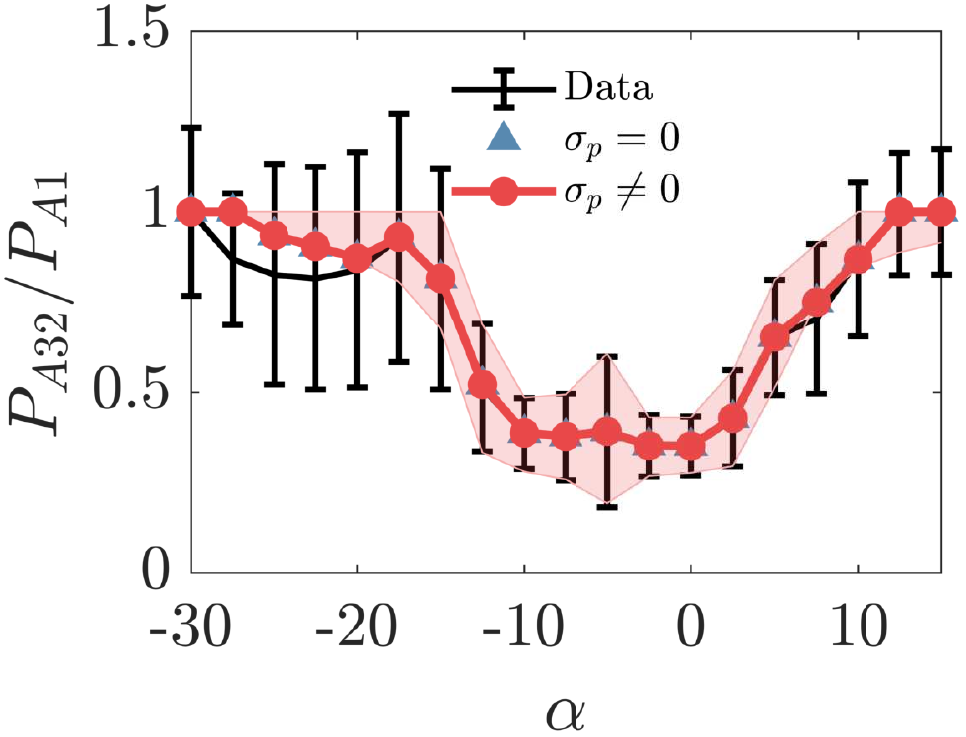} &
    \subfigimgtwo[height=2in,valign=t]{(b)}{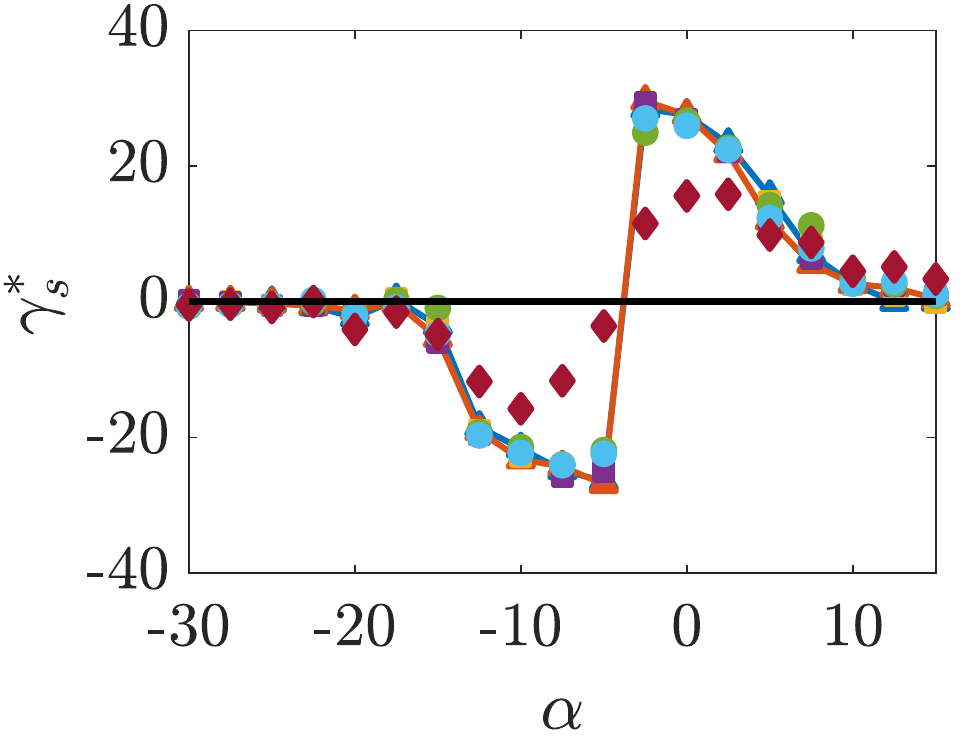}
  \end{tabular}
  \caption{(a) Normalized power production for waked turbine $A3$ in Cluster A as a function of the incident wind direction for $u=7\pm1$ m/s and $TI=5\pm2.5\%$.
  Errorbars represent one standard deviation about the mean from the one-minute averaged SCADA data.
  The shaded region corresponds to one standard deviation about the mean wake model power estimate.
  (b) Model-optimal yaw misalignment angles for turbine $A1$ given the power production distribution in (a).
  The symbols are identified in the legend of Figure \ref{fig:yaw}.
        }
    \label{fig:cluster_A}
\end{figure}

Several wake steering case studies are performed to examine the influence of the set-point optimization methodology.
The wake model described in \S \ref{sec:model} is used for the case studies.
SCADA data is used to select instances of time such that the wind speed, direction, and turbulence intensity are within wind conditions of interest.
The longest, continuous time series of such instances is selected for each test case.
The case study data differs from the training data which is used for optimal yaw set-point calculation, ensuring there is not a bias in using the same data for training and testing.
Quality filters are used on the SCADA data to ensure all wind turbines of interest are operating normally with no curtailments or error codes.
In the case study tests, the one-minute averaged SCADA data samples are provided to the wake model parameter estimation (\S \ref{sec:enkf}), and the optimal parameters for the specific instances are computed which minimize the wake model fitting error.
The yaw set-points are obtained from the pre-calculated lookup table (\S \ref{sec:param_pdf}) and applied in the wake model, which models the power production given the computed optimal parameters, the one-minute averaged wind speed and direction, and the different yaw misalignment strategies for each one-minute instance.
Yaw controllers for utility-scale turbines modify yaw with a speed around $0.5^\circ$/s.
Large, one-minute changes in yaw misalignment ($\gamma>30^\circ$), which would not be feasible, occur less than $0.5\%$ of the time in these case studies, and therefore the yaw modifications were not restricted for controller simplicity.
Since the case studies use one-minute averaged data, higher frequency variations in the power production, model parameters, and wind conditions are not included in this numerical experiment, and their influence is considered in LES (\S \ref{sec:les}).

\begin{figure}
    \centering
    \includegraphics[width=\linewidth]{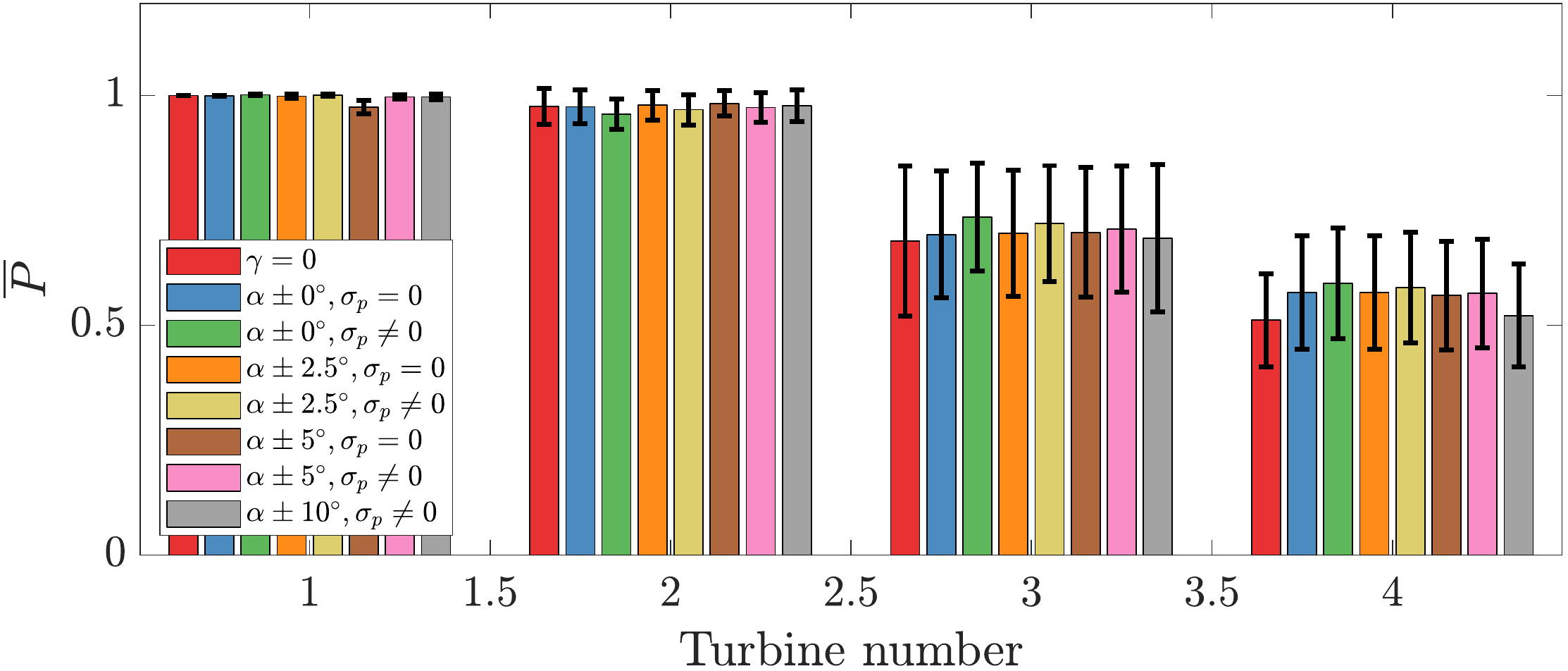}
    \caption{Time averaged power production results from Case 1. The errorbars denote one standard deviation about the mean values.
    The results of this case are summarized in Table \ref{table:power}.}
    \label{fig:power}
\end{figure}

The results from five wake steering case studies are shown in Table \ref{table:power}.
For Cluster B at a low incident turbulence intensity (Case 1), deterministically optimized yaw set-point increases power $2.3\%$, while considering parameter uncertainty increases power $3.8\%$ which is statistically significantly higher.
Incorporating wind direction variability of $\pm 2.5^\circ$ does not significantly impact the results.
The power production results for each optimization strategy in Case 1 are shown in Figure \ref{fig:power}.
Higher wind direction variabilities of $\pm 5^\circ$ or $\pm 10^\circ$ reduce the power significantly in this case by reducing the set-point optimal yaw angles (see Figure \ref{fig:yaw}).
It is worth noting that these case studies only consider frequencies lower than $f_c=1/60$ Hz due to one-minute averaging, while the wind conditions have higher frequency content \cite{rott2018robust}. 
The higher frequencies are included in the empirical wind condition probability distributions in Eq. \ref{eq:expected_value_param_uncertainty} in the LES experiments in \S \ref{sec:les}.

With increasing turbulence intensity, variability in the wind conditions and power production increases.
For the higher turbulence intensity cases, the power increase due to wake steering decreases, which is expected due to the increased mixing and reduced wake interactions.
As the turbulence intensity increases in Cases 3 and 4, increased wind direction variability in the set-point optimization leads to improved results, compared to deterministic or low variability optimization.
Incorporation of variability in the set-point optimization does not significantly influence Cluster A results in Case 5, except for $\alpha \pm 10^\circ$, which produces significantly less power than the other cases.
The relatively small impact of parameter uncertainty in Cluster A is likely due to the small spacing between turbines, low turbulence intensity inflow, and low power production standard deviation (Figure \ref{fig:cluster_A}).

\begin{threeparttable}[htb]
    \caption{Wake steering control numerical experiment power production results provided as a change with respect to baseline yaw aligned control.
    Red and green cases have statistically significantly lower and higher power than $\alpha\pm0.0^\circ$, $\sigma_p=0$, by a two-sided two-sample Kolmogorov Smirnov test at $5\%$ significance level, respectively.
    Deterministic model parameter set-point optimization is represented by $\sigma_p=0$.}
\label{table:power}
    \small
    \setlength\tabcolsep{0pt}
\begin{tabular*}{\linewidth}{@{\extracolsep{\fill}} l cc cc c @{}}
\Xhline{2\arrayrulewidth}
    \toprule
&   \multicolumn{5}{c}{Case} \\
        \cmidrule{2-6}
        \hline
Model uncertainty & (1)\tnote{a} & (2)\tnote{b} & (3)\tnote{c} & (4)\tnote{d} & (5)\tnote{e} \\ 
        \midrule
        \hline
$\alpha\pm0.0^\circ$, $\sigma_p=0$ & $2.3\%$  & $2.2\%$  & $0.4\%$ & $-0.3\%$ & $8.7\%$ \\
$\alpha\pm0.0^\circ$, $\sigma_p\neq0$ & {\color{dartmouthgreen}$3.8\%$}  & {\color{dartmouthgreen}$2.3\%$}  & $0.0\%$ & $-0.1\%$ & $8.8\%$ \\
$\alpha\pm2.5^\circ$, $\sigma_p=0$ & $2.5\%$  & $2.3\%$  & $0.2\%$ & $-0.5\%$ & $8.6\%$ \\
$\alpha\pm2.5^\circ$, $\sigma_p\neq0$  & {\color{dartmouthgreen}$3.3\%$}  & $2.4\%$  & $0.1\%$ & $-0.2\%$ & $8.8\%$ \\
$\alpha\pm5.0^\circ$, $\sigma_p=0$ & {\color{brickred}$1.7\%$}  & $2.1\%$  & $-0.2\%$ & $-0.3\%$ & $8.3\%$ \\
$\alpha\pm5.0^\circ$, $\sigma_p\neq0$ & $2.6\%$  & $2.3\%$  & $0.0\%$ & $0.0\%$ & $8.5\%$ \\
$\alpha\pm10^\circ$, $\sigma_p\neq0$ & {\color{brickred}$0.5\%$}  & $1.3\%$  & {\color{brickred}$-0.3\%$} & $0.1\%$ & {\color{brickred}$4.2\%$} \\
        \bottomrule
        \Xhline{2\arrayrulewidth}
\end{tabular*}
\begin{tablenotes}[para,flushleft,small]
\small
\item[a] (1): Cluster B, $u=7\pm1$ m/s, $TI_u=5\pm2.5\%$, $n=73$\\
\item[b] (2): Cluster B, $u=7\pm2$ m/s, $TI_u=5\pm5\%$, $n=144$ \\
\item[c] (3): Cluster B, $u=7\pm2$ m/s, $TI_u=10\pm2.5\%$, $n=51$ \\
\item[d] (4): Cluster B, $u=9\pm2$ m/s, $TI_u=10\pm2.5\%$, $n=32$ \\
\item[e] (5): Cluster A, $u=7\pm2$ m/s, $TI_u=5\pm2.5\%$, $n=115$ \\
\end{tablenotes}
\end{threeparttable}

In summary, the wake model experiments in this section demonstrate the utility of the yaw set-point optimization under parameter uncertainty approach in an idealized numerical setting with no structural modeling error.
These results also demonstrate that the incorporation of wake model parameter uncertainty in the stochastic programming improves the robustness of the optimal set-points $\gamma_s^*$ to overfitting to the training data.
In the timeseries numerical experiments, the wind conditions and temporally local wake model parameters in the wake model vary.
The deterministic wake model only produces a single set of wake model parameters based on the training data to optimize $\gamma_s$ in the lookup table.
In these numerical examples, the wake steering test results are improved when considering a distribution of wake model parameters from the training data, rather than deterministic parameters.
The more realistic application including modeling error is described in the LES experiments in \S \ref{sec:les}.

\section{Large eddy simulations of closed-loop control in an unstable boundary layer}
\label{sec:les}

In this section, large eddy simulations of a $9$ turbine model wind farm are performed in unstable, convective ABL conditions using closed-loop wake steering control.
In these numerical experiments, the LES wind farm will represent a utility-scale wind farm which is using closed-loop wake steering control to increase power as a function of time in the transient unstable ABL.
An unstable ABL test case was selected, instead of the standard, idealized neutral\cite{doekemeijer2020closed} or conventionally neutral ABL\cite{howland2020optimal}, due to its inherent wind condition variability and its prevalence in a utility-scale wind farm setting, occurring during daytime operation.
The wake model will be used for yaw set-point optimization, as in \S \ref{sec:contol}.
Therefore, this case represents a more realistic numerical experiment where wake modeling error is present, such that the wake model does not resolve all physical phenomena in the LES wind farm.
The LES and closed-loop control setup are described in \S \ref{sec:les_setup} and the results are presented in \S \ref{sec:les_results}.

\subsection{LES and closed-loop control setup}
\label{sec:les_setup}

Large eddy simulations are performed using the open-source pseudo-spectral code {\it Pad{\'e}Ops} (\url{https://github.com/FPAL-Stanford-University/PadeOps}).
The code has previously been used for a variety of LES studies including modeling turbulence in the planetary boundary layer \cite{ghate2017subfilter}, Coriolis effects in the ABL \cite{howland2018influence, howland2020influence, howland2020coriolis}, and multi-rotor turbines \cite{ghaisas2020effect}.
{\it Pad{\'e}Ops} has also been used to perform closed-loop wake steering simulations in the conventionally neutral ABL \cite{howland2020optimal, howland2021influence}.
Full details of the numerical setup are given in Ghate \& Lele (2017) \cite{ghate2017subfilter} and Howland {\it et al.} (2020) \cite{howland2020optimal}, and key details are restated here for completeness.

The flow is forced using a geostrophic pressure gradient corresponding to a geostrophic wind speed magnitude $G=8$ m/s.
The geostrophic wind direction is westerly, aligning with the $x$ axis in the computational domain.
Coriolis effects are included with the traditional approximation enforced \cite{howland2020influence} to enable a validation of the convective ABL case with reference LES data from literature \cite{kumar2006large}.
Wind turbines are represented by the actuator disk model \cite{calaf2010large} described by Howland {\it et al.} (2020) \cite{howland2020optimal}.
The wind turbines have a hub height of $100$ meters and a diameter of $D=126$ meters.
The cosine factor for $C_p$ is set to the conservative value of $P_p=2.5$ since underestimates of $P_p$ significantly degrade wake steering performance\cite{howland2020optimal}.

\begin{figure}
    \centering
    \includegraphics[width=0.65\linewidth]{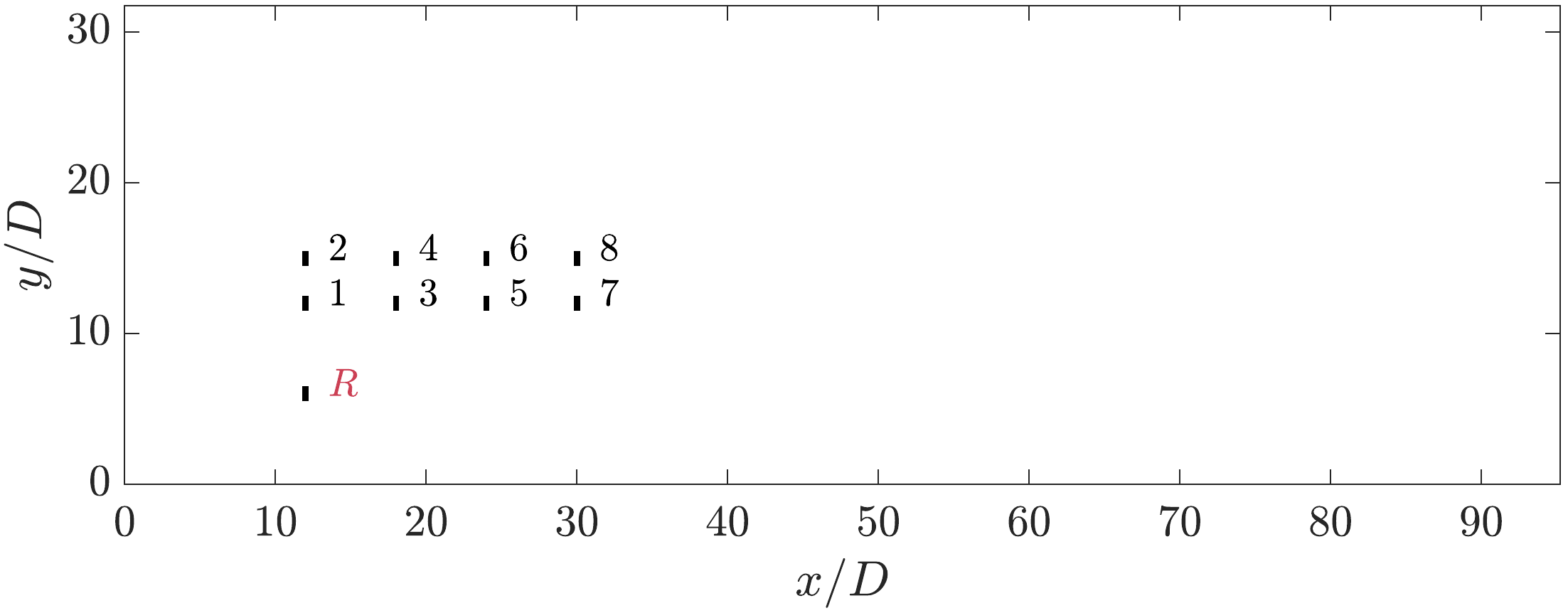}
    \caption{LES wind farm layout with the coordinates normalized by the wind turbine diameter.
    The geostrophic wind direction is in the $x$ direction.
    Turbine $R$, shown in red, is the reference turbine and is yaw aligned in all LES cases.}
    \label{fig:les_layout}
\end{figure}

\begin{figure}
  \centering
  \begin{tabular}{@{}p{0.33\linewidth}@{\quad}p{0.33\linewidth}@{\quad}p{0.33\linewidth}@{}}
    \subfigimgtwo[width=\linewidth,valign=t]{(a)}{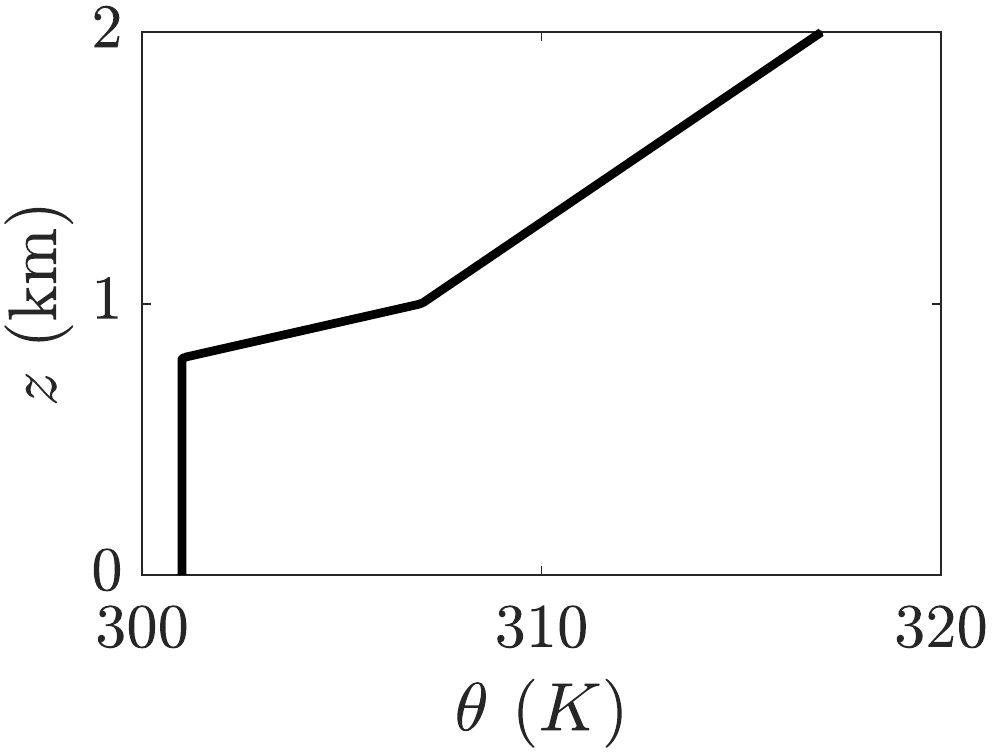} &
    \subfigimgtwo[width=\linewidth,valign=t]{(b)}{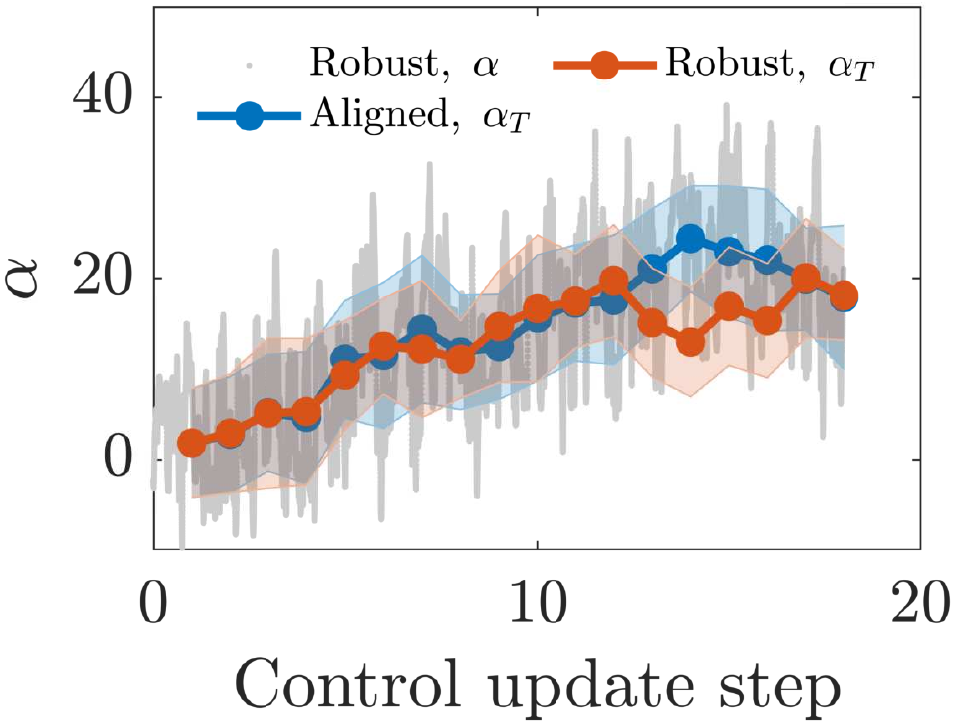} &
    \subfigimgtwo[width=\linewidth,valign=t]{(c)}{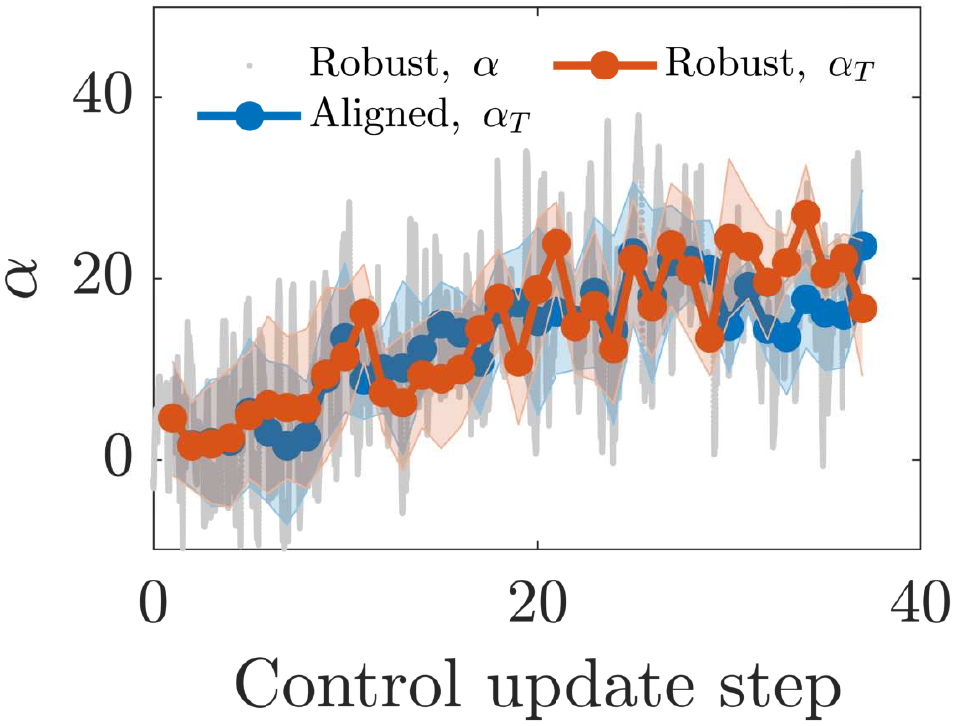}
  \end{tabular}
  \caption{(a) Unstable LES initial potential temperature profile $\theta$ as a function of height $z$.
  (b) Unstable LES wind direction as measured by the reference turbine $R$ (Figure \ref{fig:les_layout}). 
  The finite time averaged wind direction $\alpha_T$ with $T=25$ minutes is shown as a function of the control update steps for the aligned and robust optimization LES cases.
  The shaded regions correspond to one standard deviation around the mean.
  The instantaneous wind direction for the optimization under uncertainty (robust) LES case is shown as $\alpha$.
  (c) Same as (b) for $T=12.5$ minutes.
        }
    \label{fig:les_setup}
\end{figure}

An unstable, convective ABL is simulated using LES.
The domain is $12$ km by $4$ km by $2$ km with 480, 320, and 320 grid points in the $x$, $y$, and $z$ directions, respectively.
The Rossby number based on the initial boundary layer height of $\delta = 1$ km and the geostrophic wind speed is $Ro=110$ and the Froude number is $Fr=0.08$.
The concurrent precursor methodology is used \cite{munters2016turbulent}.
Fringe regions to force the primary simulation to the precursor inflow are used in the last $25\%$ of the computational domain to ensure no upstream contamination of the solution in the region of interest due to the fringe \cite{ghate2018interaction, howland2020coriolis}.
A sponge region is used in the top $25\%$ of the vertical direction \cite{howland2020optimal}.
The ABL is initialized with $u=G$, and all other velocity is zero, and the potential temperature $\theta$ profile provided in Figure \ref{fig:les_setup}(a).
The surface heat flux is prescribed as $\left< w^\prime \theta^\prime \right>=0.1 \  \mathrm{K}\cdot\mathrm{m/s}$, corresponding to unstable ABL conditions.
The simulation is run for one hour to remove startup transience.
The wind farm layout in the primary domain is shown in Figure \ref{fig:les_layout}.
The turbines of interest are $1$ through $8$ and turbine $R$ is used as a yaw aligned reference turbine in all cases.

Within the convective, unstable ABL, three separate yaw set-point optimization cases are run in addition to baseline, yaw aligned control.
For baseline, yaw aligned control, the wind turbines in the wind farm update their nacelle orientation according to a predefined control update period $T$.
The nacelle position is updated to the value of the turbine specific mean wind direction measured during the previous finite time average horizon corresponding to the control update period, i.e. at time $t$, the nacelle position is set to equal $\alpha_T = \frac{1}{T}\int_{t^-T}^{t}\alpha(t^\prime)\mathrm{d}t^\prime$.
While more sophisticated controllers could be implemented (e.g. the yaw control acts if the moving average of instantaneous yaw exceeds a deadband threshold) this approximate yaw alignment controller is used as a representative baseline, yaw aligned control case.

For the wake steering cases, the closed-loop control developed in Howland {\it et al.} (2020) \cite{howland2020optimal} is used, and is briefly described here for completeness.
Using the same control update period as the baseline, yaw aligned control, the nacelle position is updated according to the mean wind direction measured by each turbine plus an additional yaw misalignment offset corresponding to the set-point optimization result $\alpha_T + \gamma_s^*$.
The yaw misalignment set-point is optimized differently in three distinct cases.
In deterministic set-point optimization, the mean wind conditions over the previous control update period measurements are used to optimize the deterministic wake model, with no variability or uncertainty in the wind conditions or wake model parameters.
For the stochastic wind conditions case, the stochastic programming formulation in Eq. \ref{eq:opti_param} is used for variable wind conditions but for deterministic wake model parameters.
The wind condition probability distributions are constructed using the measurements collected over the previous control update window $t-T \rightarrow t$.
The wake model parameters are fit using the standard EnKF, using the mean power production to produce only one set of wake model parameters.
Finally, the full stochastic programming framework is used where the variable wind conditions are used in addition to the uncertain wake model parameter methodology described in \S \ref{sec:contol}.
Each of the four cases, the single yaw aligned case and the three wake steering cases, represents a separate LES simulation.
In all cases, the reference turbine $R$ uses yaw aligned control throughout the simulation.

The four simulations are run for three different control update periods of $T=12.5$, $T=18.75$, and $25$ minutes.
The performance of wake steering in transient flow depends on the frequency of the controller update; with shorter control update periods the yaw controller can react to high frequency variations in the wind conditions \cite{kanev2020dynamic}.
However, following the central limit theorem, for reduced control update periods $T$, there will be increased variability and uncertainty in the finite time averaged quantities, such as power production, which are used to inform the set-point optimization.
Conversely, for longer update periods, the mean flow state may evolve, as can be seen in the wind direction in Figure \ref{fig:les_setup}(b), for example.
The implications of the length of $T$ on finite time averaged statistics, such as $\alpha_T$, is shown in Figure \ref{fig:les_setup}(b,c), for $T=12.5$ and $25$ minutes.
While future work should investigate the optimal update period $T$ as a function of the ABL flow state, in this study we present results for selected, predefined values of $T$.

\begin{figure}
    \centering
    \includegraphics[width=0.75\linewidth]{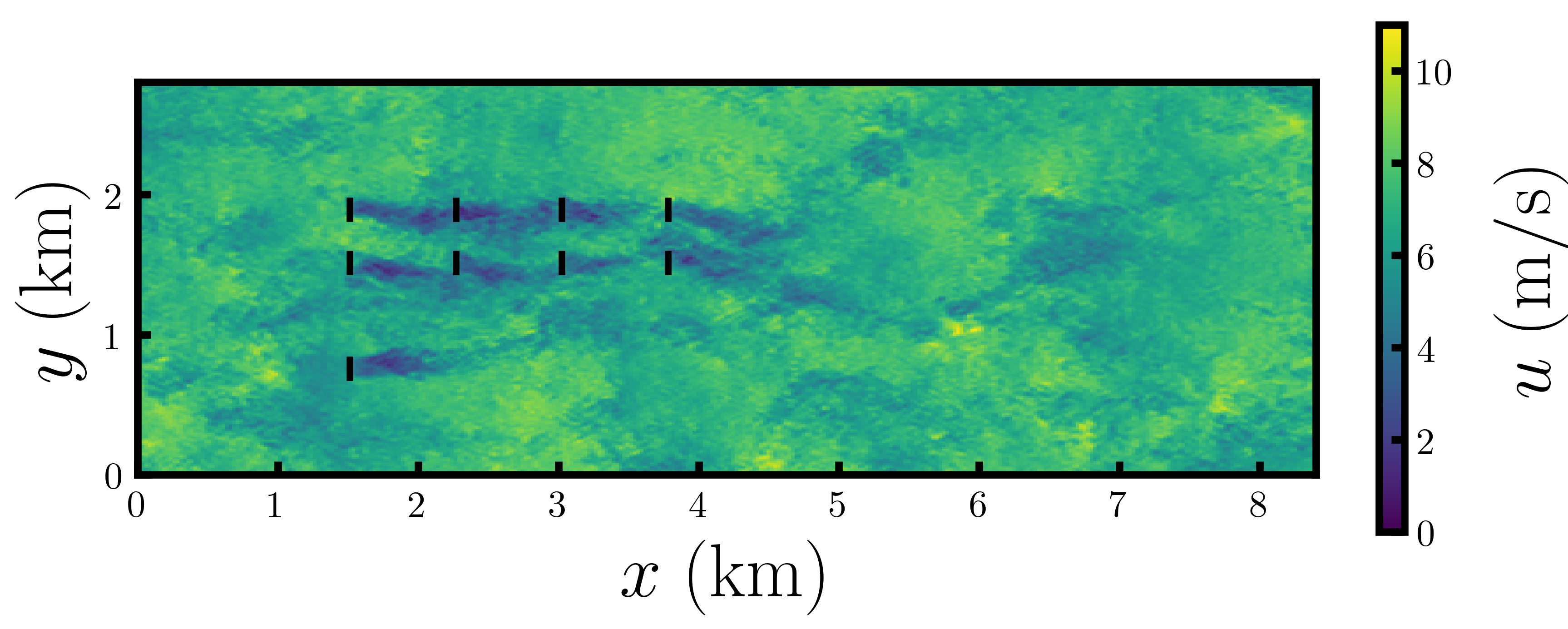}
    \caption{Instantaneous streamwise velocity $u$ flow field at the wind turbine hub-height for the yaw aligned case.}
    \label{fig:les_snapshot}
\end{figure}

\subsection{LES results}
\label{sec:les_results}

Within the convective ABL, there is a high magnitude of turbulence intensity and wind condition variability, compared to the standard neutral or conventionally neutral test cases which have previously been used for closed-loop control numerical experiments \cite{howland2020optimal, doekemeijer2020closed}.
The streamwise turbulence intensity at the wind turbine hub height is approximately $TI=15\%$, although the specific value changes during the transient simulation.
An instantaneous visualization of the streamwise velocity flow field at the wind turbine hub height is shown in Figure \ref{fig:les_snapshot} for the yaw aligned case after the first yaw control update. 
In general, convective ABL conditions produce less significant wake interactions between wind turbines compared to neutral or stable states of stratification\cite{wharton2012atmospheric} due to high turbulence intensity and convective plumes with large length scales.

The wind directions measured by the yaw aligned reference turbine in the robust, stochastic conditions and parameters simulation and the aligned simulation are shown in Figure \ref{fig:les_setup}(b) for $T=25$ minutes and Figure \ref{fig:les_setup}(c) for $T=12.5$ minutes.
The mean wind directions averaged over the control update step are shown in addition to the instantaneous wind direction measurements.
The two cases are initialized from identical states, but the wind direction measurements diverge due to the chaotic nature of turbulent flow.
For all control update steps, the mean wind direction measurements are within one standard deviation of one another, but the magnitude of the variations in the instantaneous wind direction data are significant.
Given the exponential divergence of the states in the chaotic flow and the high magnitude of variability in the convective ABL, integrated energy measures will be used to characterize the performance of each case. 
The integral time scales of the reference turbine power and wind direction measurements are approximately 60 seconds and 90 seconds, respectively.

\begin{figure}
  \centering
  \begin{tabular}{@{}p{0.33\linewidth}@{\quad}p{0.33\linewidth}@{\quad}p{0.33\linewidth}@{}}
    \subfigimgtwo[width=\linewidth,valign=t]{(a)}{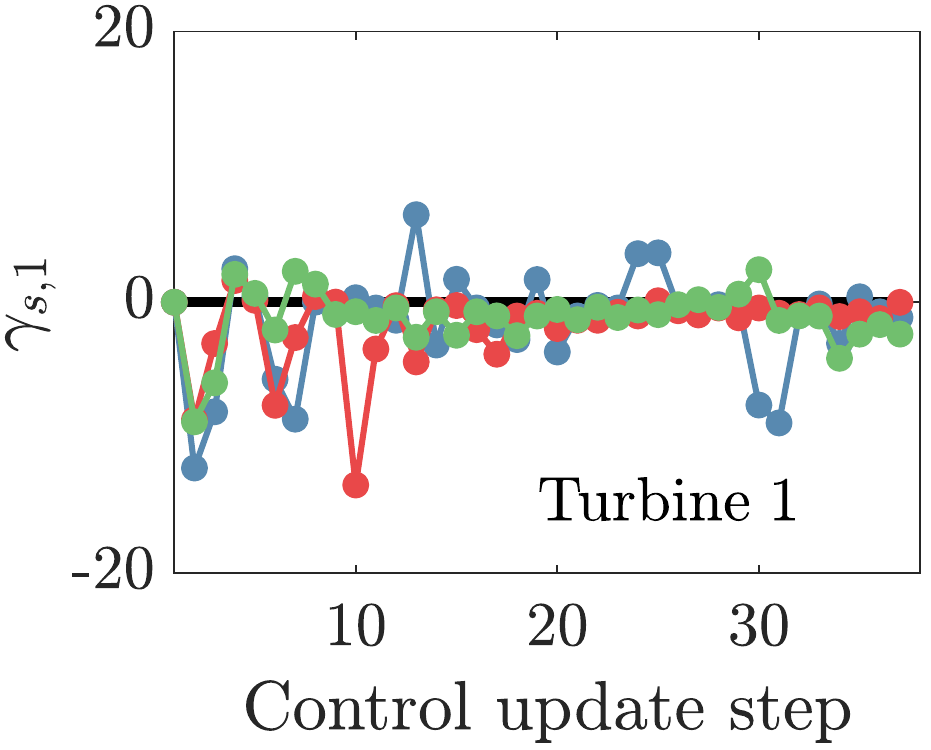} &
    \subfigimgtwo[width=\linewidth,valign=t]{(b)}{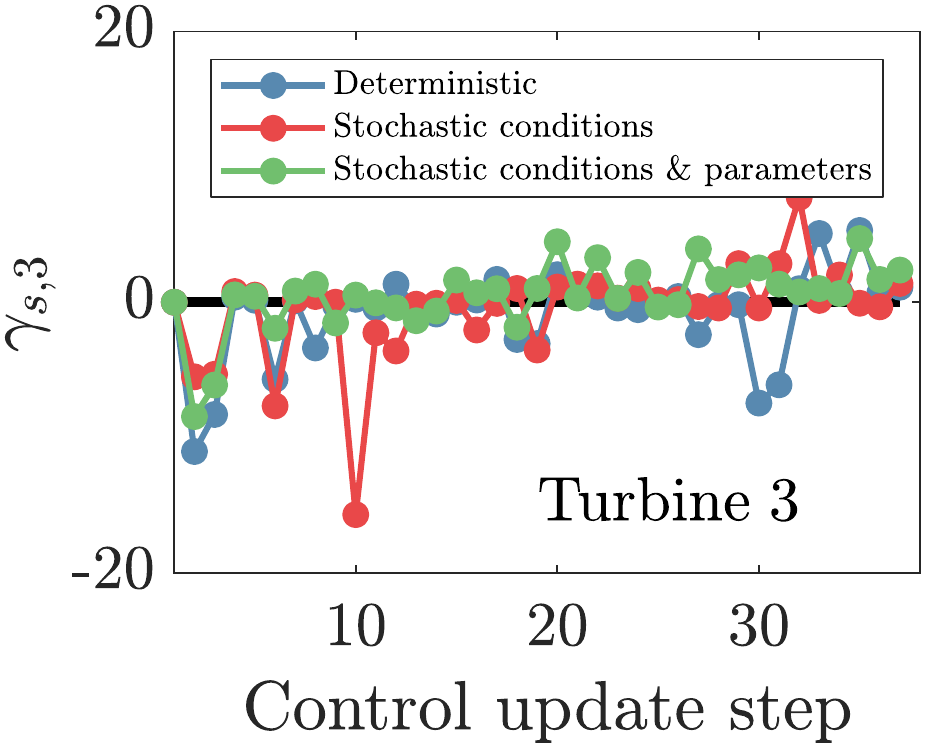} &
    \subfigimgtwo[width=\linewidth,valign=t]{(c)}{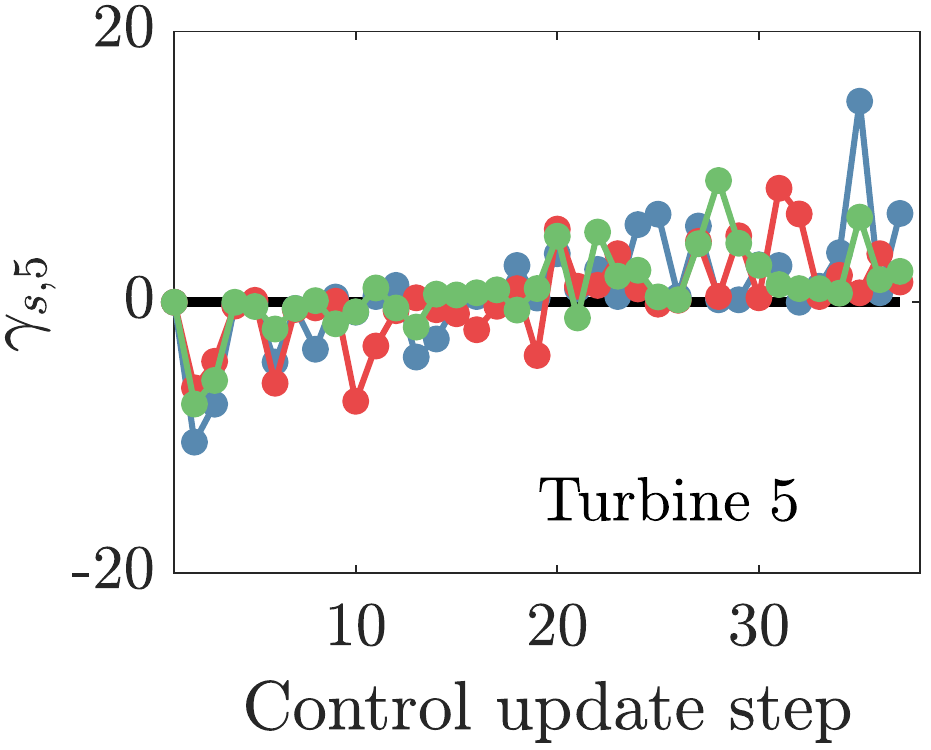}
  \end{tabular}
  \caption{Yaw misalignment set-points as a function of the control update step for $T=12.5$ minutes for turbines (a) $1$, (b) $3$, and (c) $5$.
        }
    \label{fig:T_800}
\end{figure}

The yaw misalignment set-points as a function of the control update step for the control update period of $T=12.5$ minutes are shown in Figure \ref{fig:T_800} for turbines $1$, $3$, and $5$.
The deterministic yaw set-point optimization exhibits significant variability in the set-point value as a function of the control update step.
The deterministic set-point optimization is sensitive to uncertainty in the limited time averaged statistics, since the yaw misalignment decision is based on the singular valued mean wind conditions and wake model parameters which are subject to the internal variability of the ABL flow (e.g. the averaged wind direction $\alpha_T$ will have more inherent variability for $T=12.5$ minutes than $T=25$ minutes as shown in Figure \ref{fig:les_setup}(c)).
Therefore, between control update steps, the finite time averaged statistics may change dramatically, resulting in significant changes in $\gamma_s^*$.

The yaw misalignment set-points for the stochastic programming considering wind condition variability but with deterministic wake model parameters are also shown in Figure \ref{fig:T_800} for $T=12.5$ minutes. 
The incorporation of the stochastic wind conditions in the optimization reduced the magnitude of the yaw misalignment values and the variability of the yaw values between control update steps.
Comparing the optimization with and without parameter uncertainty demonstrates that incorporating the parameter uncertainty mitigates the variations $\gamma_s^*$ as a function of the control update steps.
Provided with limited information in the finite time average over length $T$, the optimization under parameter uncertainty also suppresses overfitting to the mean power production values.

\begin{figure}
  \centering
  \begin{tabular}{@{}p{0.33\linewidth}@{\quad}p{0.33\linewidth}@{\quad}p{0.33\linewidth}@{}}
    \subfigimgtwo[width=\linewidth,valign=t]{(a)}{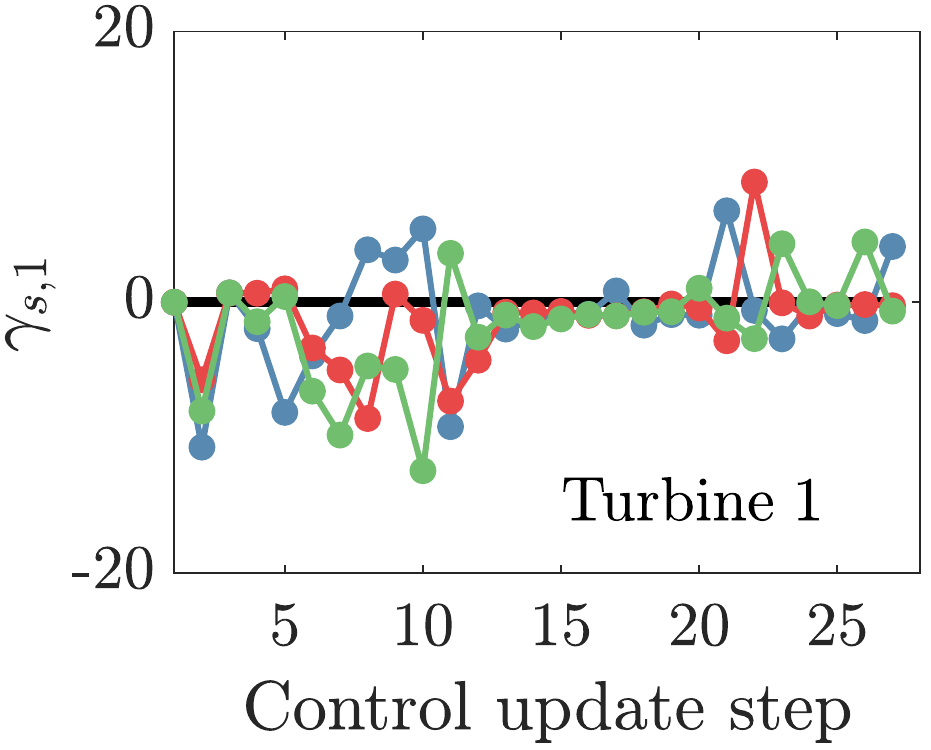} &
    \subfigimgtwo[width=\linewidth,valign=t]{(b)}{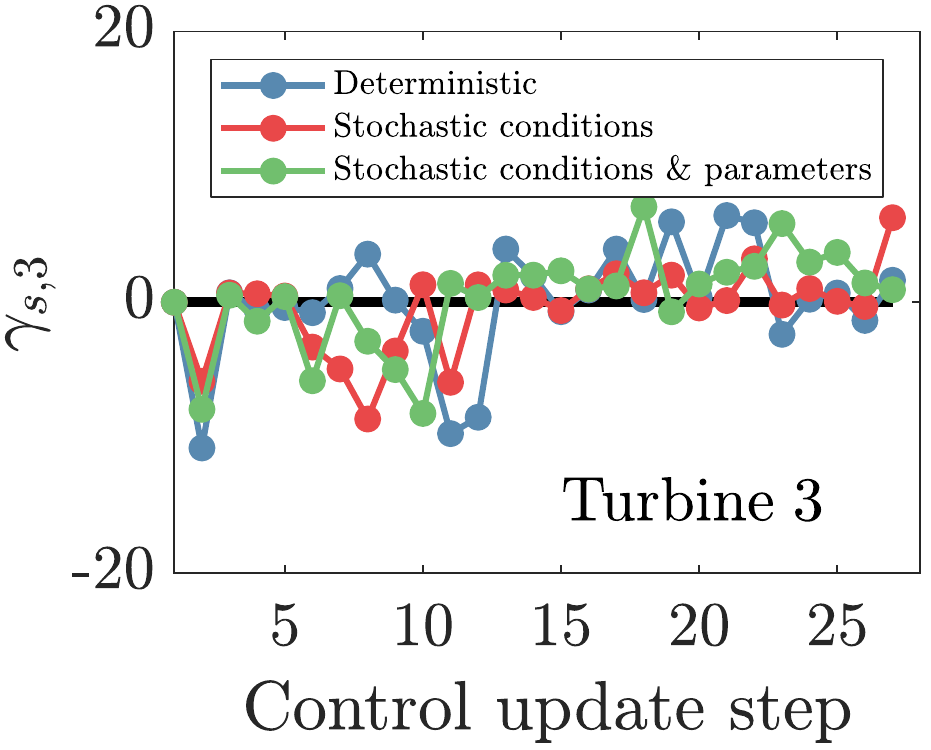} &
    \subfigimgtwo[width=\linewidth,valign=t]{(c)}{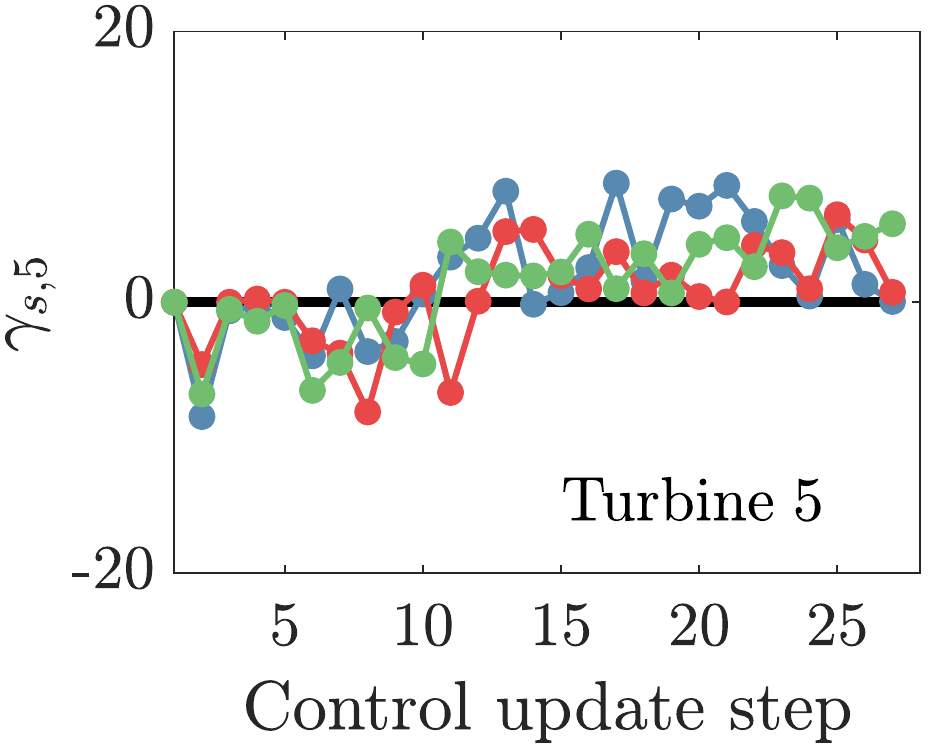}
  \end{tabular}
  \caption{Yaw misalignment set-points as a function of the control update step for $T=18.75$ minutes for turbines (a) $1$, (b) $3$, and (c) $5$.
        }
    \label{fig:T_1200}
\end{figure}

\begin{figure}
  \centering
  \begin{tabular}{@{}p{0.33\linewidth}@{\quad}p{0.33\linewidth}@{\quad}p{0.33\linewidth}@{}}
    \subfigimgtwo[width=\linewidth,valign=t]{(a)}{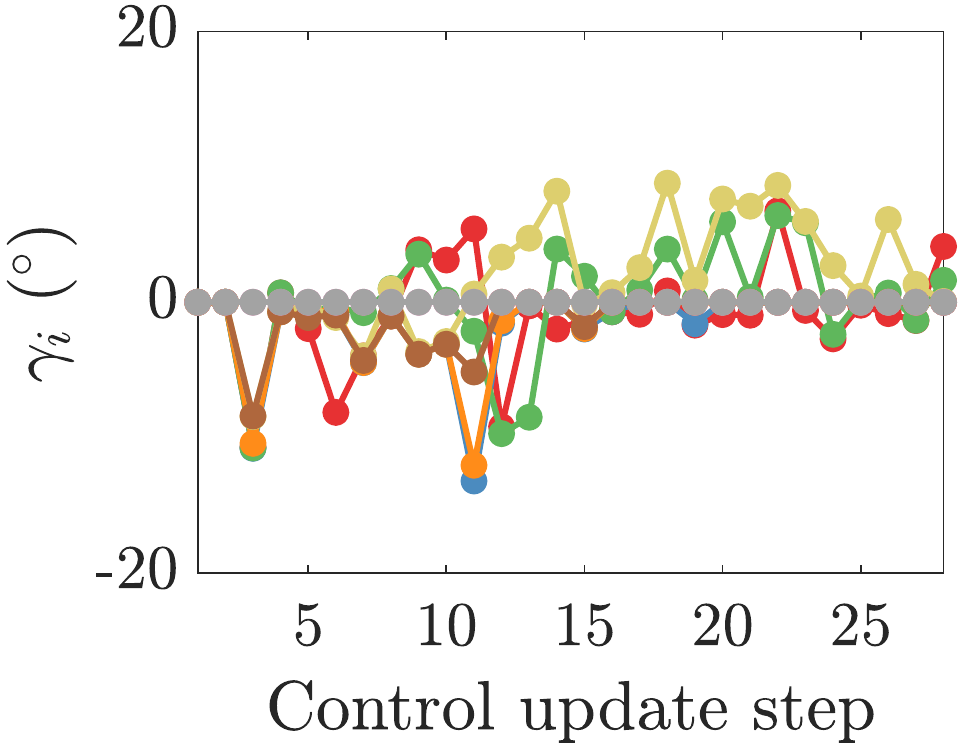} &
    \subfigimgtwo[width=\linewidth,valign=t]{(b)}{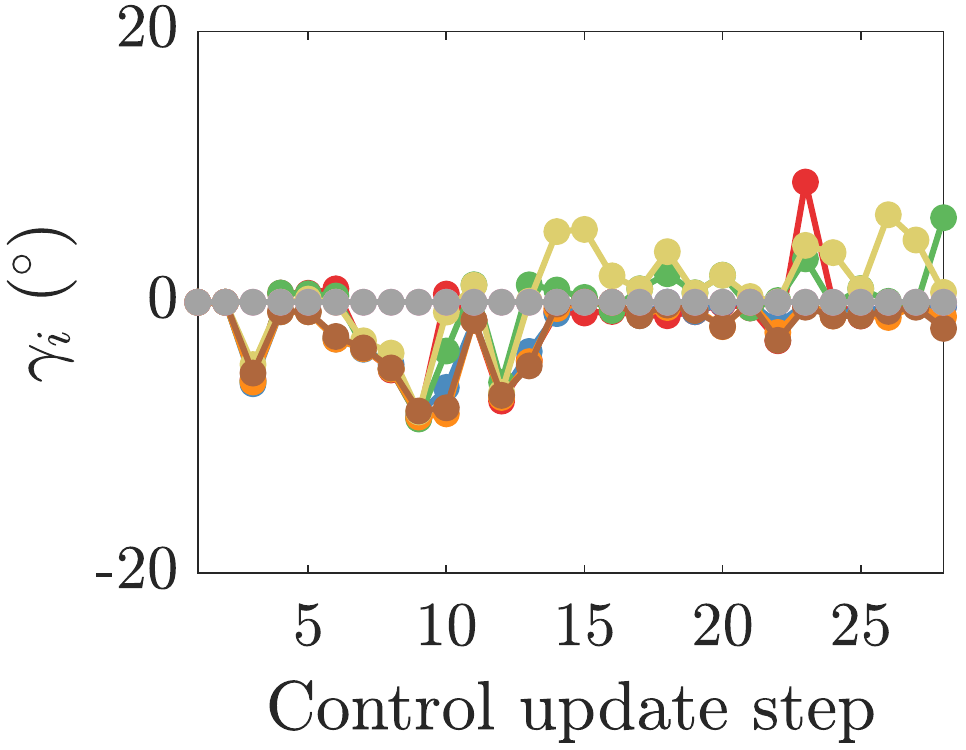} &
    \subfigimgtwo[width=\linewidth,valign=t]{(c)}{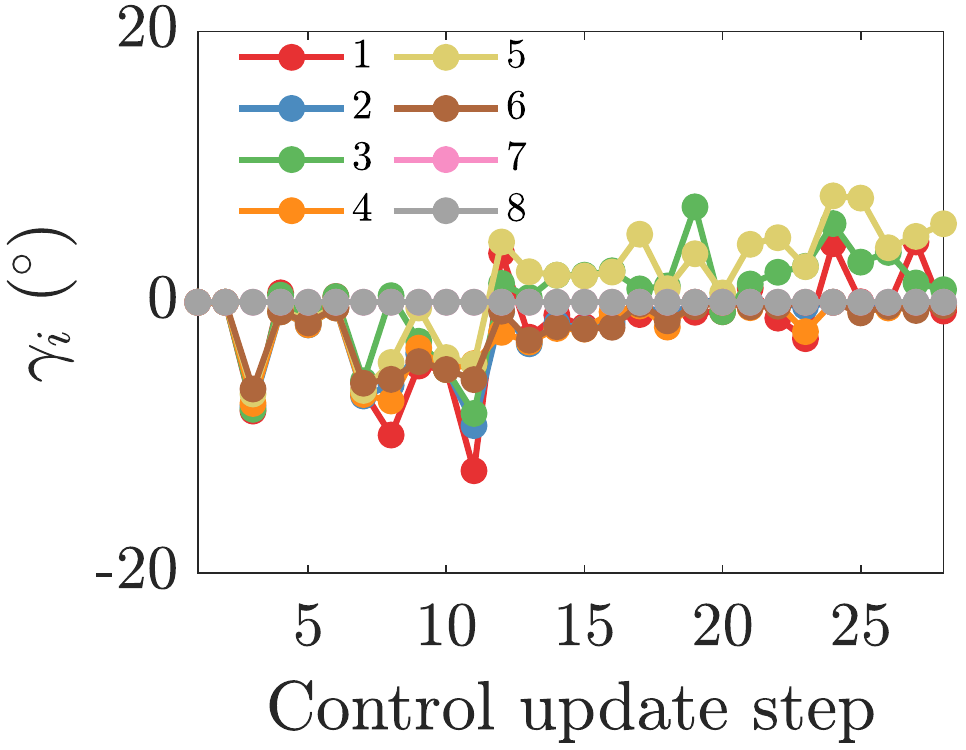}
  \end{tabular}
  \caption{Yaw misalignment set-points as a function of the control update step for $T=18.75$ minutes. 
  (a) Deterministic yaw set-point optimization, (b) yaw set-point optimization with variable wind conditions, and (c) yaw set-point optimization with variable wind conditions and uncertain wake model parameters.
        }
    \label{fig:T_1200_yaw_all}
\end{figure}

\begin{figure}
  \centering
  \begin{tabular}{@{}p{0.33\linewidth}@{\quad}p{0.33\linewidth}@{\quad}p{0.33\linewidth}@{}}
    \subfigimgtwo[width=\linewidth,valign=t]{(a)}{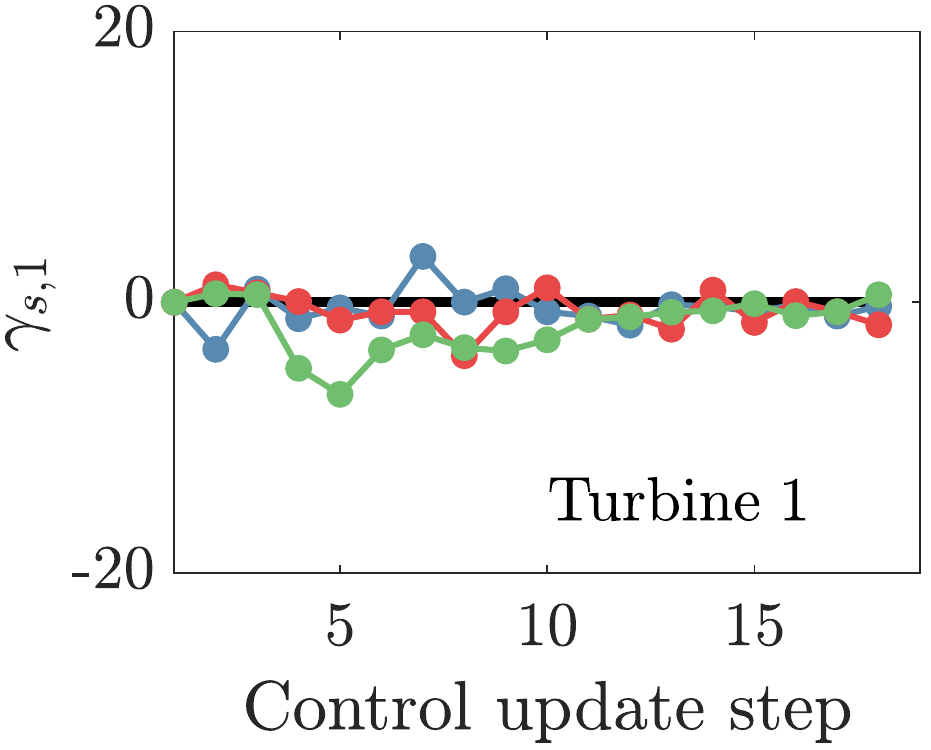} &
    \subfigimgtwo[width=\linewidth,valign=t]{(b)}{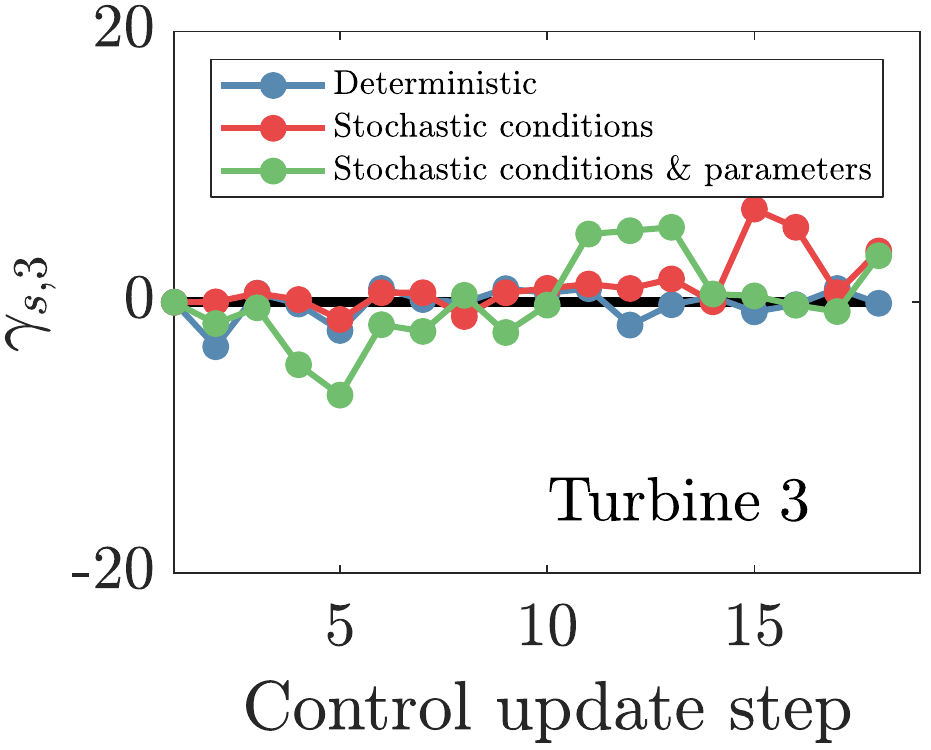} &
    \subfigimgtwo[width=\linewidth,valign=t]{(c)}{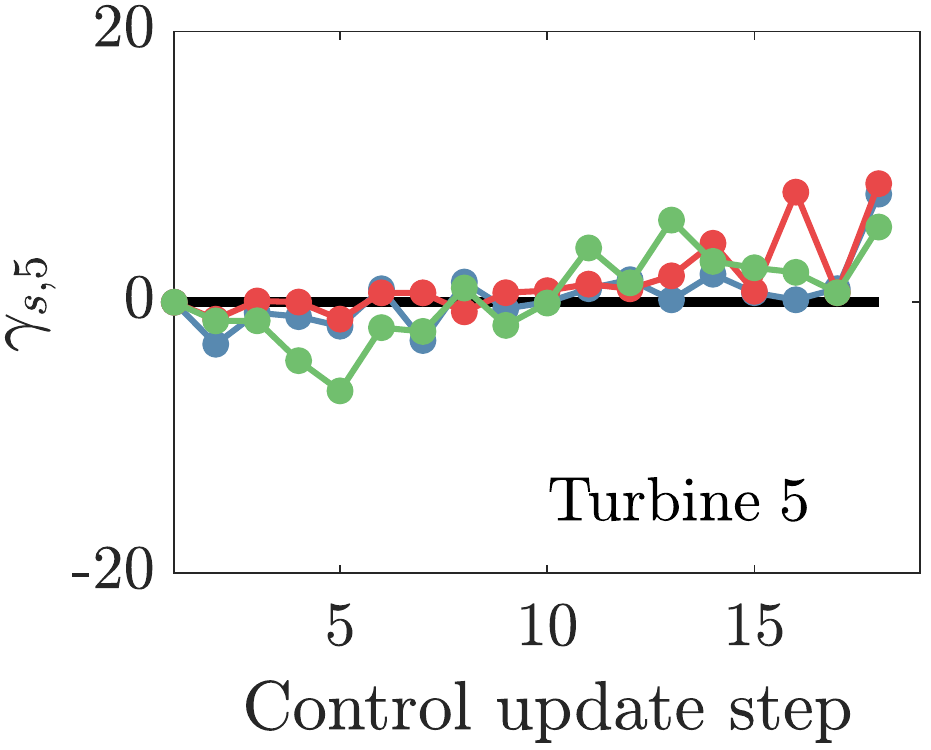}
  \end{tabular}
  \caption{Yaw misalignment set-points as a function of the control update step for $T=25$ minutes for turbines (a) $1$, (b) $3$, and (c) $5$.
        }
    \label{fig:T_1600}
\end{figure}

The yaw set-points for $T=18.75$ minutes are shown in Figures \ref{fig:T_1200} and \ref{fig:T_1200_yaw_all}.
The yaw misalignment values for all eight turbines in the LES for $T=18.75$ are shown individually in Appendix \ref{sec:yaw_appendix}.
For the $T=25$ minutes simulations, the yaw set-points are shown in Figure \ref{fig:T_1600}.
The set-point variability results are similar to $T=12.5$ and $T=18.75$ minutes.
However, contrarily to $T=12.5$ minutes, the stochastic conditions and parameters case resulted in the largest absolute value of yaw misalignment with $T=25$ minutes.
With a larger $T$, the states recorded over the time window $t-T \rightarrow t$ used to inform the yaw set-point optimization at time $t$ have a larger variability.
Depending on the joint distributions of the wind conditions and wake model parameters, the optimization under parameter uncertainty may result in larger absolute values of yaw misalignment set-points due to their nonlinear influence on wind farm power production in the wake model.
In this particular numerical example, this occurred for $T=25$ minutes, but not for $T=12.5$ minutes, indicating the complex relationship between the wind conditions and wake model parameters and the optimal yaw set-points.

The robustness of the yaw misalignment set-point values to the control update period is tested by comparing the set-points for different update period selections with fixed optimization methods.
For example, the yaw set-points for deterministic optimization will be compared for $T=12.5$ and $T=18.75$ minutes.
The different simulations are run over the same physical time window of $T_f\approx9$ hours.
The lower control update periods result in more frequent yaw updates.
For comparison, the larger control update period yaw set-points are linearly interpolated to match the discrete instances of set-points for the more frequent yaw update case.
The upsampled low frequency set-points are denoted by $\tilde{\gamma}_s$.
The resulting discrepancy between the two different control update periods is denoted by a mean squared deviation 
\begin{equation}
\Delta \gamma_s = \frac{1}{N_t} \sum_{i=1}^{N_t} \left<(\gamma_{s,T_{\mathrm{low},i}} - \tilde{\gamma}_{s,T_{\mathrm{high},i}})^2\right>,
\label{eq:yaw_robustness}
\end{equation}
where $\left<\cdot\right>$ denotes the time average. 
Turbines $1$ through $6$ are considered in the summation.
The reference turbine and turbines $7$ and $8$ are not considered since their yaw set-points are zero for all cases and for all times.
The results are shown in Table \ref{table:robustness}.
Considering uncertainty in the wind conditions and wake model parameters improved the robustness of the wake steering approach to decreases in $T$.

The performance of each case is characterized using an integrated energy ratio
\begin{equation}
E_r = \frac{ \sum_{i=1}^{N_t} \int_0^{T_f} P_i(t) \mathrm{d}t}{N_t \int_0^{T_f} P^{ref}(t) \mathrm{d}t},
\label{eq:er}
\end{equation}
which quantifies the wind farm performance compared to the reference turbine $P^{ref}$ (turbine $R$, Figure \ref{fig:les_layout}).
The energy ratio $E_r$ is, in practice, bounded by the limits $[0,1]$, where $0$ corresponds to the wind farm turbines producing zero power and $1$ indicating zero wake losses in the wind farm on average.
The energy ratio metrics are integrated for approximately $T_f=9$ hours of physical wind farm operation; $T_f$ is fixed for all cases.


\begin{threeparttable}[htb]
    \caption{Unstable ABL LES closed-loop wake steering yaw misalignment set-point $\gamma_{s}$ robustness with respect to the control update period.
    The set-points for each control update period simulation are compared.
    The set-point results are compared to different update period settings with fixed optimization methodology.
    The lower frequency (larger update period) set-point values are linearly interpolated to match the higher frequency in the comparison (denoted $\tilde{\gamma}_s$). 
    The mean squared deviation $\Delta \gamma_s$ is computed using Eq. \ref{eq:yaw_robustness}.
    }
\label{table:robustness}
    \small
    \setlength\tabcolsep{0pt}
\begin{tabular*}{\linewidth}{@{\extracolsep{\fill}} l c cc @{}}
\Xhline{2\arrayrulewidth}
    \toprule
&   \multicolumn{3}{c}{Case} \\
        \cmidrule{2-4}
        \hline
Yaw update periods & Deterministic & Stochastic conditions & \begin{tabular}{@{}c@{}}Stochastic conditions \\ \& parameters\end{tabular}\\ 
        \midrule
        \hline
$\left<(\gamma_{s,12.5 \ \mathrm{min}} - \tilde{\gamma}_{s,18.75 \ \mathrm{min}})^2\right>$  & $11.8$  & $8.7$ & $6.9$ \\
$\left<(\gamma_{s,18.75 \ \mathrm{min}} - \tilde{\gamma}_{s,25 \ \mathrm{min}})^2\right>$  & $13.9$  & $8.5$ & $7.8$ \\
$\left<(\gamma_{s,12.5 \ \mathrm{min}} - \tilde{\gamma}_{s,25 \ \mathrm{min}})^2\right>$  & $12.2$  & $9.9$ & $8.0$ \\
        \bottomrule
        \Xhline{2\arrayrulewidth}
\end{tabular*}
\end{threeparttable}

The $E_r$ results are shown in Table \ref{table:power_les}.
Given the large natural variation in this unstable ABL, the impact of wake steering on wind farm power production is not expected to be statistically significant.
With $T=12.5$ minutes, the stochastic conditions and stochastic conditions and parameters optimization cases marginally improve mean power compared to aligned control and deterministic optimization-based control, but the differences are not significant.
For $T=25$ minutes, the stochastic conditions only optimization has a mean $E_r$ much lower than aligned control and deterministic optimization results in the highest mean $E_r$.
The stochastic parameters and conditions optimization case slightly underperforms aligned control.
An ensemble average is taken over the simulation results with different values of $T$.
Overall, the optimization under model parameter and wind condition uncertainty had the highest $E_r$, followed by the aligned control (no wake steering) case.
Wake steering with set-points optimized under wind condition uncertainty without parameter uncertainty resulted in a lower ensemble $E_r$ than aligned control.
Deterministic optimization had the lowest ensemble $E_r$, since the deterministic set-point optimization was sensitive to reductions in $T$.

\begin{threeparttable}[htb]
    \caption{Unstable ABL LES closed-loop wake steering energy ratio $E_r$ (Eq. \ref{eq:er}) results.
    The ensemble averages of the results over the different yaw update periods for fixed optimization methodologies are shown.}
\label{table:power_les}
    \small
    \setlength\tabcolsep{0pt}
\begin{tabular*}{\linewidth}{@{\extracolsep{\fill}} l cc cc @{}}
\Xhline{2\arrayrulewidth}
    \toprule
&   \multicolumn{4}{c}{Case} \\
        \cmidrule{2-5}
        \hline
Yaw update period & Aligned & Deterministic & Stochastic conditions & \begin{tabular}{@{}c@{}}Stochastic conditions \\ \& parameters\end{tabular}\\ 
        \midrule
        \hline
$T=12.5$ min & $0.841 \pm 0.174$  & $0.841 \pm 0.185$  & $0.864 \pm 0.196$ & $0.862 \pm 0.193$ \\
$T=18.75$ min & $0.842 \pm 0.208$  & $0.823 \pm 0.223$  & $0.864 \pm 0.196$ & $0.840 \pm 0.198$ \\
$T=25$ min & $0.864 \pm 0.195$  & $0.870 \pm 0.195$  & $0.817 \pm 0.176$ & $0.850 \pm 0.176$ \\
\hline
Average & $0.849$ & $0.845$ & $0.848$ & $0.851$ \\ 
        \bottomrule
        \Xhline{2\arrayrulewidth}
\end{tabular*}
\end{threeparttable}

In summary, while the closed-loop wake steering control in unstable ABL conditions did not demonstrate statistically significant differences in wind farm power production, the results demonstrated that the incorporation of stochastic wind conditions and wake model parameters in the yaw set-point optimization provides robustness to the control update and statistics integration length $T$.
The stochastic parameter optimization case also resulted in the lowest variability in the yaw misalignment set-points $\Delta \gamma_s$ as a function of the control update period.
The ensemble average of energy ratios for the three control update period simulations was highest for the optimization under parameter uncertainty.

\section{Conclusions}
\label{sec:conclusions}

This paper extends yaw misalignment set-point optimization to include model parameter uncertainty.
Previous studies have extended deterministic yaw set-point optimization to consider variability in the wind turbine yaw misalignment and incident wind direction.
Wake model parameters are inherently variable and uncertain; the approach described in this paper leverages ensemble Kalman filter parameter estimation to compute the probability distribution of the model parameters given a probability distribution of SCADA power data.
The uncertain model parameters are used in a optimization framework which also incorporates wind speed and direction, yaw misalignment, and turbulence intensity variability.

The optimization under parameter uncertainty framework is experimented in open-loop wake steering wake model-based case studies where the incorporation of variability in the wake model parameters and the wind direction has a statistically significant impact on the power production in Cluster B at low turbulence intensity, but not in Cluster A at low turbulence intensity or Cluster B at high turbulence intensity.
Notably, considering uncertainty in both wind conditions and wake model parameters in the set-point optimization had a statistically significantly higher power production than considering wind condition uncertainty alone.

The optimization under parameter uncertainty framework is also tested in closed-loop wake steering control of a nine wind turbine wind farm in large eddy simulations of convective atmospheric boundary layer conditions.
The optimization under parameter uncertainty framework improved the robustness of the closed-loop wake steering controller to the yaw misalignment update period and reduced the variability in the yaw misalignment set-points as a function of the control update steps.
However, incorporating parameter uncertainty did not have a statistically significant impact on the wind farm energy production due to the underlying variability in the power production in the convective boundary layer; notably, none of the wake steering cases had a significant power impact since the standard deviation of the baseline yaw-aligned control efficiency was around $20\%$.

Future work should consider sampling methods, such as MCMC\cite{zhang2020quantification}, to approximate the Bayesian posterior distribution used in the stochastic programming framework.
Future work should also investigate the joint-probability distributions of the various parameters of interest in the wake steering problem.
Finally, future work should consider model bias in the form of structural model form uncertainty and its influence on the Bayesian model parameter posterior distributions.

\begin{acknowledgements}
The author would like to thank John Dabiri, Aditya Ghate, and Carl Shapiro for thoughtful comments on the work and the manuscript.
All simulations were performed on Stampede2 supercomputer under the XSEDE project ATM170028.
\end{acknowledgements}


\FloatBarrier

\appendix

\section{Convective LES wind farm yaw misalignment}
\label{sec:yaw_appendix}

\begin{figure}
  \centering
  \begin{tabular}{@{}p{0.4\linewidth}@{\quad}p{0.4\linewidth}@{}}
    \subfigimgtwo[width=\linewidth,valign=t]{(a)}{figures/les/T_1200/yaw_subfigs/yaw_1-eps-converted-to.pdf} &
    \subfigimgtwo[width=\linewidth,valign=t]{(b)}{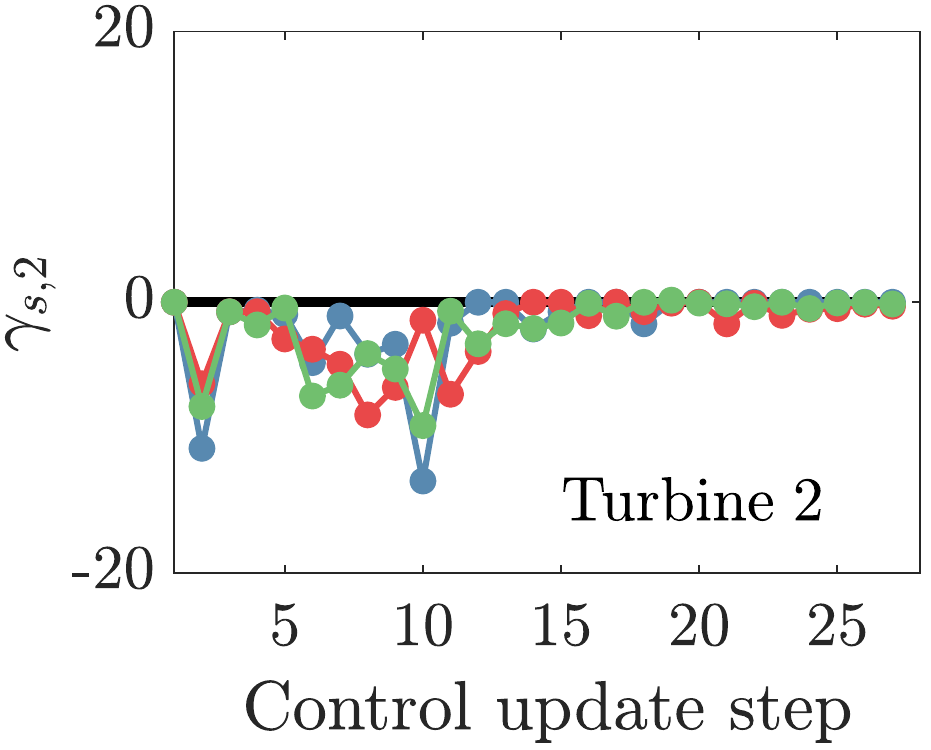} \\
    \subfigimgtwo[width=\linewidth,valign=t]{(c)}{figures/les/T_1200/yaw_subfigs/yaw_3-eps-converted-to.pdf} &
    \subfigimgtwo[width=\linewidth,valign=t]{(d)}{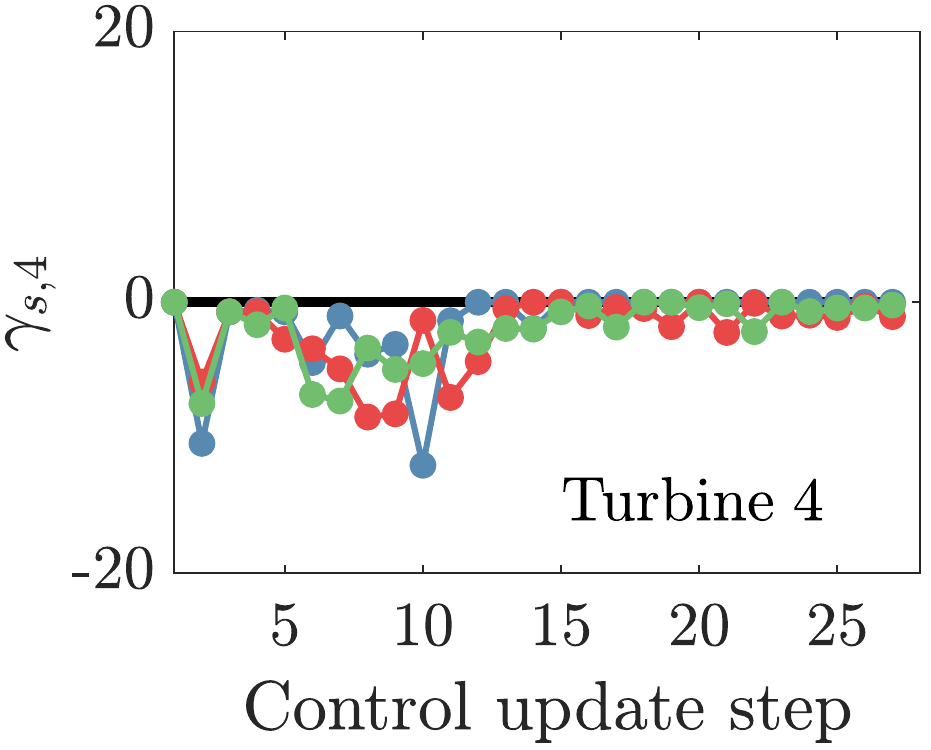} \\
    \subfigimgtwo[width=\linewidth,valign=t]{(e)}{figures/les/T_1200/yaw_subfigs/yaw_5-eps-converted-to.pdf} &
    \subfigimgtwo[width=\linewidth,valign=t]{(f)}{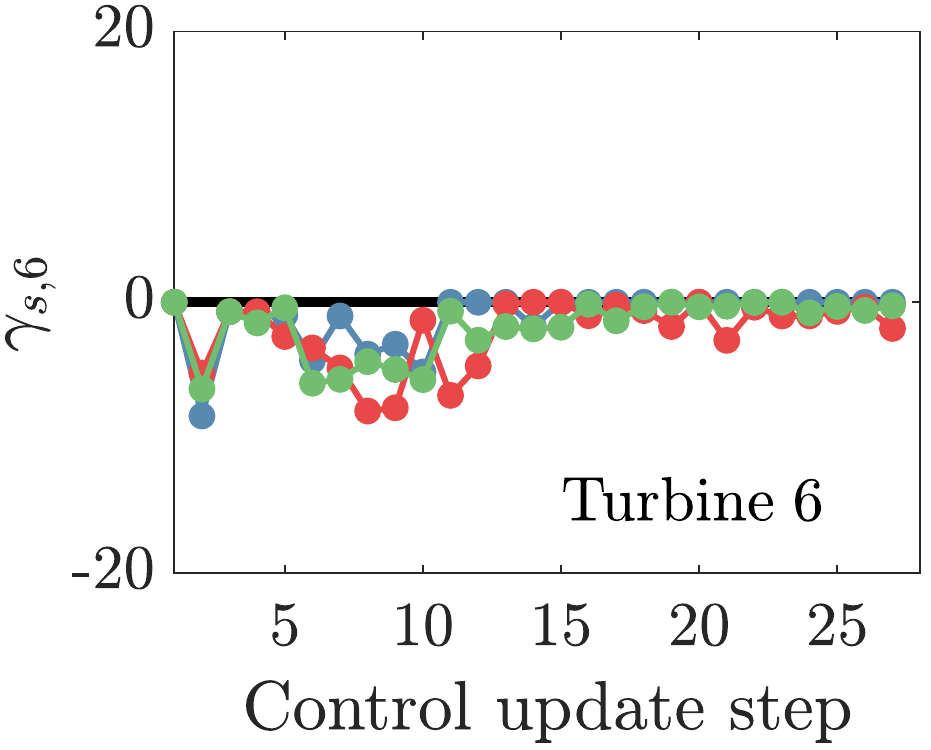} \\
    \subfigimgtwo[width=\linewidth,valign=t]{(g)}{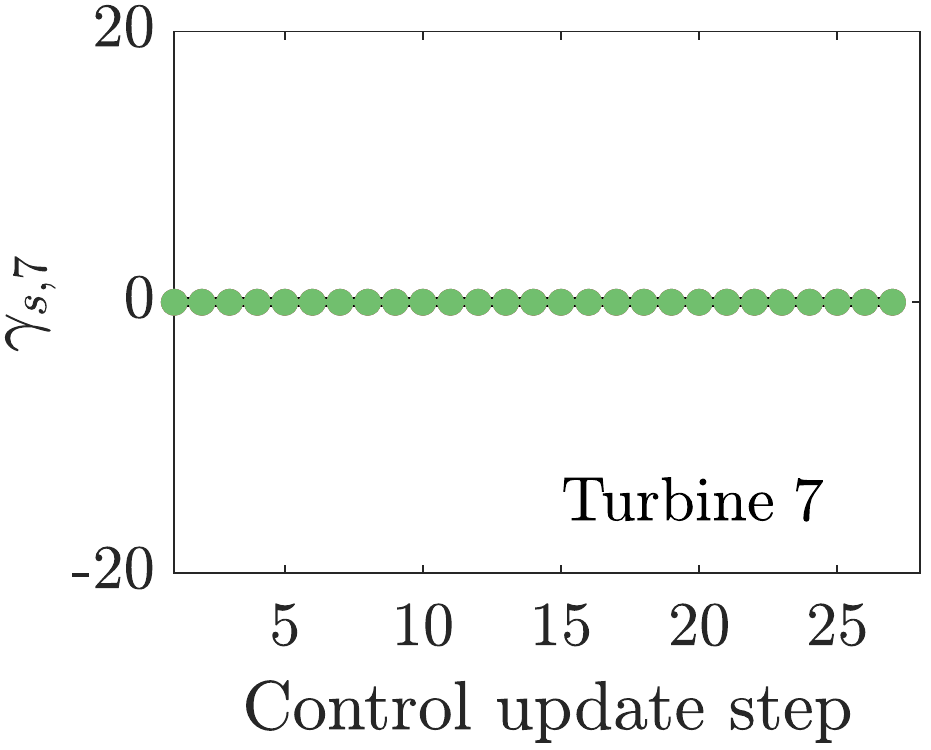} &
    \subfigimgtwo[width=\linewidth,valign=t]{(h)}{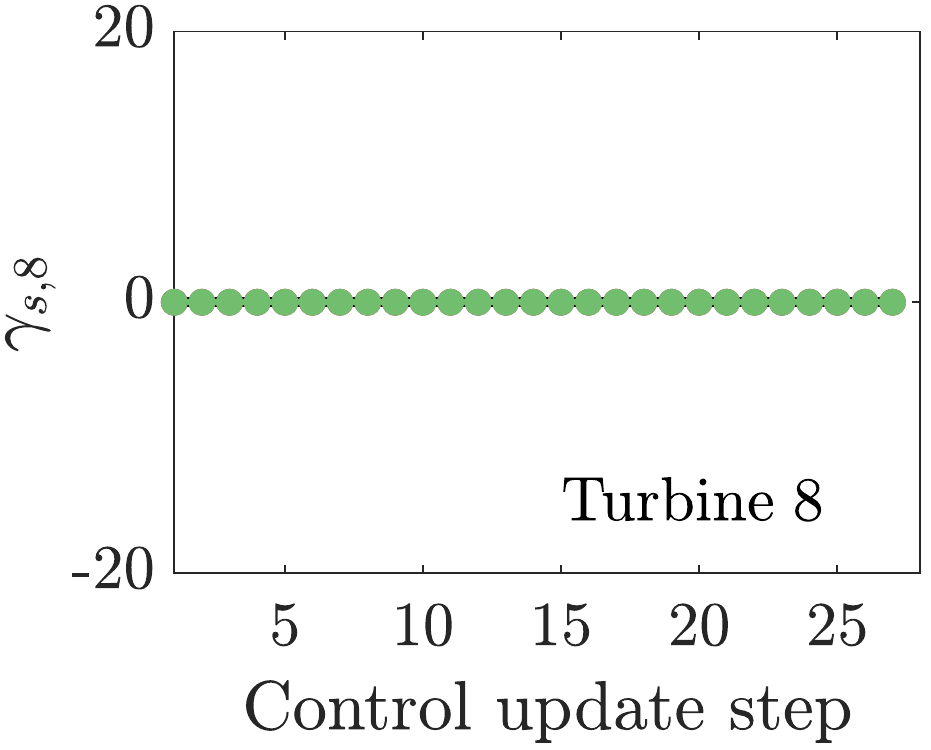}
  \end{tabular}
  \caption{Yaw misalignment set-points as a function of the control update step for $T=18.75$ minutes for all turbines in the LES model wind farm.
  The turbine layout and labels are provided in Figure \ref{fig:les_layout}.
        }
    \label{fig:yaw_all_T_1200}
\end{figure}

The yaw misalignment values for all 8 turbines considered in the unstable ABL LES closed-loop control experiments (\S \ref{sec:les}) for the $T=18.75$ minutes case are shown in Figure \ref{fig:yaw_all_T_1200}.

\FloatBarrier

\bibliography{main}

\begin{thebibliography}{48}%
\makeatletter
\providecommand \@ifxundefined [1]{%
 \@ifx{#1\undefined}
}%
\providecommand \@ifnum [1]{%
 \ifnum #1\expandafter \@firstoftwo
 \else \expandafter \@secondoftwo
 \fi
}%
\providecommand \@ifx [1]{%
 \ifx #1\expandafter \@firstoftwo
 \else \expandafter \@secondoftwo
 \fi
}%
\providecommand \natexlab [1]{#1}%
\providecommand \enquote  [1]{``#1''}%
\providecommand \bibnamefont  [1]{#1}%
\providecommand \bibfnamefont [1]{#1}%
\providecommand \citenamefont [1]{#1}%
\providecommand \href@noop [0]{\@secondoftwo}%
\providecommand \href [0]{\begingroup \@sanitize@url \@href}%
\providecommand \@href[1]{\@@startlink{#1}\@@href}%
\providecommand \@@href[1]{\endgroup#1\@@endlink}%
\providecommand \@sanitize@url [0]{\catcode `\\12\catcode `\$12\catcode
  `\&12\catcode `\#12\catcode `\^12\catcode `\_12\catcode `\%12\relax}%
\providecommand \@@startlink[1]{}%
\providecommand \@@endlink[0]{}%
\providecommand \url  [0]{\begingroup\@sanitize@url \@url }%
\providecommand \@url [1]{\endgroup\@href {#1}{\urlprefix }}%
\providecommand \urlprefix  [0]{URL }%
\providecommand \Eprint [0]{\href }%
\providecommand \doibase [0]{http://dx.doi.org/}%
\providecommand \selectlanguage [0]{\@gobble}%
\providecommand \bibinfo  [0]{\@secondoftwo}%
\providecommand \bibfield  [0]{\@secondoftwo}%
\providecommand \translation [1]{[#1]}%
\providecommand \BibitemOpen [0]{}%
\providecommand \bibitemStop [0]{}%
\providecommand \bibitemNoStop [0]{.\EOS\space}%
\providecommand \EOS [0]{\spacefactor3000\relax}%
\providecommand \BibitemShut  [1]{\csname bibitem#1\endcsname}%
\let\auto@bib@innerbib\@empty
\bibitem [{\citenamefont {Howland}\ \emph {et~al.}(2016)\citenamefont
  {Howland}, \citenamefont {Bossuyt}, \citenamefont {Mart{\'\i}nez-Tossas},
  \citenamefont {Meyers},\ and\ \citenamefont {Meneveau}}]{howland2016wake}%
  \BibitemOpen
  \bibfield  {author} {\bibinfo {author} {\bibfnamefont {M.~F.}\ \bibnamefont
  {Howland}}, \bibinfo {author} {\bibfnamefont {J.}~\bibnamefont {Bossuyt}},
  \bibinfo {author} {\bibfnamefont {L.~A.}\ \bibnamefont
  {Mart{\'\i}nez-Tossas}}, \bibinfo {author} {\bibfnamefont {J.}~\bibnamefont
  {Meyers}}, \ and\ \bibinfo {author} {\bibfnamefont {C.}~\bibnamefont
  {Meneveau}},\ }\bibfield  {title} {\enquote {\bibinfo {title} {Wake structure
  in actuator disk models of wind turbines in yaw under uniform inflow
  conditions},}\ }\href@noop {} {\bibfield  {journal} {\bibinfo  {journal} {J.
  Renew. Sustain. Energy}\ }\textbf {\bibinfo {volume} {8}},\ \bibinfo {pages}
  {043301} (\bibinfo {year} {2016})}\BibitemShut {NoStop}%
\bibitem [{\citenamefont {Gebraad}\ \emph {et~al.}(2016)\citenamefont
  {Gebraad}, \citenamefont {Teeuwisse}, \citenamefont {Van~Wingerden},
  \citenamefont {Fleming}, \citenamefont {Ruben}, \citenamefont {Marden},\ and\
  \citenamefont {Pao}}]{gebraad2016wind}%
  \BibitemOpen
  \bibfield  {author} {\bibinfo {author} {\bibfnamefont {P.}~\bibnamefont
  {Gebraad}}, \bibinfo {author} {\bibfnamefont {F.}~\bibnamefont {Teeuwisse}},
  \bibinfo {author} {\bibfnamefont {J.}~\bibnamefont {Van~Wingerden}}, \bibinfo
  {author} {\bibfnamefont {P.~A.}\ \bibnamefont {Fleming}}, \bibinfo {author}
  {\bibfnamefont {S.}~\bibnamefont {Ruben}}, \bibinfo {author} {\bibfnamefont
  {J.}~\bibnamefont {Marden}}, \ and\ \bibinfo {author} {\bibfnamefont
  {L.}~\bibnamefont {Pao}},\ }\bibfield  {title} {\enquote {\bibinfo {title}
  {Wind plant power optimization through yaw control using a parametric model
  for wake effects—a {CFD} simulation study},}\ }\href@noop {} {\bibfield
  {journal} {\bibinfo  {journal} {Wind Energy}\ }\textbf {\bibinfo {volume}
  {19}},\ \bibinfo {pages} {95--114} (\bibinfo {year} {2016})}\BibitemShut
  {NoStop}%
\bibitem [{\citenamefont {Fleming}\ \emph {et~al.}(2017)\citenamefont
  {Fleming}, \citenamefont {Annoni}, \citenamefont {Shah}, \citenamefont
  {Wang}, \citenamefont {Ananthan}, \citenamefont {Zhang}, \citenamefont
  {Hutchings}, \citenamefont {Wang}, \citenamefont {Chen},\ and\ \citenamefont
  {Chen}}]{fleming2017field}%
  \BibitemOpen
  \bibfield  {author} {\bibinfo {author} {\bibfnamefont {P.}~\bibnamefont
  {Fleming}}, \bibinfo {author} {\bibfnamefont {J.}~\bibnamefont {Annoni}},
  \bibinfo {author} {\bibfnamefont {J.~J.}\ \bibnamefont {Shah}}, \bibinfo
  {author} {\bibfnamefont {L.}~\bibnamefont {Wang}}, \bibinfo {author}
  {\bibfnamefont {S.}~\bibnamefont {Ananthan}}, \bibinfo {author}
  {\bibfnamefont {Z.}~\bibnamefont {Zhang}}, \bibinfo {author} {\bibfnamefont
  {K.}~\bibnamefont {Hutchings}}, \bibinfo {author} {\bibfnamefont
  {P.}~\bibnamefont {Wang}}, \bibinfo {author} {\bibfnamefont {W.}~\bibnamefont
  {Chen}}, \ and\ \bibinfo {author} {\bibfnamefont {L.}~\bibnamefont {Chen}},\
  }\bibfield  {title} {\enquote {\bibinfo {title} {Field test of wake steering
  at an offshore wind farm},}\ }\href@noop {} {\bibfield  {journal} {\bibinfo
  {journal} {Wind Energy Science}\ }\textbf {\bibinfo {volume} {2}},\ \bibinfo
  {pages} {229--239} (\bibinfo {year} {2017})}\BibitemShut {NoStop}%
\bibitem [{\citenamefont {Fleming}\ \emph {et~al.}(2019)\citenamefont
  {Fleming}, \citenamefont {King}, \citenamefont {Dykes}, \citenamefont
  {Simley}, \citenamefont {Roadman}, \citenamefont {Scholbrock}, \citenamefont
  {Murphy}, \citenamefont {Lundquist}, \citenamefont {Moriarty}, \citenamefont
  {Fleming} \emph {et~al.}}]{fleming2019initial}%
  \BibitemOpen
  \bibfield  {author} {\bibinfo {author} {\bibfnamefont {P.}~\bibnamefont
  {Fleming}}, \bibinfo {author} {\bibfnamefont {J.}~\bibnamefont {King}},
  \bibinfo {author} {\bibfnamefont {K.}~\bibnamefont {Dykes}}, \bibinfo
  {author} {\bibfnamefont {E.}~\bibnamefont {Simley}}, \bibinfo {author}
  {\bibfnamefont {J.}~\bibnamefont {Roadman}}, \bibinfo {author} {\bibfnamefont
  {A.}~\bibnamefont {Scholbrock}}, \bibinfo {author} {\bibfnamefont
  {P.}~\bibnamefont {Murphy}}, \bibinfo {author} {\bibfnamefont {J.~K.}\
  \bibnamefont {Lundquist}}, \bibinfo {author} {\bibfnamefont {P.}~\bibnamefont
  {Moriarty}}, \bibinfo {author} {\bibfnamefont {K.}~\bibnamefont {Fleming}},
  \emph {et~al.},\ }\bibfield  {title} {\enquote {\bibinfo {title} {Initial
  results from a field campaign of wake steering applied at a commercial wind
  farm--part 1},}\ }\href@noop {} {\bibfield  {journal} {\bibinfo  {journal}
  {Wind Energy Science}\ }\textbf {\bibinfo {volume} {4}} (\bibinfo {year}
  {2019})}\BibitemShut {NoStop}%
\bibitem [{\citenamefont {Howland}, \citenamefont {Lele},\ and\ \citenamefont
  {Dabiri}(2019)}]{howland2019wind}%
  \BibitemOpen
  \bibfield  {author} {\bibinfo {author} {\bibfnamefont {M.~F.}\ \bibnamefont
  {Howland}}, \bibinfo {author} {\bibfnamefont {S.~K.}\ \bibnamefont {Lele}}, \
  and\ \bibinfo {author} {\bibfnamefont {J.~O.}\ \bibnamefont {Dabiri}},\
  }\bibfield  {title} {\enquote {\bibinfo {title} {Wind farm power optimization
  through wake steering},}\ }\href@noop {} {\bibfield  {journal} {\bibinfo
  {journal} {Proceedings of the National Academy of Sciences}\ }\textbf
  {\bibinfo {volume} {116}},\ \bibinfo {pages} {14495--14500} (\bibinfo {year}
  {2019})}\BibitemShut {NoStop}%
\bibitem [{\citenamefont {Doekemeijer}\ \emph {et~al.}(2021)\citenamefont
  {Doekemeijer}, \citenamefont {Kern}, \citenamefont {Maturu}, \citenamefont
  {Kanev}, \citenamefont {Salbert}, \citenamefont {Schreiber}, \citenamefont
  {Campagnolo}, \citenamefont {Bottasso}, \citenamefont {Schuler},
  \citenamefont {Wilts} \emph {et~al.}}]{doekemeijer2021field}%
  \BibitemOpen
  \bibfield  {author} {\bibinfo {author} {\bibfnamefont {B.~M.}\ \bibnamefont
  {Doekemeijer}}, \bibinfo {author} {\bibfnamefont {S.}~\bibnamefont {Kern}},
  \bibinfo {author} {\bibfnamefont {S.}~\bibnamefont {Maturu}}, \bibinfo
  {author} {\bibfnamefont {S.}~\bibnamefont {Kanev}}, \bibinfo {author}
  {\bibfnamefont {B.}~\bibnamefont {Salbert}}, \bibinfo {author} {\bibfnamefont
  {J.}~\bibnamefont {Schreiber}}, \bibinfo {author} {\bibfnamefont
  {F.}~\bibnamefont {Campagnolo}}, \bibinfo {author} {\bibfnamefont {C.~L.}\
  \bibnamefont {Bottasso}}, \bibinfo {author} {\bibfnamefont {S.}~\bibnamefont
  {Schuler}}, \bibinfo {author} {\bibfnamefont {F.}~\bibnamefont {Wilts}},
  \emph {et~al.},\ }\bibfield  {title} {\enquote {\bibinfo {title} {Field
  experiment for open-loop yaw-based wake steering at a commercial onshore wind
  farm in italy},}\ }\href@noop {} {\bibfield  {journal} {\bibinfo  {journal}
  {Wind Energy Science}\ }\textbf {\bibinfo {volume} {6}},\ \bibinfo {pages}
  {159--176} (\bibinfo {year} {2021})}\BibitemShut {NoStop}%
\bibitem [{\citenamefont {van Wingerden}\ \emph {et~al.}(2020)\citenamefont
  {van Wingerden}, \citenamefont {Fleming}, \citenamefont {G{\"o}{\c{c}}men},
  \citenamefont {Eguinoa}, \citenamefont {Doekemeijer}, \citenamefont {Dykes},
  \citenamefont {Lawson}, \citenamefont {Simley}, \citenamefont {King},
  \citenamefont {Astrain} \emph {et~al.}}]{van2020expert}%
  \BibitemOpen
  \bibfield  {author} {\bibinfo {author} {\bibfnamefont {J.}~\bibnamefont {van
  Wingerden}}, \bibinfo {author} {\bibfnamefont {P.}~\bibnamefont {Fleming}},
  \bibinfo {author} {\bibfnamefont {T.}~\bibnamefont {G{\"o}{\c{c}}men}},
  \bibinfo {author} {\bibfnamefont {I.}~\bibnamefont {Eguinoa}}, \bibinfo
  {author} {\bibfnamefont {B.}~\bibnamefont {Doekemeijer}}, \bibinfo {author}
  {\bibfnamefont {K.}~\bibnamefont {Dykes}}, \bibinfo {author} {\bibfnamefont
  {M.}~\bibnamefont {Lawson}}, \bibinfo {author} {\bibfnamefont
  {E.}~\bibnamefont {Simley}}, \bibinfo {author} {\bibfnamefont
  {J.}~\bibnamefont {King}}, \bibinfo {author} {\bibfnamefont {D.}~\bibnamefont
  {Astrain}},  \emph {et~al.},\ }\bibfield  {title} {\enquote {\bibinfo {title}
  {Expert elicitation on wind farm control},}\ }\href@noop {} {\bibfield
  {journal} {\bibinfo  {journal} {arXiv preprint arXiv:2006.07598}\ } (\bibinfo
  {year} {2020})}\BibitemShut {NoStop}%
\bibitem [{\citenamefont {Wyngaard}(2010)}]{wyngaard2010turbulence}%
  \BibitemOpen
  \bibfield  {author} {\bibinfo {author} {\bibfnamefont {J.~C.}\ \bibnamefont
  {Wyngaard}},\ }\href@noop {} {\emph {\bibinfo {title} {Turbulence in the
  Atmosphere}}}\ (\bibinfo  {publisher} {Cambridge University Press},\ \bibinfo
  {year} {2010})\BibitemShut {NoStop}%
\bibitem [{\citenamefont {Quick}\ \emph {et~al.}(2017)\citenamefont {Quick},
  \citenamefont {Annoni}, \citenamefont {King}, \citenamefont {Dykes},
  \citenamefont {Fleming},\ and\ \citenamefont {Ning}}]{quick2017optimization}%
  \BibitemOpen
  \bibfield  {author} {\bibinfo {author} {\bibfnamefont {J.}~\bibnamefont
  {Quick}}, \bibinfo {author} {\bibfnamefont {J.}~\bibnamefont {Annoni}},
  \bibinfo {author} {\bibfnamefont {R.}~\bibnamefont {King}}, \bibinfo {author}
  {\bibfnamefont {K.}~\bibnamefont {Dykes}}, \bibinfo {author} {\bibfnamefont
  {P.}~\bibnamefont {Fleming}}, \ and\ \bibinfo {author} {\bibfnamefont
  {A.}~\bibnamefont {Ning}},\ }\bibfield  {title} {\enquote {\bibinfo {title}
  {Optimization under uncertainty for wake steering strategies},}\ }in\
  \href@noop {} {\emph {\bibinfo {booktitle} {J. Phys. Conf. Ser}}},\ Vol.\
  \bibinfo {volume} {854}\ (\bibinfo {year} {2017})\ p.\ \bibinfo {pages}
  {012036}\BibitemShut {NoStop}%
\bibitem [{\citenamefont {Campagnolo}\ \emph {et~al.}(2020)\citenamefont
  {Campagnolo}, \citenamefont {Weber}, \citenamefont {Schreiber},\ and\
  \citenamefont {Bottasso}}]{campagnolo2020wind}%
  \BibitemOpen
  \bibfield  {author} {\bibinfo {author} {\bibfnamefont {F.}~\bibnamefont
  {Campagnolo}}, \bibinfo {author} {\bibfnamefont {R.}~\bibnamefont {Weber}},
  \bibinfo {author} {\bibfnamefont {J.}~\bibnamefont {Schreiber}}, \ and\
  \bibinfo {author} {\bibfnamefont {C.~L.}\ \bibnamefont {Bottasso}},\
  }\bibfield  {title} {\enquote {\bibinfo {title} {Wind tunnel testing of wake
  steering with dynamic wind direction changes},}\ }\href@noop {} {\bibfield
  {journal} {\bibinfo  {journal} {Wind Energy Science}\ }\textbf {\bibinfo
  {volume} {5}},\ \bibinfo {pages} {1273--1295} (\bibinfo {year}
  {2020})}\BibitemShut {NoStop}%
\bibitem [{\citenamefont {Johnson}, \citenamefont {Fingersh},\ and\
  \citenamefont {Wright}(2005)}]{johnson2005controls}%
  \BibitemOpen
  \bibfield  {author} {\bibinfo {author} {\bibfnamefont {K.}~\bibnamefont
  {Johnson}}, \bibinfo {author} {\bibfnamefont {L.~J.}\ \bibnamefont
  {Fingersh}}, \ and\ \bibinfo {author} {\bibfnamefont {A.}~\bibnamefont
  {Wright}},\ }\href@noop {} {\enquote {\bibinfo {title} {Controls advanced
  research turbine: lessons learned during advanced controls testing},}\
  }\bibinfo {type} {Tech. Rep.}\ (\bibinfo  {institution} {National Renewable
  Energy Lab., Golden, CO (US)},\ \bibinfo {year} {2005})\BibitemShut {NoStop}%
\bibitem [{\citenamefont {Fleming}\ \emph {et~al.}(2014)\citenamefont
  {Fleming}, \citenamefont {Scholbrock}, \citenamefont {Jehu}, \citenamefont
  {Davoust}, \citenamefont {Osler}, \citenamefont {Wright},\ and\ \citenamefont
  {Clifton}}]{fleming2014field}%
  \BibitemOpen
  \bibfield  {author} {\bibinfo {author} {\bibfnamefont {P.}~\bibnamefont
  {Fleming}}, \bibinfo {author} {\bibfnamefont {A.}~\bibnamefont {Scholbrock}},
  \bibinfo {author} {\bibfnamefont {A.}~\bibnamefont {Jehu}}, \bibinfo {author}
  {\bibfnamefont {S.}~\bibnamefont {Davoust}}, \bibinfo {author} {\bibfnamefont
  {E.}~\bibnamefont {Osler}}, \bibinfo {author} {\bibfnamefont {A.~D.}\
  \bibnamefont {Wright}}, \ and\ \bibinfo {author} {\bibfnamefont
  {A.}~\bibnamefont {Clifton}},\ }\bibfield  {title} {\enquote {\bibinfo
  {title} {Field-test results using a nacelle-mounted lidar for improving wind
  turbine power capture by reducing yaw misalignment},}\ }in\ \href@noop {}
  {\emph {\bibinfo {booktitle} {Journal of Physics: Conference Series}}},\
  Vol.\ \bibinfo {volume} {524}\ (\bibinfo {organization} {IOP Publishing},\
  \bibinfo {year} {2014})\BibitemShut {NoStop}%
\bibitem [{\citenamefont {Rott}\ \emph {et~al.}(2018)\citenamefont {Rott},
  \citenamefont {Doekemeijer}, \citenamefont {Seifert}, \citenamefont
  {Wingerden},\ and\ \citenamefont {K{\"u}hn}}]{rott2018robust}%
  \BibitemOpen
  \bibfield  {author} {\bibinfo {author} {\bibfnamefont {A.}~\bibnamefont
  {Rott}}, \bibinfo {author} {\bibfnamefont {B.}~\bibnamefont {Doekemeijer}},
  \bibinfo {author} {\bibfnamefont {J.~K.}\ \bibnamefont {Seifert}}, \bibinfo
  {author} {\bibfnamefont {J.-W.~v.}\ \bibnamefont {Wingerden}}, \ and\
  \bibinfo {author} {\bibfnamefont {M.}~\bibnamefont {K{\"u}hn}},\ }\bibfield
  {title} {\enquote {\bibinfo {title} {Robust active wake control in
  consideration of wind direction variability and uncertainty},}\ }\href@noop
  {} {\bibfield  {journal} {\bibinfo  {journal} {Wind energy science}\ }\textbf
  {\bibinfo {volume} {3}},\ \bibinfo {pages} {869--882} (\bibinfo {year}
  {2018})}\BibitemShut {NoStop}%
\bibitem [{\citenamefont {Simley}, \citenamefont {Fleming},\ and\ \citenamefont
  {King}(2020)}]{simley2020design}%
  \BibitemOpen
  \bibfield  {author} {\bibinfo {author} {\bibfnamefont {E.}~\bibnamefont
  {Simley}}, \bibinfo {author} {\bibfnamefont {P.}~\bibnamefont {Fleming}}, \
  and\ \bibinfo {author} {\bibfnamefont {J.}~\bibnamefont {King}},\ }\bibfield
  {title} {\enquote {\bibinfo {title} {Design and analysis of a wake steering
  controller with wind direction variability},}\ }\href@noop {} {\bibfield
  {journal} {\bibinfo  {journal} {Wind Energy Science}\ }\textbf {\bibinfo
  {volume} {5}},\ \bibinfo {pages} {451--468} (\bibinfo {year}
  {2020})}\BibitemShut {NoStop}%
\bibitem [{\citenamefont {Quick}\ \emph {et~al.}(2020)\citenamefont {Quick},
  \citenamefont {King}, \citenamefont {King}, \citenamefont {Hamlington},\ and\
  \citenamefont {Dykes}}]{quick2020wake}%
  \BibitemOpen
  \bibfield  {author} {\bibinfo {author} {\bibfnamefont {J.}~\bibnamefont
  {Quick}}, \bibinfo {author} {\bibfnamefont {J.}~\bibnamefont {King}},
  \bibinfo {author} {\bibfnamefont {R.~N.}\ \bibnamefont {King}}, \bibinfo
  {author} {\bibfnamefont {P.~E.}\ \bibnamefont {Hamlington}}, \ and\ \bibinfo
  {author} {\bibfnamefont {K.}~\bibnamefont {Dykes}},\ }\bibfield  {title}
  {\enquote {\bibinfo {title} {Wake steering optimization under uncertainty},}\
  }\href@noop {} {\bibfield  {journal} {\bibinfo  {journal} {Wind Energy
  Science}\ }\textbf {\bibinfo {volume} {5}},\ \bibinfo {pages} {413--426}
  (\bibinfo {year} {2020})}\BibitemShut {NoStop}%
\bibitem [{\citenamefont {Niayifar}\ and\ \citenamefont
  {Port{\'e}-Agel}(2016)}]{niayifar2016analytical}%
  \BibitemOpen
  \bibfield  {author} {\bibinfo {author} {\bibfnamefont {A.}~\bibnamefont
  {Niayifar}}\ and\ \bibinfo {author} {\bibfnamefont {F.}~\bibnamefont
  {Port{\'e}-Agel}},\ }\bibfield  {title} {\enquote {\bibinfo {title}
  {Analytical modeling of wind farms: A new approach for power prediction},}\
  }\href@noop {} {\bibfield  {journal} {\bibinfo  {journal} {Energies}\
  }\textbf {\bibinfo {volume} {9}},\ \bibinfo {pages} {741} (\bibinfo {year}
  {2016})}\BibitemShut {NoStop}%
\bibitem [{\citenamefont {Zhan}, \citenamefont {Letizia},\ and\ \citenamefont
  {Iungo}(2020)}]{zhan2020optimal}%
  \BibitemOpen
  \bibfield  {author} {\bibinfo {author} {\bibfnamefont {L.}~\bibnamefont
  {Zhan}}, \bibinfo {author} {\bibfnamefont {S.}~\bibnamefont {Letizia}}, \
  and\ \bibinfo {author} {\bibfnamefont {G.~V.}\ \bibnamefont {Iungo}},\
  }\bibfield  {title} {\enquote {\bibinfo {title} {Optimal tuning of
  engineering wake models through lidar measurements},}\ }\href@noop {}
  {\bibfield  {journal} {\bibinfo  {journal} {Wind Energy Science}\ }\textbf
  {\bibinfo {volume} {5}},\ \bibinfo {pages} {1601--1622} (\bibinfo {year}
  {2020})}\BibitemShut {NoStop}%
\bibitem [{\citenamefont {Doekemeijer}, \citenamefont {Van~Wingerden},\ and\
  \citenamefont {Fleming}(2019)}]{doekemeijer2019tutorial}%
  \BibitemOpen
  \bibfield  {author} {\bibinfo {author} {\bibfnamefont {B.~M.}\ \bibnamefont
  {Doekemeijer}}, \bibinfo {author} {\bibfnamefont {J.-W.}\ \bibnamefont
  {Van~Wingerden}}, \ and\ \bibinfo {author} {\bibfnamefont {P.~A.}\
  \bibnamefont {Fleming}},\ }\bibfield  {title} {\enquote {\bibinfo {title} {A
  tutorial on the synthesis and validation of a closed-loop wind farm
  controller using a steady-state surrogate model},}\ }in\ \href@noop {} {\emph
  {\bibinfo {booktitle} {2019 American Control Conference (ACC)}}}\ (\bibinfo
  {organization} {IEEE},\ \bibinfo {year} {2019})\ pp.\ \bibinfo {pages}
  {2825--2836}\BibitemShut {NoStop}%
\bibitem [{\citenamefont {Teng}\ and\ \citenamefont
  {Markfort}(2020)}]{teng2020calibration}%
  \BibitemOpen
  \bibfield  {author} {\bibinfo {author} {\bibfnamefont {J.}~\bibnamefont
  {Teng}}\ and\ \bibinfo {author} {\bibfnamefont {C.~D.}\ \bibnamefont
  {Markfort}},\ }\bibfield  {title} {\enquote {\bibinfo {title} {A calibration
  procedure for an analytical wake model using wind farm operational data},}\
  }\href@noop {} {\bibfield  {journal} {\bibinfo  {journal} {Energies}\
  }\textbf {\bibinfo {volume} {13}},\ \bibinfo {pages} {3537} (\bibinfo {year}
  {2020})}\BibitemShut {NoStop}%
\bibitem [{\citenamefont {Howland}\ and\ \citenamefont
  {Dabiri}(2019)}]{howland2019wake}%
  \BibitemOpen
  \bibfield  {author} {\bibinfo {author} {\bibfnamefont {M.~F.}\ \bibnamefont
  {Howland}}\ and\ \bibinfo {author} {\bibfnamefont {J.~O.}\ \bibnamefont
  {Dabiri}},\ }\bibfield  {title} {\enquote {\bibinfo {title} {Wind farm
  modeling with interpretable physics-informed machine learning},}\ }\href@noop
  {} {\bibfield  {journal} {\bibinfo  {journal} {Energies}\ }\textbf {\bibinfo
  {volume} {12}},\ \bibinfo {pages} {2716} (\bibinfo {year}
  {2019})}\BibitemShut {NoStop}%
\bibitem [{\citenamefont {Howland}\ \emph
  {et~al.}(2020{\natexlab{a}})\citenamefont {Howland}, \citenamefont {Ghate},
  \citenamefont {Lele},\ and\ \citenamefont {Dabiri}}]{howland2020optimal}%
  \BibitemOpen
  \bibfield  {author} {\bibinfo {author} {\bibfnamefont {M.~F.}\ \bibnamefont
  {Howland}}, \bibinfo {author} {\bibfnamefont {A.~S.}\ \bibnamefont {Ghate}},
  \bibinfo {author} {\bibfnamefont {S.~K.}\ \bibnamefont {Lele}}, \ and\
  \bibinfo {author} {\bibfnamefont {J.~O.}\ \bibnamefont {Dabiri}},\ }\bibfield
   {title} {\enquote {\bibinfo {title} {Optimal closed-loop wake steering--part
  1: Conventionally neutral atmospheric boundary layer conditions},}\
  }\href@noop {} {\bibfield  {journal} {\bibinfo  {journal} {Wind Energy
  Science}\ }\textbf {\bibinfo {volume} {5}},\ \bibinfo {pages} {1315--1338}
  (\bibinfo {year} {2020}{\natexlab{a}})}\BibitemShut {NoStop}%
\bibitem [{\citenamefont {Zhang}\ and\ \citenamefont
  {Zhao}(2020)}]{zhang2020quantification}%
  \BibitemOpen
  \bibfield  {author} {\bibinfo {author} {\bibfnamefont {J.}~\bibnamefont
  {Zhang}}\ and\ \bibinfo {author} {\bibfnamefont {X.}~\bibnamefont {Zhao}},\
  }\bibfield  {title} {\enquote {\bibinfo {title} {Quantification of parameter
  uncertainty in wind farm wake modeling},}\ }\href@noop {} {\bibfield
  {journal} {\bibinfo  {journal} {Energy}\ }\textbf {\bibinfo {volume} {196}},\
  \bibinfo {pages} {117065} (\bibinfo {year} {2020})}\BibitemShut {NoStop}%
\bibitem [{\citenamefont {Schreiber}\ \emph {et~al.}(2020)\citenamefont
  {Schreiber}, \citenamefont {Bottasso}, \citenamefont {Salbert},\ and\
  \citenamefont {Campagnolo}}]{schreiber2020improving}%
  \BibitemOpen
  \bibfield  {author} {\bibinfo {author} {\bibfnamefont {J.}~\bibnamefont
  {Schreiber}}, \bibinfo {author} {\bibfnamefont {C.~L.}\ \bibnamefont
  {Bottasso}}, \bibinfo {author} {\bibfnamefont {B.}~\bibnamefont {Salbert}}, \
  and\ \bibinfo {author} {\bibfnamefont {F.}~\bibnamefont {Campagnolo}},\
  }\bibfield  {title} {\enquote {\bibinfo {title} {Improving wind farm flow
  models by learning from operational data},}\ }\href@noop {} {\bibfield
  {journal} {\bibinfo  {journal} {Wind Energy Science}\ }\textbf {\bibinfo
  {volume} {5}},\ \bibinfo {pages} {647--673} (\bibinfo {year}
  {2020})}\BibitemShut {NoStop}%
\bibitem [{\citenamefont {Birge}\ and\ \citenamefont
  {Louveaux}(2011)}]{birge2011introduction}%
  \BibitemOpen
  \bibfield  {author} {\bibinfo {author} {\bibfnamefont {J.~R.}\ \bibnamefont
  {Birge}}\ and\ \bibinfo {author} {\bibfnamefont {F.}~\bibnamefont
  {Louveaux}},\ }\href@noop {} {\emph {\bibinfo {title} {Introduction to
  stochastic programming}}}\ (\bibinfo  {publisher} {Springer Science \&
  Business Media},\ \bibinfo {year} {2011})\BibitemShut {NoStop}%
\bibitem [{\citenamefont {Brooks}\ \emph {et~al.}(2011)\citenamefont {Brooks},
  \citenamefont {Gelman}, \citenamefont {Jones},\ and\ \citenamefont
  {Meng}}]{brooks2011handbook}%
  \BibitemOpen
  \bibfield  {author} {\bibinfo {author} {\bibfnamefont {S.}~\bibnamefont
  {Brooks}}, \bibinfo {author} {\bibfnamefont {A.}~\bibnamefont {Gelman}},
  \bibinfo {author} {\bibfnamefont {G.}~\bibnamefont {Jones}}, \ and\ \bibinfo
  {author} {\bibfnamefont {X.-L.}\ \bibnamefont {Meng}},\ }\href@noop {} {\emph
  {\bibinfo {title} {Handbook of markov chain monte carlo}}}\ (\bibinfo
  {publisher} {CRC press},\ \bibinfo {year} {2011})\BibitemShut {NoStop}%
\bibitem [{\citenamefont {Shapiro}, \citenamefont {Gayme},\ and\ \citenamefont
  {Meneveau}(2018)}]{shapiro2018modelling}%
  \BibitemOpen
  \bibfield  {author} {\bibinfo {author} {\bibfnamefont {C.~R.}\ \bibnamefont
  {Shapiro}}, \bibinfo {author} {\bibfnamefont {D.~F.}\ \bibnamefont {Gayme}},
  \ and\ \bibinfo {author} {\bibfnamefont {C.}~\bibnamefont {Meneveau}},\
  }\bibfield  {title} {\enquote {\bibinfo {title} {Modelling yawed wind turbine
  wakes: a lifting line approach},}\ }\href@noop {} {\bibfield  {journal}
  {\bibinfo  {journal} {J. Fluid Mech.}\ }\textbf {\bibinfo {volume} {841}}
  (\bibinfo {year} {2018})}\BibitemShut {NoStop}%
\bibitem [{\citenamefont {Howland}\ and\ \citenamefont
  {Dabiri}(2021)}]{howland2021influence}%
  \BibitemOpen
  \bibfield  {author} {\bibinfo {author} {\bibfnamefont {M.~F.}\ \bibnamefont
  {Howland}}\ and\ \bibinfo {author} {\bibfnamefont {J.~O.}\ \bibnamefont
  {Dabiri}},\ }\bibfield  {title} {\enquote {\bibinfo {title} {Influence of
  wake model superposition and secondary steering on model-based wake steering
  control with {SCADA} data assimilation},}\ }\href@noop {} {\bibfield
  {journal} {\bibinfo  {journal} {Energies}\ }\textbf {\bibinfo {volume}
  {14}},\ \bibinfo {pages} {52} (\bibinfo {year} {2021})}\BibitemShut {NoStop}%
\bibitem [{\citenamefont {Zong}\ and\ \citenamefont
  {Port{\'e}-Agel}(2020)}]{zong2020momentum}%
  \BibitemOpen
  \bibfield  {author} {\bibinfo {author} {\bibfnamefont {H.}~\bibnamefont
  {Zong}}\ and\ \bibinfo {author} {\bibfnamefont {F.}~\bibnamefont
  {Port{\'e}-Agel}},\ }\bibfield  {title} {\enquote {\bibinfo {title} {A
  momentum-conserving wake superposition method for wind farm power
  prediction},}\ }\href@noop {} {\bibfield  {journal} {\bibinfo  {journal}
  {Journal of Fluid Mechanics}\ }\textbf {\bibinfo {volume} {889}} (\bibinfo
  {year} {2020})}\BibitemShut {NoStop}%
\bibitem [{\citenamefont {Howland}\ \emph
  {et~al.}(2020{\natexlab{b}})\citenamefont {Howland}, \citenamefont
  {Gonz{\'a}lez}, \citenamefont {Mart{\'\i}nez}, \citenamefont {Quesada},
  \citenamefont {Larranaga}, \citenamefont {Yadav}, \citenamefont {Chawla},\
  and\ \citenamefont {Dabiri}}]{howland2020yaw}%
  \BibitemOpen
  \bibfield  {author} {\bibinfo {author} {\bibfnamefont {M.~F.}\ \bibnamefont
  {Howland}}, \bibinfo {author} {\bibfnamefont {C.~M.}\ \bibnamefont
  {Gonz{\'a}lez}}, \bibinfo {author} {\bibfnamefont {J.~J.~P.}\ \bibnamefont
  {Mart{\'\i}nez}}, \bibinfo {author} {\bibfnamefont {J.~B.}\ \bibnamefont
  {Quesada}}, \bibinfo {author} {\bibfnamefont {F.~P.}\ \bibnamefont
  {Larranaga}}, \bibinfo {author} {\bibfnamefont {N.~K.}\ \bibnamefont
  {Yadav}}, \bibinfo {author} {\bibfnamefont {J.~S.}\ \bibnamefont {Chawla}}, \
  and\ \bibinfo {author} {\bibfnamefont {J.~O.}\ \bibnamefont {Dabiri}},\
  }\bibfield  {title} {\enquote {\bibinfo {title} {Influence of atmospheric
  conditions on the power production of utility-scale wind turbines in yaw
  misalignment},}\ }\href@noop {} {\bibfield  {journal} {\bibinfo  {journal}
  {Journal of Renewable and Sustainable Energy}\ }\textbf {\bibinfo {volume}
  {12}},\ \bibinfo {pages} {063307} (\bibinfo {year}
  {2020}{\natexlab{b}})}\BibitemShut {NoStop}%
\bibitem [{\citenamefont {Bossuyt}\ \emph {et~al.}(2017)\citenamefont
  {Bossuyt}, \citenamefont {Howland}, \citenamefont {Meneveau},\ and\
  \citenamefont {Meyers}}]{bossuyt2017measurement}%
  \BibitemOpen
  \bibfield  {author} {\bibinfo {author} {\bibfnamefont {J.}~\bibnamefont
  {Bossuyt}}, \bibinfo {author} {\bibfnamefont {M.~F.}\ \bibnamefont
  {Howland}}, \bibinfo {author} {\bibfnamefont {C.}~\bibnamefont {Meneveau}}, \
  and\ \bibinfo {author} {\bibfnamefont {J.}~\bibnamefont {Meyers}},\
  }\bibfield  {title} {\enquote {\bibinfo {title} {Measurement of unsteady
  loading and power output variability in a micro wind farm model in a wind
  tunnel},}\ }\href@noop {} {\bibfield  {journal} {\bibinfo  {journal} {Exp.
  Fluids}\ }\textbf {\bibinfo {volume} {58}},\ \bibinfo {pages} {1} (\bibinfo
  {year} {2017})}\BibitemShut {NoStop}%
\bibitem [{\citenamefont {Evensen}(2003)}]{evensen2003ensemble}%
  \BibitemOpen
  \bibfield  {author} {\bibinfo {author} {\bibfnamefont {G.}~\bibnamefont
  {Evensen}},\ }\bibfield  {title} {\enquote {\bibinfo {title} {The ensemble
  kalman filter: Theoretical formulation and practical implementation},}\
  }\href@noop {} {\bibfield  {journal} {\bibinfo  {journal} {Ocean dynamics}\
  }\textbf {\bibinfo {volume} {53}},\ \bibinfo {pages} {343--367} (\bibinfo
  {year} {2003})}\BibitemShut {NoStop}%
\bibitem [{\citenamefont {Shapiro}\ \emph {et~al.}(2019)\citenamefont
  {Shapiro}, \citenamefont {Starke}, \citenamefont {Meneveau},\ and\
  \citenamefont {Gayme}}]{shapiro2019wake}%
  \BibitemOpen
  \bibfield  {author} {\bibinfo {author} {\bibfnamefont {C.~R.}\ \bibnamefont
  {Shapiro}}, \bibinfo {author} {\bibfnamefont {G.~M.}\ \bibnamefont {Starke}},
  \bibinfo {author} {\bibfnamefont {C.}~\bibnamefont {Meneveau}}, \ and\
  \bibinfo {author} {\bibfnamefont {D.~F.}\ \bibnamefont {Gayme}},\ }\bibfield
  {title} {\enquote {\bibinfo {title} {A wake modeling paradigm for wind farm
  design and control},}\ }\href@noop {} {\bibfield  {journal} {\bibinfo
  {journal} {Energies}\ }\textbf {\bibinfo {volume} {12}},\ \bibinfo {pages}
  {2956} (\bibinfo {year} {2019})}\BibitemShut {NoStop}%
\bibitem [{\citenamefont {Doekemeijer}\ \emph {et~al.}(2017)\citenamefont
  {Doekemeijer}, \citenamefont {Boersma}, \citenamefont {Pao},\ and\
  \citenamefont {van Wingerden}}]{doekemeijer2017ensemble}%
  \BibitemOpen
  \bibfield  {author} {\bibinfo {author} {\bibfnamefont {B.}~\bibnamefont
  {Doekemeijer}}, \bibinfo {author} {\bibfnamefont {S.}~\bibnamefont
  {Boersma}}, \bibinfo {author} {\bibfnamefont {L.~Y.}\ \bibnamefont {Pao}}, \
  and\ \bibinfo {author} {\bibfnamefont {J.-W.}\ \bibnamefont {van
  Wingerden}},\ }\bibfield  {title} {\enquote {\bibinfo {title} {Ensemble
  kalman filtering for wind field estimation in wind farms},}\ }in\ \href@noop
  {} {\emph {\bibinfo {booktitle} {2017 American Control Conference (ACC)}}}\
  (\bibinfo {organization} {IEEE},\ \bibinfo {year} {2017})\ pp.\ \bibinfo
  {pages} {19--24}\BibitemShut {NoStop}%
\bibitem [{\citenamefont {Schillings}\ and\ \citenamefont
  {Stuart}(2017)}]{schillings2017analysis}%
  \BibitemOpen
  \bibfield  {author} {\bibinfo {author} {\bibfnamefont {C.}~\bibnamefont
  {Schillings}}\ and\ \bibinfo {author} {\bibfnamefont {A.~M.}\ \bibnamefont
  {Stuart}},\ }\bibfield  {title} {\enquote {\bibinfo {title} {Analysis of the
  ensemble kalman filter for inverse problems},}\ }\href@noop {} {\bibfield
  {journal} {\bibinfo  {journal} {SIAM Journal on Numerical Analysis}\ }\textbf
  {\bibinfo {volume} {55}},\ \bibinfo {pages} {1264--1290} (\bibinfo {year}
  {2017})}\BibitemShut {NoStop}%
\bibitem [{\citenamefont {Stevens}, \citenamefont {Gayme},\ and\ \citenamefont
  {Meneveau}(2015)}]{stevens2015coupled}%
  \BibitemOpen
  \bibfield  {author} {\bibinfo {author} {\bibfnamefont {R.~J.}\ \bibnamefont
  {Stevens}}, \bibinfo {author} {\bibfnamefont {D.~F.}\ \bibnamefont {Gayme}},
  \ and\ \bibinfo {author} {\bibfnamefont {C.}~\bibnamefont {Meneveau}},\
  }\bibfield  {title} {\enquote {\bibinfo {title} {Coupled wake boundary layer
  model of wind-farms},}\ }\href@noop {} {\bibfield  {journal} {\bibinfo
  {journal} {J. Renew. Sustain. Energy}\ }\textbf {\bibinfo {volume} {7}},\
  \bibinfo {pages} {023115} (\bibinfo {year} {2015})}\BibitemShut {NoStop}%
\bibitem [{\citenamefont {Starke}\ \emph {et~al.}(2020)\citenamefont {Starke},
  \citenamefont {Meneveau}, \citenamefont {King},\ and\ \citenamefont
  {Gayme}}]{starke2020area}%
  \BibitemOpen
  \bibfield  {author} {\bibinfo {author} {\bibfnamefont {G.~M.}\ \bibnamefont
  {Starke}}, \bibinfo {author} {\bibfnamefont {C.}~\bibnamefont {Meneveau}},
  \bibinfo {author} {\bibfnamefont {J.~R.}\ \bibnamefont {King}}, \ and\
  \bibinfo {author} {\bibfnamefont {D.~F.}\ \bibnamefont {Gayme}},\ }\bibfield
  {title} {\enquote {\bibinfo {title} {The area localized coupled model for
  analytical mean flow prediction in arbitrary wind farm geometries},}\
  }\href@noop {} {\bibfield  {journal} {\bibinfo  {journal} {arXiv preprint
  arXiv:2009.13666}\ } (\bibinfo {year} {2020})}\BibitemShut {NoStop}%
\bibitem [{\citenamefont {Doekemeijer}, \citenamefont {van~der Hoek},\ and\
  \citenamefont {van Wingerden}(2020)}]{doekemeijer2020closed}%
  \BibitemOpen
  \bibfield  {author} {\bibinfo {author} {\bibfnamefont {B.~M.}\ \bibnamefont
  {Doekemeijer}}, \bibinfo {author} {\bibfnamefont {D.}~\bibnamefont {van~der
  Hoek}}, \ and\ \bibinfo {author} {\bibfnamefont {J.-W.}\ \bibnamefont {van
  Wingerden}},\ }\bibfield  {title} {\enquote {\bibinfo {title} {Closed-loop
  model-based wind farm control using {FLORIS} under time-varying inflow
  conditions},}\ }\href@noop {} {\bibfield  {journal} {\bibinfo  {journal}
  {Renewable Energy}\ }\textbf {\bibinfo {volume} {156}},\ \bibinfo {pages}
  {719--730} (\bibinfo {year} {2020})}\BibitemShut {NoStop}%
\bibitem [{\citenamefont {Ghate}\ and\ \citenamefont
  {Lele}(2017)}]{ghate2017subfilter}%
  \BibitemOpen
  \bibfield  {author} {\bibinfo {author} {\bibfnamefont {A.~S.}\ \bibnamefont
  {Ghate}}\ and\ \bibinfo {author} {\bibfnamefont {S.~K.}\ \bibnamefont
  {Lele}},\ }\bibfield  {title} {\enquote {\bibinfo {title} {Subfilter-scale
  enrichment of planetary boundary layer large eddy simulation using discrete
  fourier--gabor modes},}\ }\href@noop {} {\bibfield  {journal} {\bibinfo
  {journal} {J. Fluid Mech.}\ }\textbf {\bibinfo {volume} {819}},\ \bibinfo
  {pages} {494--539} (\bibinfo {year} {2017})}\BibitemShut {NoStop}%
\bibitem [{\citenamefont {Howland}, \citenamefont {Ghate},\ and\ \citenamefont
  {Lele}(2018)}]{howland2018influence}%
  \BibitemOpen
  \bibfield  {author} {\bibinfo {author} {\bibfnamefont {M.~F.}\ \bibnamefont
  {Howland}}, \bibinfo {author} {\bibfnamefont {A.~S.}\ \bibnamefont {Ghate}},
  \ and\ \bibinfo {author} {\bibfnamefont {S.~K.}\ \bibnamefont {Lele}},\
  }\bibfield  {title} {\enquote {\bibinfo {title} {Influence of the horizontal
  component of earth’s rotation on wind turbine wakes},}\ }in\ \href@noop {}
  {\emph {\bibinfo {booktitle} {J. of Phys.: Conf. Series}}},\ Vol.\ \bibinfo
  {volume} {1037}\ (\bibinfo {organization} {IOP Publishing},\ \bibinfo {year}
  {2018})\ p.\ \bibinfo {pages} {072003}\BibitemShut {NoStop}%
\bibitem [{\citenamefont {Howland}, \citenamefont {Ghate},\ and\ \citenamefont
  {Lele}(2020{\natexlab{a}})}]{howland2020influence}%
  \BibitemOpen
  \bibfield  {author} {\bibinfo {author} {\bibfnamefont {M.~F.}\ \bibnamefont
  {Howland}}, \bibinfo {author} {\bibfnamefont {A.~S.}\ \bibnamefont {Ghate}},
  \ and\ \bibinfo {author} {\bibfnamefont {S.~K.}\ \bibnamefont {Lele}},\
  }\bibfield  {title} {\enquote {\bibinfo {title} {Influence of the geostrophic
  wind direction on the atmospheric boundary layer flow},}\ }\href@noop {}
  {\bibfield  {journal} {\bibinfo  {journal} {J. Fluid Mech.}\ }\textbf
  {\bibinfo {volume} {883}} (\bibinfo {year} {2020}{\natexlab{a}})}\BibitemShut
  {NoStop}%
\bibitem [{\citenamefont {Howland}, \citenamefont {Ghate},\ and\ \citenamefont
  {Lele}(2020{\natexlab{b}})}]{howland2020coriolis}%
  \BibitemOpen
  \bibfield  {author} {\bibinfo {author} {\bibfnamefont {M.~F.}\ \bibnamefont
  {Howland}}, \bibinfo {author} {\bibfnamefont {A.~S.}\ \bibnamefont {Ghate}},
  \ and\ \bibinfo {author} {\bibfnamefont {S.~K.}\ \bibnamefont {Lele}},\
  }\bibfield  {title} {\enquote {\bibinfo {title} {Coriolis effects within and
  trailing a large finite wind farm},}\ }in\ \href@noop {} {\emph {\bibinfo
  {booktitle} {AIAA Scitech 2020 Forum}}}\ (\bibinfo {year} {2020})\ p.\
  \bibinfo {pages} {0994}\BibitemShut {NoStop}%
\bibitem [{\citenamefont {Ghaisas}, \citenamefont {Ghate},\ and\ \citenamefont
  {Lele}(2020)}]{ghaisas2020effect}%
  \BibitemOpen
  \bibfield  {author} {\bibinfo {author} {\bibfnamefont {N.~S.}\ \bibnamefont
  {Ghaisas}}, \bibinfo {author} {\bibfnamefont {A.~S.}\ \bibnamefont {Ghate}},
  \ and\ \bibinfo {author} {\bibfnamefont {S.~K.}\ \bibnamefont {Lele}},\
  }\bibfield  {title} {\enquote {\bibinfo {title} {Effect of tip spacing,
  thrust coefficient and turbine spacing in multi-rotor wind turbines and
  farms},}\ }\href@noop {} {\bibfield  {journal} {\bibinfo  {journal} {Wind
  Energy Science}\ }\textbf {\bibinfo {volume} {5}},\ \bibinfo {pages} {51--72}
  (\bibinfo {year} {2020})}\BibitemShut {NoStop}%
\bibitem [{\citenamefont {Kumar}\ \emph {et~al.}(2006)\citenamefont {Kumar},
  \citenamefont {Kleissl}, \citenamefont {Meneveau},\ and\ \citenamefont
  {Parlange}}]{kumar2006large}%
  \BibitemOpen
  \bibfield  {author} {\bibinfo {author} {\bibfnamefont {V.}~\bibnamefont
  {Kumar}}, \bibinfo {author} {\bibfnamefont {J.}~\bibnamefont {Kleissl}},
  \bibinfo {author} {\bibfnamefont {C.}~\bibnamefont {Meneveau}}, \ and\
  \bibinfo {author} {\bibfnamefont {M.~B.}\ \bibnamefont {Parlange}},\
  }\bibfield  {title} {\enquote {\bibinfo {title} {Large-eddy simulation of a
  diurnal cycle of the atmospheric boundary layer: Atmospheric stability and
  scaling issues},}\ }\href@noop {} {\bibfield  {journal} {\bibinfo  {journal}
  {Water Resources Research}\ }\textbf {\bibinfo {volume} {42}} (\bibinfo
  {year} {2006})}\BibitemShut {NoStop}%
\bibitem [{\citenamefont {Calaf}, \citenamefont {Meneveau},\ and\ \citenamefont
  {Meyers}(2010)}]{calaf2010large}%
  \BibitemOpen
  \bibfield  {author} {\bibinfo {author} {\bibfnamefont {M.}~\bibnamefont
  {Calaf}}, \bibinfo {author} {\bibfnamefont {C.}~\bibnamefont {Meneveau}}, \
  and\ \bibinfo {author} {\bibfnamefont {J.}~\bibnamefont {Meyers}},\
  }\bibfield  {title} {\enquote {\bibinfo {title} {Large eddy simulation study
  of fully developed wind-turbine array boundary layers},}\ }\href@noop {}
  {\bibfield  {journal} {\bibinfo  {journal} {Phys. Fluids}\ }\textbf {\bibinfo
  {volume} {22}},\ \bibinfo {pages} {015110} (\bibinfo {year}
  {2010})}\BibitemShut {NoStop}%
\bibitem [{\citenamefont {Munters}, \citenamefont {Meneveau},\ and\
  \citenamefont {Meyers}(2016)}]{munters2016turbulent}%
  \BibitemOpen
  \bibfield  {author} {\bibinfo {author} {\bibfnamefont {W.}~\bibnamefont
  {Munters}}, \bibinfo {author} {\bibfnamefont {C.}~\bibnamefont {Meneveau}}, \
  and\ \bibinfo {author} {\bibfnamefont {J.}~\bibnamefont {Meyers}},\
  }\bibfield  {title} {\enquote {\bibinfo {title} {Turbulent inflow precursor
  method with time-varying direction for large-eddy simulations and
  applications to wind farms},}\ }\href@noop {} {\bibfield  {journal} {\bibinfo
   {journal} {Boundary-Layer Meteorol.}\ }\textbf {\bibinfo {volume} {159}},\
  \bibinfo {pages} {305--328} (\bibinfo {year} {2016})}\BibitemShut {NoStop}%
\bibitem [{\citenamefont {Ghate}\ \emph {et~al.}(2018)\citenamefont {Ghate},
  \citenamefont {Ghaisas}, \citenamefont {Lele},\ and\ \citenamefont
  {Towne}}]{ghate2018interaction}%
  \BibitemOpen
  \bibfield  {author} {\bibinfo {author} {\bibfnamefont {A.~S.}\ \bibnamefont
  {Ghate}}, \bibinfo {author} {\bibfnamefont {N.}~\bibnamefont {Ghaisas}},
  \bibinfo {author} {\bibfnamefont {S.~K.}\ \bibnamefont {Lele}}, \ and\
  \bibinfo {author} {\bibfnamefont {A.}~\bibnamefont {Towne}},\ }\bibfield
  {title} {\enquote {\bibinfo {title} {Interaction of small scale homogenenous
  isotropic turbulence with an actuator disk},}\ }in\ \href@noop {} {\emph
  {\bibinfo {booktitle} {2018 Wind Energy Symposium}}}\ (\bibinfo {year}
  {2018})\ p.\ \bibinfo {pages} {0753}\BibitemShut {NoStop}%
\bibitem [{\citenamefont {Kanev}(2020)}]{kanev2020dynamic}%
  \BibitemOpen
  \bibfield  {author} {\bibinfo {author} {\bibfnamefont {S.}~\bibnamefont
  {Kanev}},\ }\bibfield  {title} {\enquote {\bibinfo {title} {Dynamic wake
  steering and its impact on wind farm power production and yaw actuator
  duty},}\ }\href@noop {} {\bibfield  {journal} {\bibinfo  {journal} {Renewable
  Energy}\ }\textbf {\bibinfo {volume} {146}},\ \bibinfo {pages} {9--15}
  (\bibinfo {year} {2020})}\BibitemShut {NoStop}%
\bibitem [{\citenamefont {Wharton}\ and\ \citenamefont
  {Lundquist}(2012)}]{wharton2012atmospheric}%
  \BibitemOpen
  \bibfield  {author} {\bibinfo {author} {\bibfnamefont {S.}~\bibnamefont
  {Wharton}}\ and\ \bibinfo {author} {\bibfnamefont {J.~K.}\ \bibnamefont
  {Lundquist}},\ }\bibfield  {title} {\enquote {\bibinfo {title} {Atmospheric
  stability affects wind turbine power collection},}\ }\href@noop {} {\bibfield
   {journal} {\bibinfo  {journal} {Environmental Research Letters}\ }\textbf
  {\bibinfo {volume} {7}},\ \bibinfo {pages} {014005} (\bibinfo {year}
  {2012})}\BibitemShut {NoStop}%
\end{thebibliography}%

\end{document}